\documentclass[aps,twocolumn,showpacs,preprintnumbers,nofootinbib,prd,superscriptaddress,groupedaddress,10pt]{revtex4-1}

\makeatletter
\def\l@subsubsection#1#2{}
\def\l@subsubsubsection#1#2{}
\makeatother

\usepackage{amsmath}
\usepackage{graphicx,amssymb,amsmath,amsthm,amsfonts,epsfig,epsf}
\usepackage[usenames,dvipsnames,svgnames,table]{xcolor}
\usepackage{epstopdf}
\definecolor{darkred}{rgb}{0.5,0,0}
\usepackage{aas_macros}
\usepackage{bm}
\usepackage{dcolumn}
\usepackage[utf8]{inputenc}
\usepackage{latexsym}
\usepackage{rotating}
\usepackage{longtable}

\usepackage{enumerate}
\usepackage{tensor,multirow}
\usepackage{url}
\usepackage[linktocpage,hidelinks]{hyperref}

\def\nn{\nonumber}

\def\be{\begin{equation}}
\def\ee{\end{equation}}
\newcommand{\beq}{\begin{eqnarray}}
\newcommand{\eeq}{\end{eqnarray}}

\newcommand{\nocontentsline}[3]{}
\newcommand{\tocless}[2]{\bgroup\let\addcontentsline=\nocontentsline#1{#2}\egroup}
\def\ba{\begin{align}}
\def\ea{\end{align}}

\newcommand{\warn}[1]{{\textcolor{red}{\sf{[IN PROGRESS]}} }}

\begin{document}

\title{The response of ultralight dark matter to supermassive black holes and binaries\\
}

\author{
Lorenzo Annulli$^{1}$,
Vitor Cardoso$^{1}$,
Rodrigo Vicente$^{1}$
}

\affiliation{${^1}$
Centro de Astrof\'{\i}sica e Gravita\c c\~ao - CENTRA, Departamento de F\'{\i}sica, Instituto
  Superior T\'ecnico - IST, Universidade de Lisboa - UL, Avenida Rovisco Pais 1, 1049-001 Lisboa, Portugal}

\begin{abstract}
Scalar fields can give rise to confined structures, such as boson stars or Q-balls. These objects are interesting hypothetical new ``dark matter stars,'' but also good descriptions of dark matter haloes when the fields are ultralight.
Here, we study the dynamical response of such confined bosonic structures when excited by external matter (stars, planets or black holes) in their vicinities. Such perturbers can either be plunging through the bosonic configuration or undergoing periodic motion around its center. 
Our setup can also efficiently describe the interaction between a moving, massive black hole and the surrounding environment. 
It also depicts dark matter depletion as a reaction to an inspiralling binary within the halo.
We calculate total energy loss, and linear and angular momenta radiated during these processes, and perform the first self-consistent calculation of dynamical friction acting on moving bodies in these backgrounds. We show that the gravitational collapse to a supermassive black hole
at the center of a Newtonian boson star (NBS) is accompanied by a small change in the surrounding core. The NBS eventually gets accreted, but only on times larger than a Hubble scale
for astrophysical parameters.
Stellar or supermassive binaries are able to ``stir'' the NBS and lead to scalar radiation.
For binaries in the LIGO or LISA band, close to coalescence, scalar emission affects the waveform at leading $-6$ PN order with respect to the dominant quadrupolar term; the coefficient is too small to allow detection by next-generation interferometers. 
Our results provide a complete picture of the interaction between black holes or stars and the ultralight dark matter environment they live in. 
\end{abstract}

\maketitle

\tableofcontents

\section{Introduction}\label{Introduction}
The existence, stability and dynamical behavior of ``objects'' in a given theory is relevant for a wide range of topics, from planetary science to a description of fundamental particles. 
Taking as starting point a theory of a scalar field in flat space, it can be shown that localized time-independent solutions cannot exist~\cite{Derrick:1964ww}. This powerful result limits the ability of fundamental scalars to describe possible novel objects where the scalar is confined. A promising way to circumvent such no-go result is to consider time-dependent fields.
Within this more general framework, it can be shown that black holes (BHs) can stimulate the growth of structures in their vicinities~\cite{Herdeiro:2014goa,Brito:2015oca},
and that new self-gravitating solutions are possible. Such objects can describe dark stars which have so far gone undetected~\cite{Barack:2018yly,Cardoso:2019rvt,Giudice:2016zpa,Ellis:2017jgp}. Surprisingly, the simplest such solutions also seem to be a good description of structures we know to exist: dark matter (DM) cores in haloes. These are often referred to as fuzzy DM models, and require ultralight bosonic fields (we refer the reader to Refs.~\cite{Robles:2012uy,Hui:2016ltb,Bar:2019bqz,Bar:2018acw,Desjacques:2019zhf,Davoudiasl:2019nlo}, but the literature on the subject is very large and growing).

In this work, we consider two different theories of scalar fields, yielding localized objects with a static energy-density profile, but with a time-periodic scalar. The first theory describes a self-gravitating massive scalar, and the resulting objects are known as boson stars~\cite{Kaup:1968zz,Ruffini:1969qy,Liebling:2012fv}. Newtonian boson stars (NBS) made of very light fields (in particular, bosons with a mass~$\sim 10^{-22}\,{\rm eV}$) are good descriptions of most cores of DM haloes; thus, this is an especially exciting simple theory to consider. 
The second theory describes a nonlinearly-interacting scalar in flat space, yielding solutions known as Q-balls: non-topological solitons which arise in a large family of field theories admitting a conserved charge $Q$, associated with some continuous internal symmetry~\cite{Coleman:1985ki}. Q-balls are not particularly well motivated as a DM candidate, but serve as an additional example of a scalar configuration to which our formalism can be directly applied.

%

\noindent{\bf Stirring-up DM.}
The study of the dynamics of such objects is interesting for a number of reasons. As DM candidates, it is important to understand the stability of such configurations, and the way they interact with surrounding bodies (stars, BHs, etc)~\cite{Macedo:2013qea,Khlopov:1985}. For example, the mere {\it presence} of a star or planet will change the local DM density. In which way?
The motion of a compact binary can, in principle, stir the surrounding DM to such an extent that a substantial emission of scalars takes place. How much, and how is it dependent on the binary parameters? 
When a star crosses one of these extended bosonic configurations, it may change its properties to the extent that the configuration simply collapses or disperses; in the eventuality that it settles down to a new configuration, it is important to understand the timescales involved. Such processes are specially interesting in the context of the growth of DM haloes and supermassive BHs. Baryonic matter, in fact, tends to slowly accumulate near the center of a DM structure, where it may eventually collapse to a massive BH. Gravitational collapse can impart a recoil velocity $v_{\rm recoil}$ to the BH of the order of $300\,{\rm km/s}$~\cite{1973ApJ...183..657B}, leaving the BH in an damped oscillatory motion through the DM halo, with respect to its center, with a crossing timescale
%
\be
\tau_{\rm cross}=\sqrt{\frac{3\pi}{G\rho}}\sim 1.4\times 10^6 \,{\rm yr}\sqrt{\frac{10^3M_{\odot}\,{\rm pc}^{-3}}{\rho}}\,,
\ee
and an amplitude
\be
{\cal A}\sim 69\,{\rm pc} \sqrt{\frac{10^3M_{\odot}{\rm pc}^{-3}}{\rho}}\frac{v_{\rm recoil}}{300 \, {\rm km/s}}\,.
\ee
The damping is due to dynamical friction caused by stars and DM; our results suggest that the DM effects may be comparable to the one of stars in galactic cores.
Finally, massive objects traveling through scalar media can deposit energy and momentum in the surrounding scalar field due to gravitational interaction~\cite{Hui:2016ltb,Bernard:2019nkv,Cardoso:2019dte}.
Thus, it is important to quantify the gravitational drag that bodies are subjected to when immersed in scalar structures, and to confirm existing estimates~\cite{Hui:2016ltb}. 

All of this applies also in the context where scalar structures are viewed as compact, and potentially strong, gravitational-wave (GW) sources, when they could mimic BHs, or simply be new sources on their own right~\cite{Cardoso:2019rvt,Palenzuela:2017kcg}. Additionally, we expect some of these findings to be also valid in theories with a massive vector or tensor.

\noindent{\bf Gravitational-wave astronomy and DM.}
Understanding the behavior of DM when moving perturbers drift by, or when a binary inspirals within a DM medium
is crucial for attempts at detecting DM via GWs. In the presence of a nontrivial environment accretion, gravitational drag and the self-gravity of the medium all contribute to a small, but potentially observable, change of the GW phase~\cite{Eda:2013gg,Macedo:2013qea,Barausse:2014tra,Hannuksela:2018izj,Cardoso:2019rou,Baumann:2019ztm,Kavanagh:2020cfn}. Understanding the backreaction on the environment seems to be one crucial ingredient in this endeavour, at least for equal-mass mergers and when the Compton wavelength of DM is very small~\cite{Kavanagh:2020cfn}.

\noindent{\bf Screening mechanisms.} Our results and methods can be of direct interest also for theories with screening mechanisms, where new degrees of freedom -- usually scalars --
are screened, via nonlinearities, on some scales~\cite{Babichev:2013usa}. Such mechanisms do give rise to nontrivial profiles for the new degrees of freedom,
for which many of the tools we use here should apply (see also Ref.~\cite{Brito:2014ifa}).

\begin{figure}	
\includegraphics[width=8.3cm,keepaspectratio]{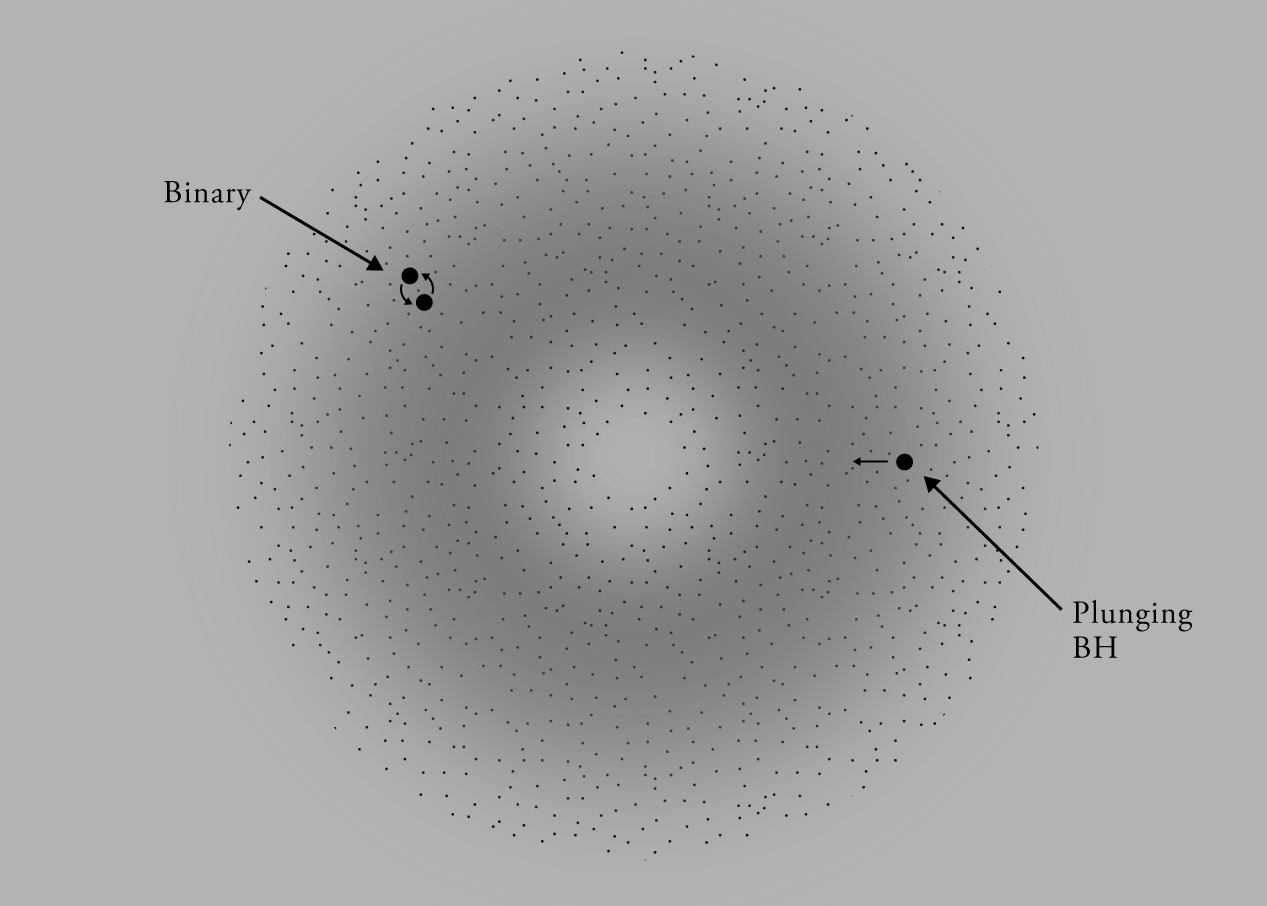} 
\caption{An equatorial slice of our setup, where a binary of two BHs or stars is orbiting inside an NBS, and a single BH is plunging through it.
Our formalism is able to accommodate both scenarios, and others.
The NBS scalar field is pictured in gray dots, and forms a large spherical configuration. The motion of the binary or of the plunging BH or star stirs the scalar profile, excites the NBS modes and may
eject some scalar field. All these quantities are computed in the main body of this work.}\label{fig:anatomy}
\end{figure}
Here, we wish to provide the answers to these questions. 
This work studies the response of localized scalar configurations to bodies moving in their vicinities. The setup is depicted in Fig.~\ref{fig:anatomy}. The moving external bodies are modelled as point-like.
Such approximation is a standard and successful tool in BH perturbation theory~\cite{Zerilli:1971wd,Davis:1971gg,Barack:2018yvs}, in seismology~\cite{Ari} or in calculations of gravitational drag by fluids~\cite{Ostriker:1998fa,Vicente:2019ilr}. In this approximation one loses
small-scale information. 
For light fields -- those we focus on -- the Compton wavelength of the field is much larger
than the size of stars, planets or BHs. In other words, we do not expect to lose important details of the physics at play. The extrapolation of our results to moving BHs or BH binaries should yield sensible answers.

A summary of our findings is reported in a recent Letter~\cite{Annulli:2020ilw}.
For readers wishing to skip the technical details, our main results are described there, and also discussed in Sections~\ref{sec_sitting_bs}-\ref{BS_binaries} for NBSs, which are being advocated as good descriptions of the solitonic cores of galaxies and in Sections~\ref{sec:Plunging_particle}-\ref{sec:Orbiting_particle} for Q-balls (no gravity).
We will use units where the speed of light, Newton's constant and reduced Planck's constant are all set to unity, $c=G=\hbar=1$. 

\section{Framework}\label{Setup}
\subsection{The theory}\label{The theory}
We consider a general $U(1)$-invariant, self-interacting, complex scalar field $\Phi(x^\mu)$ minimally coupled to gravity described by the action
\be
\mathcal{S}\equiv \int d^4x \sqrt{-g}\left(\frac{R}{16 \pi}-\frac{1}{2} g^{\mu\nu}\partial_{\mu}\Phi\partial_{\nu}\Phi^*-\mathcal{U}\right)\,,\label{action}
\ee
where $R$ is the Ricci scalar of the spacetime metric $g_{\mu \nu}$, $g\equiv \det(g_{\mu \nu})$ is the metric determinant, and $\mathcal{U}(|\Phi|^2)$ is a real-valued, $U(1)$-invariant, self-interaction potential. 
For a weak scalar field $|\Phi| \ll 1$, the self-interaction potential is $\mathcal{U}\sim \mu^2|\Phi|^2/2+{\cal O} (|\Phi|^4)$, where $\mu$ is the scalar field mass. Our methods are applicable, in principle, to any nonlinear potential.

By virtue of Noether's theorem, this theory admits the conserved current
\be
j_\mu =- \frac{i}{2}\left(\Phi^*\partial_\mu \Phi-\Phi\partial_\mu \Phi^*\right)\,, \label{NoetherCurrent}
\ee
and the associated conserved charge
\be 
Q = -\int d^3x \sqrt{h} \,j_t\,,\label{NoetherCharge}
\ee
where the last integration is performed over a spacelike hypersurface of constant time coordinate $t$, with $h\equiv \det(g_{\mu \nu})$ the determinant of the induced metric $h_{\mu \nu}=g_{\mu \nu}-\delta_\mu^0 \delta_\nu^0$. We shall interpret this charge as the number of bosonic particles in the system. 

The scalar field stress-energy tensor is
\begin{equation}\label{StressEnergy}
T^S_{\mu \nu }=\partial_{(\mu}\Phi^* \partial_{\nu)}\Phi-\frac{1}{2}g_{\mu \nu}\left[\partial_\alpha \Phi^* \partial^\alpha \Phi +2\mathcal{U}(|\Phi|^2) \right] \,,
\end{equation}
and its energy within some spatial region at an instant $t$ is given by
\begin{equation} \label{Energy}
E= \int d^3x \sqrt{h}\, T^S_{t t}\,.
\end{equation}
%

\subsection{The objects}\label{The objects}
We are interested in spherically symmetric, time-periodic, localized solutions of the field equations. These will be describing, for example, new DM stars or the core of DM halos.
We take the following ansatz for the scalar in such a configuration,
\begin{equation} 
\Phi_0=\Psi_0(r)e^{-i\Omega t}\,,\label{BKG_ansatz}
\end{equation}
where $\Psi_0$ is a real-function satisfying $\partial_r \Psi_0(0)=0$ and $\lim_{r\to \infty} \Psi_0=0$. 

Our primary target are self-gravitating solutions. When gravity is included, a simple minimally coupled massive field is able to self-gravitate. Thus, we consider minimal boson stars -- self-gravitating configurations of scalar field in curved spacetime with a simple mass term potential
\begin{equation} 
\mathcal{U}_{\rm NBS}=\frac{\mu^2}{2}|\Phi|^2\,.\label{Potential_BS}
\end{equation}
In this work, for simplicity, we restrict to the Newtonian limit of these objects, where gravity is not very strong. So, we study NBSs.

However, many of the technical issues of dealing with NBS are present as well in a simple theory in Minkowski background.
Thus, we will also consider Q-balls~\cite{Coleman:1985ki} -- objects made of a nonlinearly-interacting scalar field in {\it flat space}. For these objects, we use the Minkowski spacetime metric $\eta_{\mu \nu}$ and restrict to the class of nonlinear potentials 
\begin{equation} 
\mathcal{U}_{\rm Q}=\frac{\mu^2}{2}|\Phi|^2\left(1-\frac{|\Phi|^2}{\Phi_c^2}\right)^2\,,\label{Potential_Qball}
\end{equation}
where $\Phi_c$ is a real free parameter of the theory. 


We are ultimately interested not in the objects {\it per se}, but rather on their dynamical response to external agents.
The response to external perturbers is taken into account, by linearizing against the spherically symmetric, stationary background,
\begin{equation} 
\Phi=\left[\Psi_0(r)+\delta \Psi(t,r,\theta, \varphi)\right] e^{-i \Omega t}\,,\label{Perturbation}
\end{equation}
with the assumption $|\delta \Psi|\ll 1$, where $\Psi_0$ is the radial profile of the unperturbed object, and $\theta$ and $\varphi$ are coordinates used to parametrize the 2-sphere. Then, the perturbation $\delta \Psi$ allows us to obtain all the physical quantities of interest, like the modes of vibration of the object, and the energy, linear and angular momenta radiated in a given process. This approach has a range of validity, $|\delta \Psi|\ll 1$, which can be controlled by selecting the perturber. As we show below, $\delta \Psi \propto m_p \mu$, where $m_p$ is the rest mass or a mass-related parameter of the external perturber. Since our results scale simply with $m_p$, it is always possible to find an external source whose induced dynamics always fall in our perturbative scheme.

For a generic point-like perturber, the stress-energy tensor is given by
\begin{equation}
T_p^{\mu \nu}=m_p \frac{u^\mu u^\nu}{u^0} \frac{\delta\left(r-r_p(t)\right)}{r^2}\frac{\delta\left(\theta-\theta_p(t)\right)}{\sin\theta} \delta\left(\varphi-\varphi_p(t)\right) \,,\label{Stress_energy_particle}
\end{equation}
where $u^\mu\equiv dx_p^\mu/d\tau$ is the perturber's 4-velocity and $x_p^\mu(t)=(t,r_p(t),\theta_p(t),\varphi_p(t))$ a parametrization of its worldline in spherical coordinates. 
\subsection{The fluxes}\label{The fluxes}
The energy, linear and angular momenta contained in the radiated scalar can be obtained by computing the flux of certain currents through a 2-sphere at infinity. These currents are derived from the stress-energy tensor of the scalar. Since we are not aware of literature where such important quantities are shown or derived for scalar fields, we present them below.

First, we decompose the fluctuations as
\begin{equation} 
\delta \Psi=\sum_{l,m}\int \frac{d \omega}{\sqrt{2 \pi} r} \left[Z_1^{\omega l m }Y_l^m e^{-i\omega t}+\left(Z_2^{\omega l m }\right)^*\left(Y_l^m\right)^* e^{i\omega t}\right]\label{Decomposition}
\end{equation}
where $Y_l^m(\theta,\varphi)$ is the spherical harmonic function of degree $l$ and order $m$, and $Z_1(r)$ and $Z_2(r)$ are radial complex-functions.~\footnote{It should be noted that~$Z_1$ and~$Z_2$ are not linearly independent. In particular, for the setups considered in this work, we find~$Z_1(\omega,l,m;r)=(-1)^mZ_2(-\omega,l,-m;r)^*$. For generality, we do not impose any constraint on the relation between these functions.} This decomposition can be rewritten in the equivalent form
\begin{align} 
\delta \Psi&=\sum_{l,m}\int \frac{d \omega}{\sqrt{2 \pi} r}Y_l^m e^{-i \omega t} \nonumber\\
&\times\left[Z_1(\omega,l,m;r)+(-1)^m Z_2(-\omega,l,-m;r)^*\right]\,.\label{Decompositionv2}
\end{align}
Unless strictly needed, hereafter, we omit the labels $\omega$, $l$ and $m$ in the functions $Z_1^{\omega l m}(r)$ and $Z_2^{\omega l m}(r)$ to simplify the notation.
For a source vanishing at spatial infinity, we will see that one has the asymptotic fields
\beq 
Z_1(r \to \infty) &\sim& Z_1^\infty e^{i \epsilon_1 \left(\sqrt{\left(\omega+\Omega\right)^2-\mu^2}\right) r}\nonumber\,,\\
Z_2(r \to \infty) &\sim& Z_2^\infty e^{i \epsilon_2 \left(\sqrt{\left(\omega-\Omega\right)^2-\mu^2}\right)^* r}\,,\label{Asymptotics}
\eeq
where $\epsilon_1\equiv \text{sign}(\omega+\Omega+ \mu)$ and $\epsilon_2\equiv \text{sign}(\omega-\Omega-\mu)$, and $Z_1^\infty$ and $Z_2^\infty$ are complex amplitudes which depend on the source. We choose the signs~$\epsilon_1$ and~$\epsilon_2$ to enforce the Sommerfeld radiation condition at large distances.~\footnote{By Sommerfeld condition we mean either: (i) outgoing group velocity for propagating frequencies; or, (ii) regularity for bounded frequencies.}

Scalar field fluctuations cause a perturbation to its stress-energy tensor, which, at leading order and asymptotically, is given by
\beq
\delta T_{\mu \nu}^S(r\to \infty)&\sim&\partial_{(\mu}\delta \Phi^* \partial_{\nu)}\delta \Phi\nonumber\\
&-&\frac{1}{2}\eta_{\mu \nu}\left[\partial_\alpha \delta \Phi^* \partial^\alpha \delta \Phi +\mu^2|\delta \Phi|^2 \right] \,,
\eeq
with $\delta \Phi \equiv e^{-i \Omega t} \delta \Psi$.
Then, the (outgoing) flux of energy at an instant $t$ through a 2-sphere at infinity is
\be
\dot{E}^{\rm rad}= \lim_{r\to \infty} r^2 \int d\theta d\varphi \sin \theta\, \delta T_{r \mu}^S \xi_t^\mu  \,,\label{energy}
\ee
with the timelike Killing vector field $\boldsymbol{\xi}_t= -\partial_t$.
Plugging the asymptotic fields~\eqref{Asymptotics} in the last expression, it is straightforward to show that the total energy radiated with frequency in the range between $\omega$ and $\omega+d\omega$ is
\begin{align}
&\frac{dE^{\rm rad}}{d\omega}=\left|\omega+\Omega\right| {\rm Re}\left[\sqrt{(\omega+\Omega)^2-\mu^2}\right]  \nonumber \\
&\times\sum_{l,m}\left|Z_1^{\infty}(\omega,l,m)+(-1)^m Z_2^{\infty}\left(-\omega,l,-m\right)^*\right|^2\,.\label{Energy_flux}
\end{align}
In deriving the last expression we considered a process in which the small perturber interacts with the background configuration during a finite amount of time. In the case of a (eternal) periodic interaction (\textit{e.g.}, small particle orbiting the scalar configuration) the energy radiated is not finite. However, we can compute the average rate of energy emission in such processes, obtaining
\begin{align} 
	&\dot{E}^{\rm rad}=\int \frac{d\omega}{2\pi}\left|\omega+\Omega\right| {\rm Re}\left[\sqrt{(\omega+\Omega)^2-\mu^2}\right]  \nonumber \\
	&\times\sum_{l,m}\left|Z_1^{\infty}(\omega,l,m)+(-1)^m Z_2^{\infty}\left(-\omega,l,-m\right)^*\right|^2\,.\label{Energy_flux_rate}
\end{align}
The last expression must be used in a formal way, because, as we will see, the amplitudes~$Z_1^\infty$ and~$Z_2^\infty$ contain Dirac delta functions in frequency $\omega$. The correct way to proceed is to substitute the product of compatible delta functions by just one of them, and the incompatible by zero.~\footnote{It is easy to do a more rigorous derivation applying the formalism directly to a specific process. For generality, we let~\eqref{Energy_flux_rate} as it is.} 

The (outgoing) flux of linear momentum at instant $t$ is
\begin{equation} 
\dot{P}^{\rm rad}_i= \lim_{r\to \infty} r^2 \int  d\theta d\varphi \sin \theta\, \delta T_{r \mu}^S e_i^\mu\,,\label{momentum}
\end{equation}
with $i=\{x,y,z\}$ and where $\boldsymbol{e}_x$, $\boldsymbol{e}_y$, $\boldsymbol{e}_z$ are unit spacelike vectors in the $x$, $y$, $z$ directions, respectively. These are given by
\beq
\boldsymbol{e}_x&=& \sin \theta \cos \varphi \, \boldsymbol{e}_r+ \frac{\cos \theta \cos\varphi}{r}\, \boldsymbol{e}_\theta -\frac{\sin \varphi}{r \sin \theta}\, \boldsymbol{e}_\varphi\,, \nn \\
\boldsymbol{e}_y&=& \sin \theta \sin \varphi \, \boldsymbol{e}_r+ \frac{\cos \theta \sin\varphi}{r}\, \boldsymbol{e}_\theta +\frac{\cos \varphi}{r \sin \theta}\, \boldsymbol{e}_\varphi\,, \nn \\
\boldsymbol{e}_z&=& \cos \theta \, \boldsymbol{e}_r- \frac{\sin \theta}{r} \, \boldsymbol{e}_\theta \nn \,,
\eeq
with $e_r^\mu=\delta_r^\mu$, $e_\theta^\mu=\delta_\theta^\mu$ and $e_\varphi^\mu=\delta_\varphi^\mu$ in spherical coordinates.
For an axially symmetric process there are only modes with azimuthal number $m=0$ composing the scalar field fluctuation~\eqref{Decomposition}. In that case, using the asymptotic fields~\eqref{Asymptotics}, one can show that the total linear momentum radiated along $z$ with frequency in the range between $\omega$ and $\omega+d\omega$ is~\footnote{Additionally, it is straightforward to show that no linear momentum is radiated along $x$ and $y$ in an axially symmetric process.}
\beq
\frac{d P_z^{\rm rad}}{d \omega}&=&\sum_l\frac{2(l+1)\Theta\left[\left(\omega+\Omega\right)^2-\mu^2\right]\left|(\omega+\Omega)^2-\mu^2\right|}{\sqrt{(2l+1)(2l+3)}} \nonumber \\
&\times&\left[\Lambda_{11}(\omega,l)+2\Lambda_{12}(\omega,l)+\Lambda_{22}(\omega,l)\right]\,,\label{Momentum_flux}
\eeq
where $\Theta(x)$ is the Heaviside step function and we defined the functions
\begin{align*}
	\Lambda_{11}(\omega,l)&\equiv{\rm Re}\Big[Z_1^\infty(\omega,l,0)Z_1^\infty(\omega,l+1,0)^*\Big]\,,\\
	\Lambda_{12}(\omega,l)&\equiv{\rm Re}\Big[Z_1^\infty(\omega,l,0)Z_2^\infty(-\omega,l+1,0)\Big]\,,\\
	\Lambda_{22}(\omega,l)&\equiv{\rm Re}\Big[Z_2^\infty(-\omega,l+1,0)Z_2^\infty(-\omega,l,0)^*\Big]\,.
\end{align*}

Finally, the (outgoing) flux of angular momentum along $z$ at instant $t$ is
\be 
\dot{L}^{\rm rad}_z = \lim_{r\to \infty} r^2 \int d\theta d\varphi \sin \theta\, \delta T_{r \mu}^S e_\varphi^\mu  \,,\label{angularm}
\ee
with the spacelike Killing vector $\boldsymbol{e}_\varphi$. Plugging the asymptotic fields~\eqref{Asymptotics} in the last expression, it can be shown that the total angular momentum along $z$ radiated with frequency in the range between $\omega$ and $\omega+d\omega$ is 
\begin{align}
&\frac{d L_z^{\rm rad}}{d\omega}={\rm Re}\left[\sqrt{(\omega+\Omega)^2-\mu^2}\right]  \nonumber \\
&\times\sum_{l,m}m\left|Z_1^{\infty}(\omega,l,m)+(-1)^m Z_2^{\infty}\left(-\omega,l,-m\right)^*\right|^2\,.\label{AngularMomentum_flux}
\end{align}
In the case of a periodic interaction, the angular momentum along~$z$ is radiated at a rate given by
\begin{align}
	&\dot{L}_z^{\rm rad}=\int \frac{d \omega}{2\pi}\,{\rm Re}\left[\sqrt{(\omega+\Omega)^2-\mu^2}\right]  \nonumber \\
	&\times\sum_{l,m}m\left|Z_1^{\infty}(\omega,l,m)+(-1)^m Z_2^{\infty}\left(-\omega,l,-m\right)^*\right|^2\,.
\label{AngularMomentum_flux_rate}
\end{align}

We can also compute how many scalar particles cross the 2-sphere at infinity per unit of time. This is obtained by
\begin{equation}
	\dot{Q}^{\rm rad}= \lim_{r\to \infty} r^2 \int d\theta d\varphi \sin \theta\, \delta j_{r}  \,,\label{numberf}
\end{equation}
with
\begin{equation}
	\delta j_r(r \to \infty)\sim {\rm Im}\left(\delta \Phi^*\partial_r \delta \Phi\right)\,,
\end{equation}
at leading order. Using the asymptotic fields~\eqref{Asymptotics}, we can show that the number of particles radiated in the range between~$\omega$ and~$\omega+d\omega$ is
\begin{align}
	&\frac{d Q^{\rm rad}}{d \omega}=\epsilon_1 {\rm Re}\left[\sqrt{(\omega+\Omega)^2-\mu^2}\right]  \nonumber \\
	&\times\sum_{l,m}\left|Z_1^{\infty}(\omega,l,m)+(-1)^m Z_2^{\infty}\left(-\omega,l,-m\right)^*\right|^2\,.\label{Particles_flux}
\end{align}
This gives us a simple interpretation for expressions~\eqref{Energy_flux} and~\eqref{AngularMomentum_flux}. The spectral flux of energy is just the product between the spectral flux of particles and their individual energy $\Omega+\omega$; similarly, the spectral flux of angular momentum matches the number of particles radiated with azimuthal number~$m$ times their individual angular momentum -- which is also~$m$. For a periodic interaction, scalar particles are radiated at an average rate
\begin{align}
	&\dot{Q}^{\rm rad}=\int \frac{d \omega}{2\pi}\,{\rm Re}\left[\sqrt{(\omega+\Omega)^2-\mu^2}\right]  \nonumber \\
	&\times\sum_{l,m}\left|Z_1^{\infty}(\omega,l,m)+(-1)^m Z_2^{\infty}\left(-\omega,l,-m\right)^*\right|^2\,.
	\label{Particles_flux_rate}
\end{align}

One may wonder what is the relation between the radiated fluxes and the energy and momenta lost by the massive perturber ($E^{\rm lost}$,~$P_z^{\rm lost}$,~$L_z^{\rm lost}$). 
Noting that both the energy and momenta of the scalar configuration may change due to the interaction, by conservation of the total energy and momenta we know that
\begin{align} \label{LossRad}
	E^{\rm lost}&=\Delta E+E^{\rm rad}\,, \nn \\
	P_z^{\rm lost}&=\Delta P_z+P_z^{\rm rad}\,, \nn \\
	L_z^{\rm lost}&=\Delta L_z+L_z^{\rm rad}\,,	
\end{align}
where $\Delta E$,~$\Delta P_z$ and~$\Delta L_z$ are the changes in the energy and momenta of the configuration.
So, if we have the radiated fluxes, determining the energy and momenta loss reduces to computing the change in the respective quantities of the scalar configuration.

In a perturbation scheme it is hard to aim at a direct calculation of these changes, because in general they include second order fluctuations of the scalar -- terms mixing~$\Phi_0$ with~$\delta^2 \Phi$; this does not concern the radiated fluxes, since~$\Phi_0$ is suppressed at infinity.
However, for certain setups we can compute indirectly the change in the configuration's energy~$\Delta E$. Let us see an example.
An object interacting with the scalar only through gravitation is described by a~$U(1)$-invariant action; so, Noether's theorem implies that
\begin{align}
	\nabla_ \mu \,\delta j^\mu=0\,, 
\end{align}
with~\footnote{The cautious reader may have noticed that we are neglecting the lower order perturbation~$\delta j^\mu=\,{\rm Im}\left(\Phi_ 0^*\partial^\mu \delta \Phi+\delta \Phi^* \partial^\mu \Phi_0\right)$. This current does not contribute to a change in the number of particles in the configuration~$\Delta Q$, because it is suppressed at large distances by the factor~$\Phi_0$ (and its derivatives). In~\eqref{NoetherPert} we are also omitting the terms involving only~$\Phi_0$, since it is easy to show that they are static and, so, do not contribute to~$\Delta Q$.}
\begin{equation}\label{NoetherPert}
	\delta j^\mu= {\rm Im}\left(\delta \Phi^*\partial^\mu \delta \Phi+\Phi_0^* \partial^\mu \delta^2 \Phi+\delta^2 \Phi^* \partial^\mu \Phi_ 0  \right)\,.
\end{equation}
Using the divergence theorem, we obtain that the number of particles is conserved,
\begin{align} \label{DeltaQ}
	\Delta Q&=-\int_{t=+\infty}d^3x \sqrt{h} \,\delta j_t+\int_{t=-\infty}d^3x \sqrt{h} \,\delta j_t \nn\\
	&=-Q^{\rm rad}\,, 
\end{align}
which means that the number of particles lost by the configuration matches the number of radiated particles -- no scalar particles are created.
If, additionally, we can express the change in the configuration's mass in terms of the change in the number of particles -- as (we will show) it happens for NBS -- we are able to compute~$\Delta M$ from the number of radiated particles~$Q^{\rm rad}$; so, we obtain the energy loss of the perturber~$E^{\rm lost}$ using only radiated fluxes. The loss of momenta~$P_z^{\rm lost}$ and~$L_z^{\rm lost}$ can, then, be obtained through the energy-momenta relations; for example, a non-relativistic perturber moving along~$z$ satisfies
\begin{align} 
E^{\rm lost}&= \frac{\left(m_p v_{\rm i}\right)^2-\left(m_p v_{\rm i}-P_z^{\rm lost}\right)^2}{2 m_p} \nn \\
&=P_z^{\rm lost} v_{\rm i}-\frac{(P_z^{\rm lost})^2}{2 m_p}\,, \label{EvsPloss}
\end{align}
where~$v_{\rm i}$ is the initial velocity along~$z$.
Finally, we can compute the change in the scalar configuration momenta~$\Delta P_z$ and~$\Delta L_z$ using~\eqref{LossRad}. 

The conservation of the number of particles (\textit{i.e,} Noether's theorem) plays a key role in our scheme; it allows us to compute the change in the number of particles -- a quantity that involves the second order fluctuation~$\delta^2\Phi$ -- using only the first order fluctuation~$\delta \Phi$. When the perturber couples directly with the scalar via a scalar interaction that breaks the~$U(1)$ symmetry -- like the coupling in~\eqref{coupling_Qball} -- the number of scalar particles is not conserved; the perturber can create and absorb particles. In that case, our scheme fails and it is not obvious how to circumvent this issue to calculate of~$\Delta M$. 
In Section~\ref{sec:SmallPert} we apply explicitly the scheme described above to compute the energy and momentum loss of an object perturbing an NBS (\textit{e.g.,} a BH binary) from the radiation that reaches infinity.

\section{Newtonian boson stars}\label{BosonS}
We start with the simplest theory of a scalar giving rise to self-gravitating objects. The theory is that of a minimally coupled massive field,
or even with higher order interactions, but taken at Newtonian level. The objects themselves -- NBSs -- have been studied for decades, either as BH mimickers, as toy models for more complicated exotica that could exist, or as realistic configurations that can describe DM~\cite{Kaup:1968zz,Ruffini:1969qy,Liebling:2012fv}. Despite the intense study and the recent activity at the numerical relativity level~\cite{Cardoso:2016oxy,Helfer:2018vtq,Palenzuela:2017kcg,Sanchis-Gual:2019ljs,Bezares:2018qwa,Sanchis-Gual:2018oui,Widdicombe:2019woy}, their interaction with smaller objects (describing, for example, stars piercing through or orbiting such NBSs) has hardly been studied. The variety and disparity of scales in the problem makes it ill-suited for full-blown numerical techniques, but ideal for perturbation theory.
\subsection{Background configurations}
%
\begin{figure}	
\includegraphics[width=8.3cm,keepaspectratio]{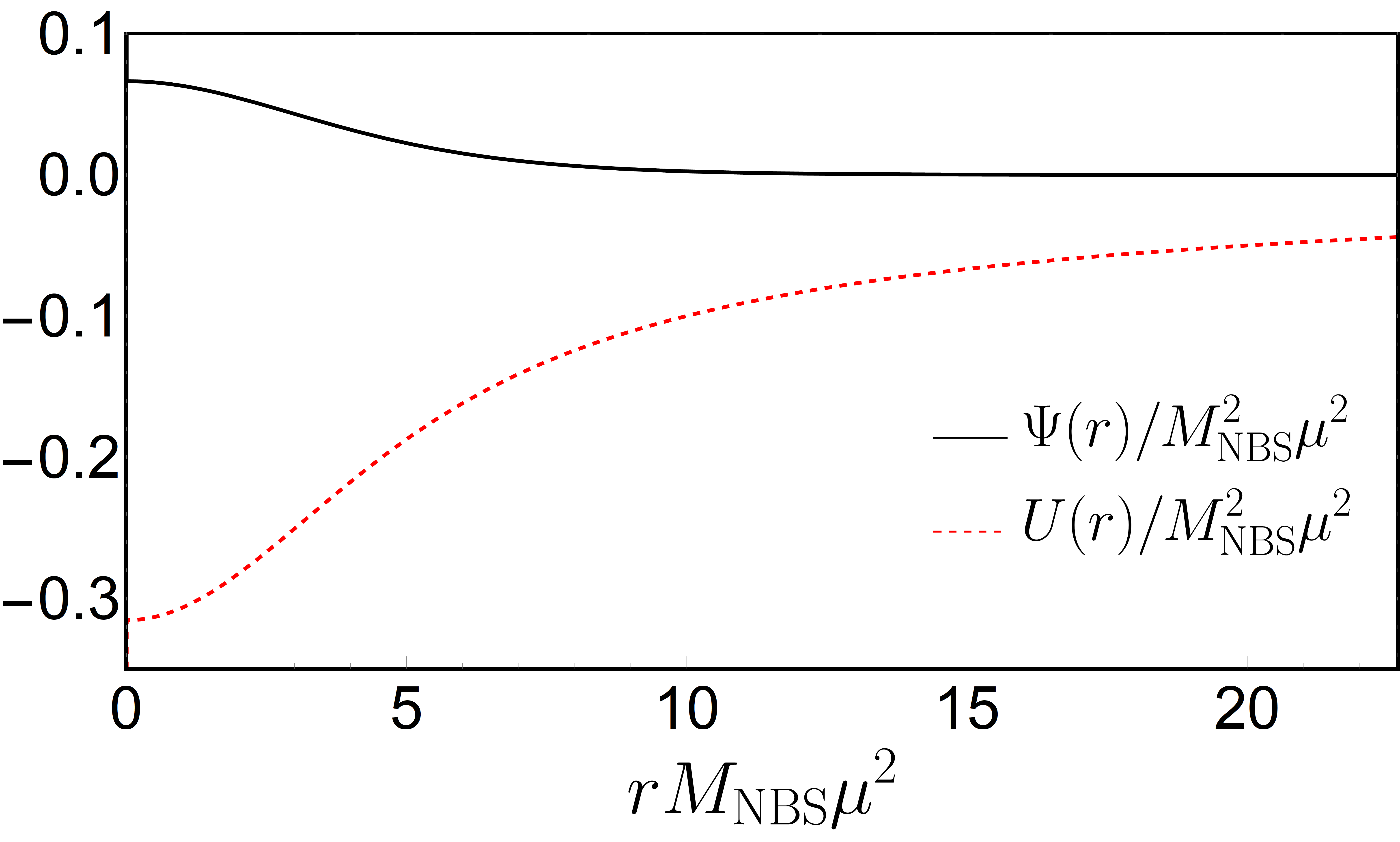} 
\caption{Universal radial profiles $\Psi(r)$ and $U(r)$ of the numerical solution of system~\eqref{EOM_BS_radial} with appropriate boundary conditions. 
Due to the scaling \eqref{eq:scaling}, this profile describes all the fundamental NBSs. They are characterized by the re-scaling invariant quantity
$\gamma/(M_{\rm NBS}^2\mu^3)\simeq0.162712$ and the mass-radius relation \eqref{eq:mass_radius_BS}.
}\label{fig:BS}
\end{figure}
The field equations for $\Phi$ and $g_{\mu \nu}$ are obtained through the variation of action~\eqref{action} with respect to $\Phi^*$ and $g_{\mu \nu}$, resulting in
\beq 
&&\frac{1}{\sqrt{-g}}\partial_{\mu}\left(\sqrt{-g}g^{\mu\nu}\partial_{\nu}\Phi\right)= \mu^2\Phi\,, \nn \\
&&R_{\mu\nu}- \frac{1}{2}R g_{\mu\nu}= 8\pi T_{\mu\nu}^S\,.\label{EOM_BosonS}
\eeq
Here, we are already using that $\mathcal{U}\sim \mu^2 \left| \Phi\right|^2/2$, since we want to consider a (Newtonian) weak scalar field $\left|\Phi\right| \ll 1$. The stress-energy tensor of the scalar $T_{\mu \nu}^S$ is given in Eq.~\eqref{StressEnergy}. We are interested in localized solutions of this model with a scalar field of the form~\eqref{BKG_ansatz}, with frequency 
\be
\Omega = \mu-\gamma\,.
\ee
in the limit $0<\gamma\ll\mu$. These are the so-called NBSs. In this case, the energy $\Omega$ of the individual scalar \textit{particles} forming the NBS is approximately given by their rest-mass energy $\mu$. In appendix~\ref{app:Newtonian} we show that, using the Newtonian spacetime metric
\begin{equation}
ds^2=-\left(1+2U\right)dt^2+dr^2+r^2\left(d\theta^2+\sin^2\theta d\varphi^2\right)\,,
\end{equation}
with a weak gravitational potential $|U(r)|\ll1$, and retaining only the leading order terms, system~\eqref{EOM_BosonS} reduces to the simpler system
\beq
i \partial_t\widetilde{\Phi}&=&-\frac{1}{2 \mu} \nabla^2 \widetilde{\Phi}+ \mu U \widetilde{\Phi}\,, \nn \\
\nabla^2 U&=& 4\pi \mu |\widetilde{\Phi}|^2\,, 
\eeq
where the Schrödinger field~$\widetilde{\Phi}$ is related with the Klein-Gordon field~$\Phi$ through 
\begin{equation}
\widetilde{\Phi}\equiv \sqrt{\mu} \, e^{i \mu t} \Phi\,.
\end{equation}
This is known as Schr\"{o}dinger-Poisson system (see, \textit{e.g.}, Ref.~\cite{Chavanis1}). 
To arrive at this description, one assumes that the scalar field $\Phi$ is non-relativistic, which implies $|\partial_t\widetilde{\Phi}| \ll \mu |\widetilde{\Phi}|$.
Using ansatz~\eqref{BKG_ansatz} for the scalar field $\Phi$, one finds 
\beq
&&\partial_r^2 \Psi+\frac{2}{r}\partial_r \Psi - 2 \mu \left(\mu U+\gamma\right) \Psi=0\,, \nn \\
&&\partial_r^2 U +\frac{2}{r} \partial_r U- 4\pi \mu^2 \Psi^2=0\,,\label{EOM_BS_radial}
\eeq
with the constraints~$0<\gamma \ll \mu$,~$|U|\ll 1$ and~$|\Psi|\ll 1$. Remarkably, this system is left invariant under the transformation
\be
(\Psi,U,\gamma) \to \lambda^2 (\Psi, U, \gamma)\,,\quad  r \to r/\lambda\,.\label{eq:scaling}
\ee
These relations imply that the NBS mass scales as $M_{\rm NBS} \to \lambda M_{\rm NBS}$ (\textit{see} Eq.~\eqref{M_NBS}).
This scale invariance is extremely useful, because it allows us to effectively ignore the constraints on $\gamma$, $U$ and $\Psi$ when solving Eq.~\eqref{EOM_BS_radial}; one can always rescale the obtained solution with a sufficiently small $\lambda$, such that the constraints (\textit{i.e.}, the Newtonian approximation) are satisfied for the rescaled solution. Even more importantly is the fact that once a fundamental (\textit{i.e.}, ground state) NBS solution is found, all other fundamental stars can be obtained through a rescaling of that solution; obviously, the same applies to any other particular excited state.

A numerical solution of system~\eqref{EOM_BS_radial}, with appropriate boundary conditions, describing all fundamental NBSs, is summarized in Fig.~\ref{fig:BS}.~\footnote{In addition to the conditions on $\Phi$ (stated in Sec.~\ref{The objects}), here we impose $\partial_r U(0)=0$ and $\lim_{r\to \infty} U=0$.} 
It is easy to see that, at large distances, the scalar decays exponentially as $\Psi \sim e^{-\sqrt{2\mu\gamma}r}/r$, whereas the Newtonian potential falls off as $-M_{\rm NBS}/r$.
Noting that the mass of an NBS is given by
\begin{equation} \label{M_NBS}
M_{\rm NBS}=4 \pi \mu^2 \int_{0}^{\infty}dr\, r^2 \left|\Psi \right|^2\,,
\end{equation}
it is possible to show that, for a fundamental NBS, 
\be
\frac{M_{\rm NBS}}{M_\odot} \simeq 3 \times 10^{12}\, \lambda\left(\frac{10^{-22}\, {\rm eV} }{ \mu}\right)\,,
\ee
with a scaling parameter $\lambda$, such that $\{\Psi,U,\gamma/\mu\} \sim \mathcal{O}(\lambda^2)$. If one is interested in describing a DM core of mass $M \sim 10^{10} M_\odot$, this can be achieved then via a fundamental NBS made of self-gravitating scalar particles of mass $\mu \sim 10^{-22} \,{\rm eV}$, with a scaling parameter $\lambda \sim  10^{-2}$, which satisfies the Newtonian constraints. 

All the fundamental NBSs satisfy the scaling-invariant mass-radius relation
\be
M_{\rm NBS}\mu=\frac{9.1}{R\mu}\,,\label{eq:mass_radius_BS}
\ee
where the NBS radius is defined as the radius of the sphere enclosing $98\%$ of its mass. This result agrees well with previous
results in the literature~\cite{Liebling:2012fv,Boskovic:2018rub,Bar:2018acw,Membrado,Chavanis1,Chavanis2}. Comparing with some relevant scales, it can be written as
\be
\frac{M_{\rm NBS}}{M_{\odot}}=9\times 10^9\,\frac{100\, {\rm pc}}{R}\,\left(\frac{10^{-22}\,{\rm eV}}{\mu}\right)^2\,.\label{eq:mass_radius_BS2}
\ee

Accurate fits for the profile of the scalar wavefunction are provided in Ref.~\cite{Kling:2017mif}. Unfortunately, these fits are defined by branches, and similar results for the gravitational potential are not discussed at length. We find that a good description of the gravitational potential of NBSs, accurate to within 1\% everywhere is the following:
\beq
U&=&\mu^2M_{\rm NBS}^2f\,,\\
f&=&\frac{a_0+11\frac{a_0}{r_1}x+\sum_{i=2}^{9}a_ix^i-x^{10}}{(x+r_1)^{11}}\,,\\
x&=&\mu^2M_{\rm NBS}r\,,r_1=1.288\,,\nonumber\\
a_0&=&-5.132\,,a_2=-143.279\,,a_3=-645.326\,,\nonumber\\
a_4&=&277.921\,, a_5=-2024.838\,,a_6=476.702\,,\nonumber\\
a_7&=&-549.051,\, a_8=-90.244\,,a_9=-13.734\,.
\eeq
The (cumbersome) functional form was chosen such that it yields the correct large-$r$ behavior,
and the correct regular behavior at the NBS center. For the scalar field, we find the following $1\%$-accurate expression inside the star,
\beq
\Psi&=&\mu^2 M_{\rm NBS}^2g\,,\\
g&=&e^{-0.570459 x}\frac{\sum_{i=0}^{8}b_ix^i+b_fx^{9.6}}{(x+r_2)^{9}}\,,\\
x&=&\mu^2M_{\rm NBS}r\,,r_2=1.182\,,\nonumber\\
b_0&=&0.298\,,b_1=2.368\,,b_2=10.095\,,\nonumber\\
b_3&=&12.552\,, b_4=51.469\,,b_5=-8.416\,,\nonumber\\
b_6&=&54.141,\, b_7=-6.167\,,b_8=8.089\,,\nonumber\\
b_f&=&0.310\,.
\eeq

Finally, for future reference, the number of particles contained in an NBS is
\begin{equation}
	Q_{\rm NBS}=4 \pi \mu \int_0^\infty dr\, r^2\left|\Psi\right|^2\,,
\end{equation}
and, then, we can write the mass as~$M_ {\rm NBS}=\mu Q_{\rm NBS}$.

\subsection{Small perturbations} \label{sec:SmallPert}
As shown in appendix~\ref{app:Newtonian}, small perturbations of the form~\eqref{Perturbation} to the scalar field, together with the NBS perturbed gravitational potential
\begin{equation}
U=U_0(r)+\delta U (t,r,\theta, \varphi)\,,
\end{equation}
satisfy the linearized system of equations
\begin{align}
&i \partial_{t} \delta \Psi=-\frac{1}{2 \mu}\nabla^2 \delta \Psi+ \left(\mu U_0 +\gamma\right) \delta \Psi+ \mu \Psi_0 \delta U\,,\label{Sourced_SP_System1}\\
&\nabla^2 \delta U=4 \pi\left[ \mu^2 \Psi_0 \left(\delta \Psi+\delta \Psi^*\right)+P\right]\,,\label{Sourced_SP_System2}
\end{align}
where $U_0$ is the gravitational potential of the unperturbed star, and we have included an external point-like perturber~\footnote{This was obtained considering a non-relativistic external perturber. Note that $P$ is just the non-relativistic limit of $T^p_{t t}$ given in~\eqref{Stress_energy_particle}.}
\be
P\equiv m_p \frac{\delta \left(r-r_p(t)\right)}{r^2} \frac{\delta\left(\theta-\theta_p(t)\right)}{\sin \theta} \delta\left(\varphi-\varphi_p(t)\right)\,.\label{source_BS}
\ee
This system of equations was derived for non-relativistic fluctuations, which satisfy $|\partial_t \delta \Psi|\ll \mu |\delta \Psi|$, and are sourced by a non-relativistic, Newtonian perturber. 
To study the sourceless case, one can simply set $m_p=0$.
As shown in detail in Appendix~\ref{app:Newtonian}, the perturber couples to the NBS through the total stress energy tensor entering Einstein's equation in~\eqref{EOM_BosonS}, which is taken to be the sum of the stress energy tensor of the scalar $T_{\mu \nu}^S$ (given in~Eq.\eqref{StressEnergy}) and of the perturber $T^p_{\mu \nu}$ (given in~Eq.\eqref{Stress_energy_particle}). We neglect the backreaction on the perturber's motion and treat its worldline as given.

Let us decompose the fluctuations of the scalar field as in~\eqref{Decomposition}, and the gravitational potential and the source, respectively, as~\footnote{Note that the perturbation $\delta U$ must be real-valued. Again, we will omit the labels $\omega$, $l$ and $m$ in the functions $u^{\omega l m }(r)$ and~$p^{\omega l m }(r)$ to simplify the notation.}
\beq
\delta U&=&\sum_{l,m}\int \frac{d \omega}{\sqrt{2 \pi} r} \left[u^{\omega l m }Y_l^m e^{-i\omega t}+\left(u^{\omega l m }\right)^*\left(Y_l^m\right)^* e^{i\omega t}\right]\,,\nn\\
P&=& \sum_{l,m} \int \frac{d \omega}{\sqrt{2 \pi} r} \left[p^{\omega l m} Y_l^m e^{-i \omega t}+\left(p^{\omega l m}\right)^* \left(Y_l^m\right)^* e^{i \omega t}\right]\,,\nonumber
\eeq
where~$p^{\omega l m}$ are radial complex-functions defined by
\begin{equation} 
	p^{\omega l m} \equiv \frac{r}{2 \sqrt{2 \pi}}\int dt d \theta d \varphi \sin \theta \,P  \left(Y_l^m\right)^* e^{i \omega t}\,.\label{p_def}
\end{equation}
From equations~\eqref{Sourced_SP_System1} and~\eqref{Sourced_SP_System2} one obtains the matrix equation
\begin{equation} 
	\partial_r \boldsymbol{X} -V_{\rm B}(r) \boldsymbol{X}= \boldsymbol{P}\,,\label{BS_Perturbation_Matrix_Sourced}
\end{equation}
with the vector $\boldsymbol{X}\equiv (Z_1, Z_2, u, \partial_r Z_1, \partial_r Z_2, \partial_r u)^T$, the matrix $V_B$ given by 
\be
\begin{pmatrix} 
0 & 0& 0 & 1 & 0 & 0 \\
0 & 0 & 0 & 0 & 1 & 0 \\
0 & 0 & 0 & 0 & 0 & 1 \\
V-2 \mu (\omega- \gamma) & 0 & 2 \mu^2\Psi_0 & 0 &0 & 0  \\
0 & V+2 \mu (\omega+ \gamma) & 2 \mu^2\Psi_0 & 0 & 0 & 0 \\
4\pi \mu^2 \Psi_0 & 4\pi \mu^2 \Psi_0 & V-2 \mu^2 U_0 & 0 & 0 & 0  
\end{pmatrix}\,.\nonumber
\ee
Here, the radial potential
\be
V(r)\equiv \frac{l(l+1)}{r^2}+2 \mu^2 U_0\,,
\ee
and the source term
\begin{equation}
\boldsymbol{P}(r)\equiv \left(0,0,0,0,0,4\pi p\right)^T\,.
\end{equation}
Note that the condition of non-relativistic fluctuations translates, here, into the simple inequality $|\omega| \ll \mu$.

As suitable boundary conditions to solve for the fluctuations, we require both regularity at the origin,
\beq
&&\boldsymbol{X}(r \to 0)\nn\\
&&\sim\left(a r^{l+1},b r^{l+1},c r^{l+1},a (l+1)r^l,b (l+1)r^l,c (l+1)r^l\right)^T\,,\nn
\eeq
with complex constants $a$, $b$ and $c$, and the Sommerfeld radiation condition at infinity,
\beq
&&\boldsymbol{X}(r \to \infty)\nn\\
&&\sim\left(Z_1^\infty e^{i k_1 r} ,Z_2^\infty e^{i k_2 r},u^\infty, i k_1 Z_1^\infty e^{i k_1r},i k_2 Z_2^\infty e^{i k_2r},0\right)^T\,,\nn\\
&& \label{BC_sommerfeld_infinity}
\eeq
with 
\beq \label{Wave_number_1}
k_1&\equiv& \sqrt{2 \mu \left(\omega-\gamma\right)}\,, \\ \label{Wave_number_2}
k_2&\equiv& -\left(\sqrt{-2 \mu \left(\omega+\gamma\right)}\right)^*\,.
\eeq
In the last expression we are using the principal complex square root.

To calculate the fluctuations we will make use of the set of independent homogeneous solutions $\{\boldsymbol{Z_{(1)}},\boldsymbol{Z_{(2)}},\boldsymbol{Z_{(3)}},\boldsymbol{Z_{(4)}},\boldsymbol{Z_{(5)}},\boldsymbol{Z_{(6)}}\}$, uniquely determined by
\beq
&&\boldsymbol{Z_{(1)}}(r \to 0)\sim \Big(r^{l+1},0,0,(l+1)r^l,0,0\Big)^T\,,\nonumber \\
&&\boldsymbol{Z_{(2)}}(r \to 0)\sim \Big(0,r^{l+1},0,0,(l+1)r^l,0\Big)^T\,,\nonumber \\
&&\boldsymbol{Z_{(3)}}(r \to 0)\sim \Big(0,0,r^{l+1},0,0,(l+1)r^l\Big)^T\,,\nonumber \\
&&\boldsymbol{Z_{(4)}}(r \to \infty)\sim \Big(e^{i k_1 r},0,0,i k_1 e^{i k_1 r},0,0\Big)^T\,,\nonumber \\
&&\boldsymbol{Z_{(5)}}(r \to \infty)\sim \Big(0,e^{i k_2 r},0,0,i k_2 e^{i k_2 r},0\Big)^T\,, \nonumber \\
&&\boldsymbol{Z_{(6)}}(r \to \infty)\sim \Big(0,0,u^\infty,0,0,0\Big)^T\,.
\label{BC_BS}
\eeq
Then, the matrix
\be
F(r)\equiv\big(\boldsymbol{Z_{(1)}},\boldsymbol{Z_{(2)}},\boldsymbol{Z_{(3)}},\boldsymbol{Z_{(4)}},\boldsymbol{Z_{(5)}},\boldsymbol{Z_{(6)}}\big)
\label{eq:fundamental_matrix}
\ee
is known as the fundamental matrix of system~\eqref{BS_Perturbation_Matrix_Sourced}. As shown in Appendix~\ref{app:detF}, the determinant of $F$ is independent of $r$.

Finally, note that system~\eqref{BS_Perturbation_Matrix_Sourced} is invariant under the re-scaling
\begin{equation}
(U_0, \Psi_0,\gamma, \omega) \to \lambda^2 (U_0, \Psi_0, \gamma, \omega)\,,\quad r \to r/\lambda\,, \label{eq:scaling2}
\end{equation}
and, so, it can always be pushed into obeying the non-relativistic constraint. Additionally, for convenience, we impose that~$\delta \Psi$ and~$\delta U$ are left invariant by the re-scaling, by performing the extra transformation
\begin{align}
(Z_{1,2}, u)\to \lambda^{-3}(Z_{1,2}, u)\,, \quad m_p\to\lambda^{-1} m_p \,.
\end{align}

For a process happening during a finite amount of time the change in the NBS energy is, at leading order,
\begin{align} \label{DeltaE}
	\Delta E_{\rm NBS}&=-\int_{t=+\infty}d^3 x \sqrt{h}\, \delta T^S_{tt}+\int_{t=-\infty}d^3 x \sqrt{h}\, \delta T^S_{tt} \nn \\
	 &=\mu \Delta Q_{\rm NBS}\,,
\end{align}
since, at leading order,
\begin{align}
	\delta T^S_{tt}=\mu^2 \left(\left|\delta \Psi\right|^2+2\Psi_0{\rm Re}(\delta^2\Psi)\right)=\mu \,\delta j_t \,,
\end{align}
where~$\delta^2\Psi$ is a second order fluctuation of the scalar and we used~\eqref{NoetherPert} for the second equality.

\subsubsection{Validity of perturbation scheme}
The perturbative scheme requires that $|\delta \Psi|\ll 1$, which can always be enforced by making $m_p$ as small as necessary.
On the other hand, the background construction neglects higher-order post-Newtonian (PN) contributions. A self-consistent perturbative expansion requires that such neglected terms (of order $\sim U_0^2$) do not affect the dynamics of small fluctuations (of order $\sim \delta U$). This imposes 
$m_p\gtrsim 10^4 M_{\odot}\,\left(\frac{M_{\rm NBS}}{10^{10}M_\odot}\right)^3\left(\frac{\mu}{10^{-22}\,{\rm eV}}\right)^2$,
which holds true for many systems of astrophysical interest. As shown in Appendix~\ref{app:Newtonian}, the scalar evolution equation \eqref{KG_all} is sourced by higher PN-order terms. However,
these are nearly static, or very low frequency terms, hence will make a negligible contribution for high-energy binaries or plunges. In other words, the previous constraint can be substantially relaxed in dynamical situations, such as the ones we focus on.
Finally, the Newtonian, non-relativistic approximation requires the source to have a small frequency $\lesssim 2\times 10^{-8}\,\left(\mu/10^{-22}{\rm eV}\right)\,{\rm Hz}$, in the case of a periodic motion. In Appendix~\ref{app:Newtonian} we show how to extend the formalism to include Newtonian but high frequency sources, and use it to calculate emission by a high frequency binary in Section~\ref{BS_binaries}. For plunges of nearly constant velocity $v$ piercing through an NBS, the Newtonian and non-relativistic approximation requires that $v\lesssim R\mu$. Fortunately, any NBS has $R \mu \gg 1$ and the latter condition is trivially verified.

\subsubsection{Sourceless perturbations}

Free oscillations of NBSs are fluctuations of the form
\beq
\delta \Psi&=& \frac{1}{\sqrt{2 \pi} r} \left[Z_1 Y_l^m e^{-i \omega t}+Z_2^*\left(Y_l^m\right)^* e^{i \omega^* t}\right]\,, \nonumber \\
\delta U&=& \frac{1}{\sqrt{2 \pi} r} \left[u Y_l^m e^{-i \omega t}+u^*\left(Y_l^m\right)^* e^{i \omega^* t}\right]\,,
\eeq
where $Z_1$, $Z_2$ and $u$ are regular solutions of system~\eqref{BS_Perturbation_Matrix_Sourced} with $P=0$, satisfying the Sommerfeld condition at infinity. These are also known as quasi-normal mode (QNM) solutions, and the corresponding frequency~$\omega$ is the QNM frequency.
Noting that the condition
\be
{\rm det}(F)=0\,,\label{eq:fundamental_matrix_condition}
\ee
holds if and only if $\omega$ is a QNM frequency, we are able to find the NBS proper oscillation modes by solving the sourceless system~\eqref{BS_Perturbation_Matrix_Sourced}, and requiring at the same time that~\eqref{eq:fundamental_matrix_condition} is verified. These frequencies are shown in Table~\ref{table:QNM_BS_invariant}.

Additionally, notice that the sourceless system~\eqref{BS_Perturbation_Matrix_Sourced} admits also the trivial solution
\begin{align} \label{HS_trivial}
\delta \Psi_\epsilon&= \epsilon\, \Psi_0 (1+i \gamma t)\,, \nonumber \\
\delta U_\epsilon&=\epsilon\, U_0\,,	
\end{align}
with a constant $\epsilon\ll1$. This solution is valid only for a certain amount of time (while the perturbation scheme holds) and it corresponds just to an infinitesimal change of the background NBS (\textit{i.e,} an infinitesimal re-scaling of the original star) by a $\lambda=1+\epsilon/2$. This perturbation causes a static change in the number of particles in the star
\begin{equation}
	\delta Q_\epsilon= \frac{\epsilon}{2} Q_{\rm NBS}\,,
\end{equation}
and in its mass
\begin{equation}
	\delta M_\epsilon= \mu\, \delta Q_\epsilon=\frac{\epsilon}{2}  M_{\rm NBS}\,.
\end{equation}

\subsubsection{External perturbers}
\label{sec:External perturbers}
In the presence of an external perturber, one needs to prescribe its motion through the source term~\eqref{source_BS}. The solution of system~\eqref{BS_Perturbation_Matrix_Sourced} which is regular at the origin and satisfies the Sommerfeld condition at infinity can be obtained through the method of variation of parameters, and it reads
\beq \label{Z1_of_r}
Z_1(r)&=& 4\pi\Bigg[\sum_{n=1}^3 F_{1,n}(r) \int_\infty^r dr' F^{-1}_{n,6}(r') p(r') \nonumber \\ 
&+&\sum_{n=4}^6 F_{1,n}(r) \int_0^r dr' F^{-1}_{n,6}(r') p(r') \Bigg]\,,\\\label{Z2_of_r}
Z_2(r)&=& 4\pi\Bigg[\sum_{n=1}^3 F_{2,n}(r) \int_\infty^r dr' F^{-1}_{n,6}(r') p(r') \nonumber \\ 
&+&\sum_{n=4}^6 F_{2,n}(r) \int_0^r dr' F^{-1}_{n,6}(r') p(r') \Bigg]\,,\\\label{u_of_r}
u(r)&=& 4\pi\Bigg[\sum_{n=1}^3 F_{3,n}(r) \int_\infty^r dr' F^{-1}_{n,6}(r') p(r') \nonumber \\ 
&+&\sum_{n=4}^6 F_{3,n}(r) \int_0^r dr' F^{-1}_{n,6}(r') p(r') \Bigg]\,,
\eeq
where $F_{i,j}$ is the $(i,j)$-component of the fundamental matrix defined in Eq.~\eqref{eq:fundamental_matrix}. To obtain the total energy, linear and angular momenta radiated during a given process, all we need are the amplitudes $Z_1^\infty$ and $Z_2^\infty$. These are given by
\beq
Z_1^\infty&=&4\pi \int_{0}^{\infty} dr' F^{-1}_{4,6}(r')p(r') \,, \label{Z1inf_BS}\\
Z_2^\infty&=&4 \pi \int_{0}^{\infty} dr' F^{-1}_{5,6}(r')p(r') \,. \label{Z2inf_BS}
\eeq
Let us now apply our framework to a few physically interesting external perturbers.
\paragraph*{\underline{Plunging particle.}}
Consider a pointlike perturber plunging into an NBS. Without loss of generality, one can assume its motion to take place in the~$z$-axis, being described by the worldline~$x^\mu(t)=(t,0,0,z_p(t))$ in Cartesian coordinates.
Neglecting the backreaction of the fluctuations on the perturber's motion,
\begin{align}
	\ddot{z}_p(t)=-\partial_z U_0(z_p)\,.
\end{align}
We consider that the perturber crosses the NBS center at~$t=0$ (\textit{i.e.,}~$z_p(0)=0$) with velocity
\begin{align}
\dot{z}_p(0)=-\sqrt{2\left(U_0(R)-U_0(0)\right)+v_R^2}\,,
\end{align}
where~$v_R$ is the velocity with which the massive object enters the NBS; in other words, it is the velocity at~$r=R$.
In spherical coordinates the source reads
\begin{align}
&P=m_p \frac{\delta(\varphi)}{r^2 \sin \theta}\nonumber \\
&\times\left[\delta\left(r-z_p(t)\right)\delta\left(\theta\right)+ \delta\left(r+z_p (t)\right)\delta\left(\theta-\pi\right)\right].
\end{align}

Here we do not want to be restricted to massive objects describing unbounded motions and, so, we consider also perturbers with small~$v_R$. These may not have sufficient energy to escape the NBS gravity, being doomed to remain in a bounded oscillatory motion (\textit{see} Section~\ref{oscillating_particle_BS}). In these cases, we want to find the energy and momentum loss in one full crossing of the NBS and, so, we shall take the above source as "active" just during that time interval, vanishing whenever else.

Using Eq.~\eqref{p_def} the function~$p$ is
\begin{align}
	p=-\frac{m_p}{\sqrt{2 \pi}}Y_l^0(0) \delta_m^0 \frac{|t'_p(r)|}{r} \left(e^{-i \omega t_p(r)}+(-1)^l e^{i \omega t_p(r)}\right) \nn\,,
\end{align}
with~$t_p(r)\geq0$ defined by~$z_p\left[t_p(r)\right]=-r$.
This can be rewritten in the form
\beq
&&p=\frac{m_p}{\sqrt{2\pi}}Y_l^0(0)\, \delta_m^0\nonumber \\
&&\times \frac{|t_p'(r)|}{r}\left(\cos\left[\omega t_p(r)\right] \delta_l^\text{even}-i \sin\left[\omega t_p(r)\right] \delta_l^\text{odd}\right).
\eeq
The property 
\be
p(\omega,l,0;r)=p(-\omega,l,0;r)^*\,,
\ee
together with the form of system~\eqref{BS_Perturbation_Matrix_Sourced}, implies that
\beq
Z_2(\omega,l,0;r)&=&Z_1(-\omega,l,0;r)^*\,,\\
Z_2^\infty(\omega,l,0)&=&Z_1^\infty(-\omega,l,0)^*\,.\label{property_Z1Z2_plunge_BS}
\eeq
So, the spectral fluxes~\eqref{Particles_flux},~\eqref{Energy_flux},~\eqref{Momentum_flux} and~\eqref{AngularMomentum_flux} become, respectively,
\begin{align}
&\frac{d Q^{\rm rad}}{d \omega}=4\, {\rm Re}\left[\sqrt{2 \mu  (\omega-\gamma)}\right] \sum_l \left|Z_1^\infty(\omega,l,0)\right|^2\,,
\label{Particles_flux_BS}
\end{align}
\begin{align}
&\frac{d E^{\rm rad}}{d \omega}= \left(\mu-\gamma+\omega\right) \frac{d Q^{\rm rad}}{d \omega}\simeq \mu \frac{d Q^{\rm rad}}{d \omega} \,,
\label{Energy_flux_BS}
\end{align}
\beq
\frac{d P_z^{\rm rad}}{d \omega}&=&\sum_{l}\frac{16\mu(l+1)}{\sqrt{(2l+1)(2l+3)}}\, \Theta\left( \omega-\gamma\right)\left|\omega-\gamma\right| \nonumber \\
&\times&\text{Re}\left[Z_1^\infty(\omega,l,0) Z_1^\infty(\omega,l+1,0)^*\right]\,,\label{Momentum_flux_BS}
\eeq
and
\be
\frac{d L_z^{\rm rad}}{d \omega}=0\,.
\ee
These expressions were derived assuming a perturber in an unbounded motion. However, these are also good estimates to the energy and momenta radiated during one full crossing of the NBS by a bounded perturber, as long as its half-period is much larger than the NBS crossing time.

To compute how much energy is lost by the perturber, we need to know the change in the NBS energy~$\Delta E_{\rm NBS}$. At leading order, this is given by
\begin{equation}
	\Delta E_{\rm NBS}=\mu\, \Delta Q_{\rm NBS}=- \mu\, Q^{\rm rad}\,, 
\end{equation}
using Eq.~\eqref{DeltaE} in the first equality and~\eqref{DeltaQ} in the second.
Conservation of total energy-momenta, expressed through Eq.~\eqref{LossRad}, implies that the perturber loses the energy
\begin{align}
	&E^{\rm lost}= \Delta E_{\rm NBS}+E^{\rm rad}=\int d\omega\,(\omega- \gamma) \frac{d Q^{\rm rad}}{d \omega}  \nn \\
	&=4\sqrt{2 \mu}\int d\omega\, {\rm Re}\left[(\omega-\gamma)^{\frac{3}{2}}\right] \sum_l \left|Z_1^\infty(\omega,l,0)\right|^2\,.
	\label{Energy_loss_BS}
\end{align}
The last expression should be understood as an order of magnitude estimate. If we had considered only the leading order contribution to~$\Delta E_{\rm NBS}$ and~$E^{\rm rad}$, we would have obtained~$E^{\rm lost}=0$. In the second equality we used higher order corrections to $E^{\rm rad}$ -- the factor~$(\omega-\gamma) \ll \mu$; but not to~$\Delta E_{\rm NBS}$. The corrections to~$\Delta E_{\rm NBS}$ may be of the same order of the corrections to~$E^{\rm rad}$ and should be included in a rigorous calculation of~$E^{\rm lost}$. We do not attempt that in this work. Interestingly, in our approximation the energy loss of the perturber matches the kinetic energy of the radiated scalar particles at infinity, as can be readily verified. The terms neglected should contain information about, for instance, the gravitational and kinetic energy of the radiated particles when they were in the unperturbed NBS. Still, we believe that Eq.~\eqref{Energy_loss_BS} is good estimate of the order of magnitude of~$E^{\rm lost}$ and that it scales correctly with the boson star and perturber's mass,~$M_{\rm NBS}$ and~$m_p$, respectively.

For a small perturber~$m_p \mu \ll v_R$, its momentum and energy loss are related through (\textit{see} Eq.~\eqref{EvsPloss})~\footnote{Using the full expression~\eqref{EvsPloss}, it is easy to see that if~$E^{\rm lost}\propto m_p^2$, then~$P^{\rm lost} \propto m_p^2$ in the limit $m_p \mu\ll v_R$. The~$E^{\rm lost}\propto m_p^2$ follows from~$Z_1^\infty \propto m_p$ (\textit{see} Eq.~\eqref{Z1inf_BS}).}
\begin{align}
	P_z^{\rm lost}\simeq-\frac{E^{\rm lost}}{v_R}\,.
\end{align}
Conservation of total momentum, as expressed in~\eqref{LossRad}, implies that the NBS acquires a momentum ~\footnote{The watchful reader may wonder why the kinetic energy associated with the momentum acquired by the boson star~$\Delta P_z$ is not included in~$\Delta E_{\rm NBS}$. Actually, this is one of the higher order corrections neglected in~\eqref{Energy_loss_BS}, but it is easy to check that it is subleading comparing with the correction of~$E^{\rm rad}$ considered.} 
\begin{align}
P_{\rm NBS}=P_z^{\rm lost}-P_z^{\rm rad}=-\frac{E^{\rm lost}}{v_R}-P_z^{\rm rad}\,.\label{eq:NBS_momentum}
\end{align}
%

\paragraph*{\underline{Orbiting particles}.}
Consider an equal-mass binary, with each component having mass $m_p$, and describing a circular orbit of radius $r_{\rm orb}$ and angular frequency $\omega_{\rm orb}$ in the equatorial plane of an NBS. The source is modelled as
\beq
P&=&\frac{m_p}{r_{\rm orb}^2} \delta(r-r_{\rm orb})\delta\left(\theta-\frac{\pi}{2}\right)\nonumber\\
&&\times\left[\delta(\varphi-\omega_{\rm orb} t)+\delta(\varphi+\pi-\omega_{\rm orb} t)\right]\,. \label{P_orbiting}
\eeq
We are assuming that the center of mass of the binary is at the center of the NBS, but in principle our results extend to all binaries sufficiently deep inside the NBS.
Also, our methods can be applied to any binary as long as a suitable source~$P$ is given.

Using Eq.~\eqref{p_def} the source above yields
\beq
p&=&m_p\sqrt{\frac{\pi}{2}} \frac{Y_l^m\left(\pi/2,0\right)}{r_{\rm orb}}\nn\\
&& \times (1+(-1)^m)  \delta\left(r-r_{\rm orb}\right)\delta\left(\omega -m \omega_{\rm orb}\right)\,.\label{p_orbiting}
\eeq
The perturber's motion is fully specified by a prescription relating $r_{\rm orb}$ and $\omega_{\rm orb}$; we consider Keplerian orbits
$r_{\rm orb}^3=M/\omega_{\rm orb}^2$, where $M=2m_p$ is the total mass. This setup describes either stellar-mass or supermassive BH binaries orbiting inside a NBS.
Alternatively, applying the transformation $m_p(1+(-1)^m)\to m_p$, we obtain a source that describes an extreme mass ratio inspiral (EMRI). This could be, for instance, a star of mass $m_p$ on a circular orbit around a central massive BH of mass $M_{\rm BH}$. In such case we consider the Keplerian prescription $r_{\rm orb}^3=M_{\rm BH}/\omega_{\rm orb}^2$. 

The symmetry
\be
p(\omega,l,m;r)=(-1)^m p(-\omega,l,m;r)^*\,,
\ee
together with the form of system~\eqref{BS_Perturbation_Matrix_Sourced} implies
\beq
Z_2(\omega,l,m;r)&=&(-1)^m Z_1(-\omega,l,-m;r)^*\,,\\
Z_2^\infty(\omega,l,m)&=&(-1)^m Z_1^\infty(-\omega,l,-m)^*\,.
\eeq
These simplify the emission rate expressions~\eqref{Particles_flux_rate},~\eqref{Energy_flux_rate} and~\eqref{AngularMomentum_flux_rate}, yielding
\begin{align}
&\dot{Q}^{\rm rad}=\frac{2}{\pi} \int d\omega\, {\rm Re}\left[\sqrt{2 \mu \left(\omega-\gamma\right)}\right]\sum_{l,m} \left|Z_1^\infty(\omega,l,m)\right|^2  \,, \nn\\
&\dot{E}^{\rm rad}=\nn \\
&\frac{2}{\pi} \int d\omega(\mu-\gamma+\omega) {\rm Re}\left[\sqrt{2 \mu \left(\omega-\gamma\right)}\right]\sum_{l,m} \left|Z_1^\infty(\omega,l,m)\right|^2  \,, \nn\\
&\dot{L}_z^{\rm rad}=\frac{2}{\pi} \int d\omega \, {\rm Re}\left[\sqrt{2 \mu \left(\omega-\gamma\right)}\right]\sum_{l,m} m \left|Z_1^\infty(\omega,l,m)\right|^2\,.\nn 
\end{align}
These can be written explicitly as
\beq
\dot{Q}^{\rm rad}&=&32 \pi \, \widetilde{p}^2 \sum_{l,m}{\rm Re}\left(\sqrt{2 \mu \left(m \omega_{\rm orb}-\gamma\right)}\right)\nonumber\\
&\times& \left|F_{4,6}^{-1}\left(m\omega_{\rm orb};\,r_{\rm orb}\right)\right|^2\,,\label{Particles_flux_orbitingBS2}\\
\dot{E}^{\rm rad}&=&32 \pi \, \widetilde{p}^2 \sum_{l,m}{\rm Re}\left(\sqrt{2 \mu \left(m \omega_{\rm orb}-\gamma\right)}\right)\nonumber\\
&\times&(\mu-\gamma+m \omega_{\rm orb}) \left|F_{4,6}^{-1}\left(m\omega_{\rm orb};\,r_{\rm orb}\right)\right|^2,\label{Energy_flux_orbitingBS2}\\
\dot{L}_z^{\rm rad}&=&32 \pi \, \widetilde{p}^2  \sum_{l,m}m {\rm Re}\left(\sqrt{2 \mu \left(m \omega_{\rm orb}-\gamma\right)}\right)\nonumber\\
&\times& \left|F_{4,6}^{-1}\left(m\omega_{\rm orb};\,r_{\rm orb}\right)\right|^2\,,\label{AngularMomentum_flux_orbitingBS}
\eeq
where we defined
\begin{equation*}
	\widetilde{p}\equiv m_p \sqrt{\frac{\pi}{2}}\frac{Y_l^m(\pi/2,0)}{r_{\rm orb}}\left(1+(-1)^m\right)\,.
\end{equation*}
Equation~\eqref{Energy_flux_orbitingBS2} can be further simplified using
\begin{equation*}
	\mu-\gamma+m\omega_{orb}\simeq \mu\,,
\end{equation*}
since we are treating the scalar fluctuations as non-relativistic; that is only valid if~$\gamma \ll \mu$ and~$\omega_{\rm orb}\ll \mu$.~\footnote{Large azimuthal numbers~$m$ do not spoil the approximation, because the emission is strongly suppressed by~$F_{4,6}^{-1}$ in that limit.} 

Now we follow the same procedure that we applied in the previous section to a \textit{plunging particle}, to estimate the rate of energy loss of the binary. We start by computing, at leading order, the change in the NBS energy per unit of time:
\begin{equation}\label{eq:circular_flux_Erad}
	\dot{E}_{\rm NBS}=\mu \dot{Q}_{\rm NBS}=-\mu \dot{Q}^{\rm rad}\,,
\end{equation}
where we used Eq.~\eqref{DeltaE} in the first equality and~\eqref{DeltaQ} in the second.~\footnote{Equations~\eqref{DeltaQ} and~\eqref{DeltaE} are easy to adapt to changes happening during a finite amount of time~$\Delta t$. To get the rates of change one just needs to divide these expressions by~$\Delta t$ and take the limit~$\Delta t\to 0$.}
Conservation of the total energy implies that the binary energy loss per unit of time is 
\begin{align} \label{eq:E_loss_circular}
	&\dot{E}^{\rm lost}=\dot{E}^{\rm rad}+ \dot{E}_ {\rm NBS}= 32 \pi \widetilde{p}^2 \sum_{l,m}\left(m \omega_{\rm orb}-\gamma\right)\nonumber \\
	&\times {\rm Re}\left(\sqrt{2 \mu \left(m \omega_{\rm orb}-\gamma\right)}\right) \left|F_{4,6}^{-1}\left(m\omega_{\rm orb};\,r_{\rm orb}\right)\right|^2\,.
\end{align}
Again, the last expression should be understood as an order of magnitude estimate~(the reason is discussed in the previous section where we considered a \textit{plunging particle}).

For a small perturber~$m_p\ll |\omega_{\rm orb}| r_{\rm orb}^2$, its angular momentum and energy loss are related through
\begin{align}
	\dot{L}_z^{\rm lost}\simeq \frac{\dot{E}^{\rm lost}}{\omega_ {\rm orb}}\,.
\end{align}
Conservation of total angular momentum, expressed through Eq.~\eqref{LossRad}, implies that per unit of time the NBS acquires the angular momentum
\begin{align}
	 \dot{L}_ {\rm NBS}=\dot{L}_z^{\rm lost}-\dot{L}_z^{\rm rad}=\frac{\dot{E}^{\rm lost}}{\omega_ {\rm orb}}-\dot{L}_z^{\rm rad}\,.
\end{align}
%

\subsection{Free oscillations}
%
\begin{table}[th] 
	\begin{tabular}{c||c}
		\hline
		\hline
		$l$ &  \multicolumn{1}{c}{$\frac{\omega^{(n)}_{\rm QNM}}{M_{\rm NBS}^2\mu^3}$} \\ 
		\hline
		\hline
		0 & $0.0682\;\,\,\,    0.121\;\,\,     0.138\;\,\,    0.146\;\,\,    0.151\;\,\,  0.154\;\,\, 0.159$\\
		1 & $\,\,0.111\,\,\;\,\,     0.134\;\,\,     0.144\;\,\,    0.149\;\,\,    0.153\;\,\,  0.157\;\,\, 0.162$\\
		2 & $\,\,0.106\,\,\;\,\,     0.131\;\,\,     0.143\;\,\,    0.149\;\,\,    0.153\;\,\,  0.156\;\,\, 0.161$\\
		\hline
		\hline
	\end{tabular} 
	\caption{Normal frequencies of an NBS of mass $M_{\rm NBS}$ for the three lowest multipoles. For each multipole $l$ we show the fundamental mode ($n=0$) and the first five overtones.
	At large overtone number the modes cluster around $\gamma\simeq0.162712 M_{\rm NBS}^2\mu^3$. The first mode for $l=0$ agrees with that of Ref.~\cite{Guzman:2004wj} when properly normalized and with an ongoing fully relativistic analysis~\cite{Caio:2020comment}. The two lowest $l=0,\,1,\,2$ modes are in good agreement with a recent time-domain analysis~\cite{Guzman:2018bmo}.}
	\label{table:QNM_BS_invariant}
\end{table}
The characteristic, non-relativistic oscillations of NBSs are regular solutions of the system~\eqref{Sourced_SP_System1}-\eqref{Sourced_SP_System2}
satisfying Sommerfeld conditions~\eqref{BC_sommerfeld_infinity} at large distances. For each angular number $l$, there seems to be an infinite, discrete set of solutions which we label with an overtone index $n$, $\omega^n_{\rm QNM}$. The first few characteristic frequencies, normalized to the NBS mass, are shown in Table~\ref{table:QNM_BS_invariant}. 
They turn out to be all {\it normal mode} solutions, confined within the NBS. The characteristic frequencies are all purely real and cluster around $\gamma$.
We highlight the fact that the numbers in Table~\ref{table:QNM_BS_invariant} are universal, they hold for any NBS. The fundamental $l=0$ mode (the first entry in the Table)
had been computed previously~\cite{Guzman:2004wj}, and agrees with our calculation to excellent precision (after proper normalization). Our results are also in very good agreement with the
frequencies of the first two modes, obtained in a recent time-domain analysis~\cite{Guzman:2018bmo}.
Modes of relativistic stars have been considered in the literature~\cite{Yoshida:1994xi,Kojima:1991np,Macedo:2013jja,Macedo:2016wgh,GRITJHU}
and should smoothly go over to the numbers in Table~\ref{table:QNM_BS_invariant}. Note that modes of relativistic BSs are damped, due to couplings between the scalar and the metric and the possibility to lose energy via gravitational waves. Such damping -- which is small for the relevant polar fluctuations~\cite{Macedo:2013jja,Macedo:2016wgh,GRITJHU} -- should get smaller as one approaches the Newtonian regime, but a full characterization of the modes of boson stars is missing.
Our results show that NBSs are linearly mode stable; it would be interesting to have a formal proof,
perhaps following the methods of Ref.~\cite{Kimura:2018eiv,Kimura:2017uor}.
We point out that the stabilization of a perturbed boson star through the emission of scalar field -- known as \textit{gravitational cooling} -- has been studied previously~\cite{Seidel1994,Balakrishna:2006ru,Guzman:2006yc}.

\subsection{A perturber sitting at the center\label{sec_sitting_bs}}
%
\begin{figure}[t]
	\includegraphics[width=8cm,keepaspectratio]{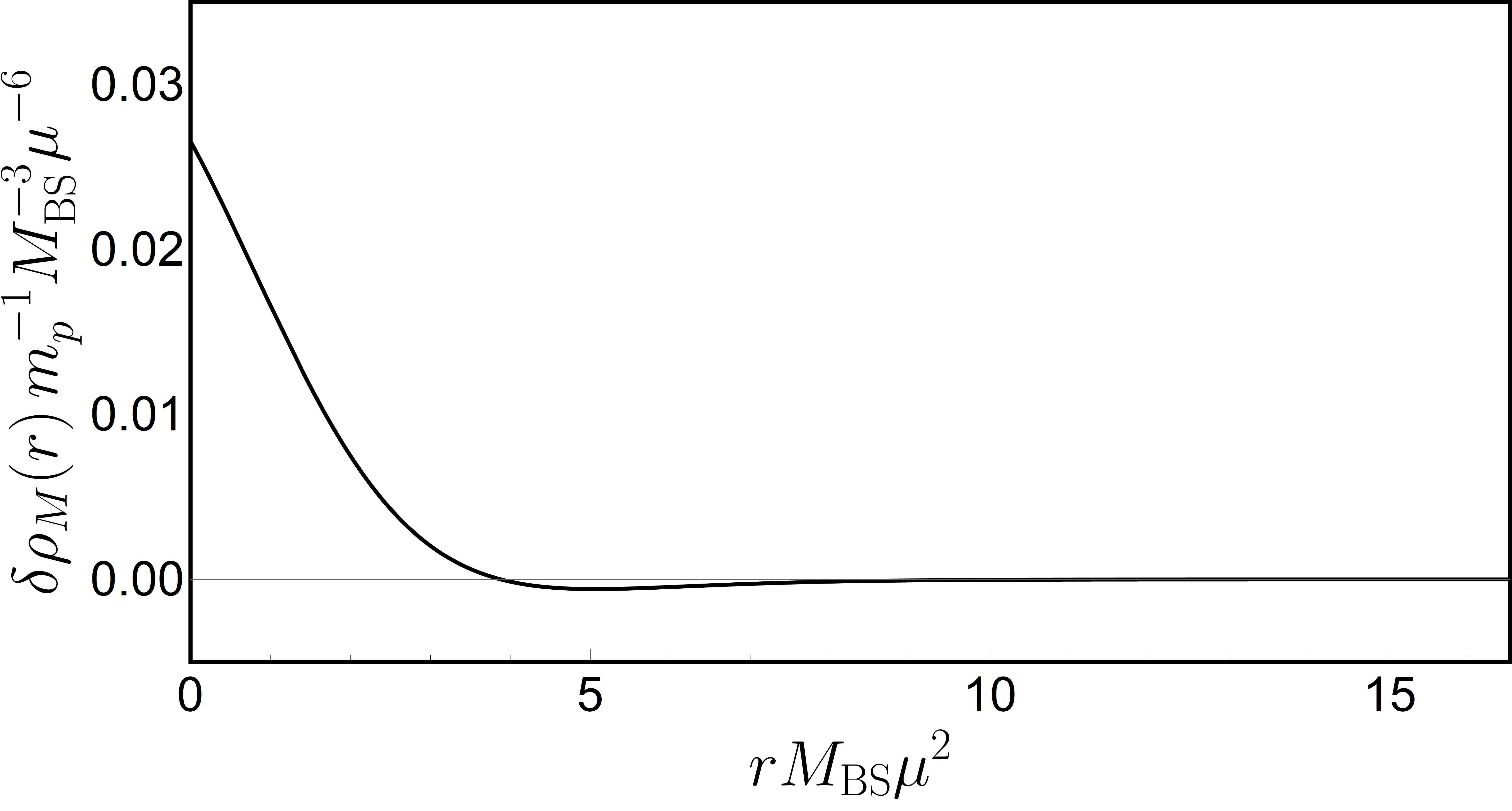} 
	\includegraphics[width=8cm,keepaspectratio]{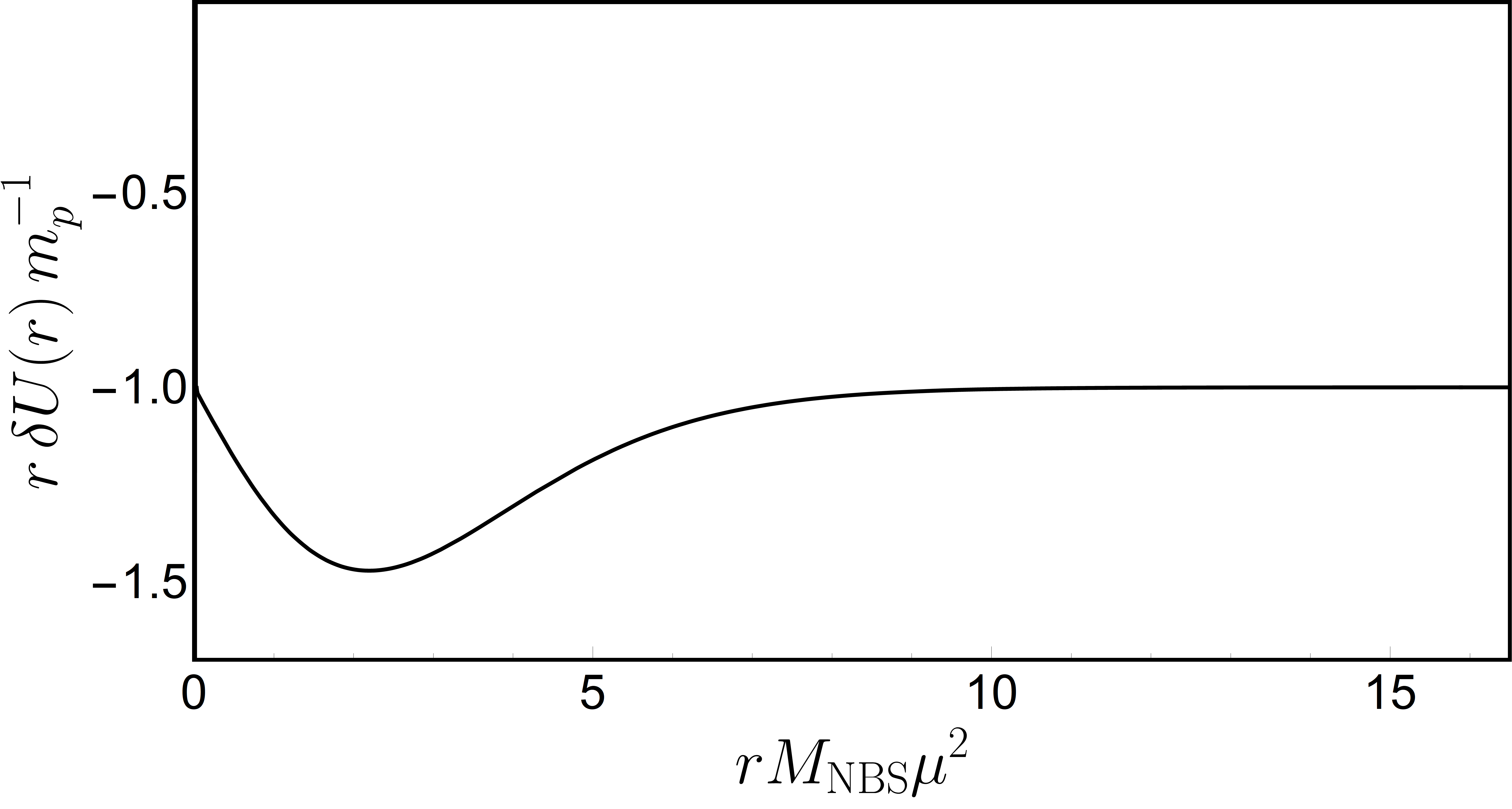}
	\caption{Universal perturbations induced a massive object, of mass $m_p$, sitting at the center of the scalar configuration. We assume that the perturber was brought adiabatically so that $\delta Q_{\rm NBS}=\delta M_{\rm NBS}=0$. Upper panel: perturbation in the mass density of the NBS obtained using Eq.~\eqref{deltarho}. 
Lower panel: perturbation in the gravitational potential $r \delta U= r \left(\delta U_p+\delta U_\epsilon\right)$. As expected, for large $r$, one recovers the Coulombian potential $U=-m_p/r$.
}\label{fig:Particlesittingcenter}
\end{figure}
Static perturbations of NBSs, or of solitonic DM cores of light fields are interesting
in their own right. For perturbers localized far away, the induced tidal effects can dissipate energy
and lead to distinct signatures, both in GW signals and in the dynamics of objects close to such configurations~\cite{Mendes:2016vdr,Cardoso:2017cfl,Sennett:2017etc}.
We will not perform a general analysis of static tidal effects and will instead focus on perturbations due to a massive object at the center of an NBS.
Such object can be taken to be a supermassive BH or a neutron star, and the induced changes are important to understand how DM distribution
is affected by baryonic ``impurities.'' 

Consider then a BH or star, described by the source \eqref{source_BS}, and inducing static, spherically symmetric, real perturbations on the scalar field and gravitational potential, respectively, $\delta \Psi_p(r)$ and $\delta U_p(r)$. Then, Eqs.~\eqref{Sourced_SP_System1} and~\eqref{Sourced_SP_System2} become
\begin{align}
	\nabla^2\delta \Psi_p&= 2\mu \left(\mu U_0+\gamma\right)\delta \Psi_p+2\mu^2\Psi_0 \delta U_p\,,  \nonumber\\
	\nabla^2\delta U_p&= 4\pi \left(2\mu^2 \Psi_0\, \delta \Psi_p+P\right) \,. 
\end{align}
In the static source limit, it is easy to show that the matter moments are given by
\begin{equation}
p=\lim_{r_p \to 0}\,\frac{1}{2 \sqrt{2}} \frac{m_p}{r_p} \,\delta_l^0 \delta_m^0 \delta(\omega) \delta(r-r_p)  \,,
\end{equation}
which, through the variation of parameters, implies that
\begin{align}
	\delta \Psi_p &=m_p \sum_{n=4}^{6} \frac{F_{1,n}(r)}{r}  \lim_{r_p \to 0} \left(\frac{F^{-1}_{n,6}(r_p)}{r_p}\right)\,, \nonumber \\
	\delta U_p &= m_p \sum_{n=4}^{6} \frac{F_{3,n}(r)}{r}  \lim_{r_p \to 0} \left(\frac{F^{-1}_{n,6}(r_p)}{r_p}\right)\,,
\end{align}
where the components of the fundamental matrix and its inverse are evaluated at $l=m=\omega=0$. Note that the change in the number of particles and mass of the NBS, respectively, $\delta Q_p$ and $\delta M_p$, is static, but non-zero in general. This is a consequence of the source being treated as if it was eternal. However, we know that if the perturber is brought in an adiabatic way to the center of the NBS there is no scalar radiation emitted, and, so, no change in the number of particles and mass of the star, $\delta Q_{\rm NBS}=\delta M_{\rm NBS}=0$. Fortunately, we are free to sum a trivial homogeneous solution~\eqref{HS_trivial} to enforce $\delta Q_{\rm NBS}=\delta M_{\rm NBS}=0$, while keeping $\delta \Psi=\delta \Psi_p+ \delta \Psi_\epsilon$ and $\delta U=\delta U_p+\delta U_\epsilon$ a solution of the inhomogeneous system. 
The perturbation induced in the density of particles is given by
\begin{equation}
	\delta \rho_Q=-\delta j_t=2\mu \Psi_0 \,{\rm Re}\left( \delta \Psi\right)=2\mu \Psi_0\left(\delta \Psi_p+\frac{\epsilon}{2} \Psi_0\right)\,, \nonumber
\end{equation}
and the one induced in the mass density by
\begin{equation} \label{deltarho}
\delta \rho_M=\delta T_{00}^S= \mu\, \delta \rho_Q= 2\mu^2 \Psi_0\left(\delta \Psi_p+\frac{\epsilon}{2} \Psi_0\right)\,, 
\end{equation}
where $j_t$ is the $t$-component of the Noether's current. The parameter $\epsilon$ associated with the trivial homogeneous solution must be chosen appropriately, so that
\begin{equation}
4\pi \int_{0}^{\infty} dr \,r^2 \delta \rho_Q= 4\pi \int_{0}^{\infty} dr\, r^2 \delta \rho_M=0\,.
\end{equation}

The perturbations in the mass density and gravitational potential of an NBS induced by a massive object sitting at its center are shown in Fig.~\ref{fig:Particlesittingcenter}.
Our results indicate that the particle attracts scalar field towards the center, where the gravitational potential corresponds solely to that of the point-like mass.
These results are consistent with those in Ref.~\cite{Bar:2018acw}. 
We find an insignificant change in the local DM mass density, when placing a point-like perturber at the center of an NBS; notice that $\delta \rho_M(0)/\rho_M(0)\sim 10\, m_p/M_{\rm NBS}$. Thus, a massive perturber will not enhance greatly the local DM density, which is smooth and flat for light scalars.  

On the other hand, studies with particle-like DM models find that its density close to supermassive BHs increases significantly~\cite{Gondolo:1999ef,Sadeghian:2013laa}.
This is in clear contrast to our results for light fields, a perturber does not significantly alter the local ambient density, since its size is much smaller than the scalar Compton wavelength.
Parenthetically, large overdensities seem to be in some tension with observations~\cite{Robles:2012uy}. Possible ways to ease the tension rely on scattering of DM by stars or BHs, or accretion by the central BH, induced by heating in its vicinities~\cite{Merritt:2002vj,Bertone:2005hw,Merritt:2003qk}.
These outcomes cannot possibly generalize to light scalars, at least not when the configuration is spherically symmetric, since there are no stationary 
BH configurations with scalar ``hair''~\cite{Herdeiro:2015waa,Cardoso:2016ryw}. But these results do prompt the questions: what happens to an NBS when a BH is placed at its center? what happens to the local scalar amplitude of an NBS when a binary is orbiting? We now turn to these issues.

\subsection{A black hole eating its host boson star\label{sec_sitting_bh}}
As we noted, there are no stationary, spherically symmetric configurations when a non-spinning BH is placed at the center.
On long timescales, the entire NBS will be accreted by the BH, a fraction dissipating to infinity.
This means, in particular, that our results cannot be extrapolated to when the point-like particle is a BH, and describe the system only at intermediate times. 
What {\it is} the lifetime of such a system, composed of a small BH sitting at the center of an NBS?
Unfortunately, most of the studies on BH growth and accretion assume a fluid-like environment~\cite{Giddings:2008gr}, an assumption that breaks down completely
here, since the Compton wavelength of the scalar is much larger than that of the BH. Exceptions to this rule exist~\cite{Clough:2019jpm,Hui:2019aqm}, but focus
on different aspects, and do not consider setups with the necessary difference in lengthscales.

The precise answer to this question requires full nonlinear simulations in a challenging regime, with proper initial conditions.
However, in the limit we are interested in, where the BH, of mass $M_{\rm BH}\ll M_{\rm NBS}$, is orders of magnitude smaller and lighter than the NBS, 
a perturbative calculation is appropriate. Consider a sphere of radius $r_+$ centred at the origin of the NBS. The NBS is stationary, and there is a flux of energy crossing such
a sphere inwards (detailed in Appendix~\ref{app:incoming_flux})
\be
\dot{E}_{\rm in}\approx 10^{-3} \mu^7 r_+^2 M_{\rm NBS}^5\,,
\ee
and the same amount crossing it outwards. If such a sphere defines the BH boundary $r_+=2M_{\rm BH}$~\footnote{Actually, such a sphere should be placed outside the effective potential for wave propagation around BHs, but the difference is not relevant here.}, a fraction will be absorbed by the BH. Because of relativistic effects, 
low-frequency waves (the scalar field frequency is $\mu$ and we are in the low frequency regime with $\mu M_{\rm BH}\ll 1$) are poorly absorbed, and one finds that the flux into the BH is~\cite{Unruh:1976fm}~\footnote{We are taking the limit $\omega \to \mu$ in the expression for the transmission. Strictly speaking, we are in the $\omega<\mu$ regime, but continuity of results should be valid.}
\beq \nonumber
\dot{E}_{\rm abs}&=&32\pi\left(M_{\rm BH}\mu \right)^3\dot{E}_{\rm in}=\frac{16\pi}{125}\frac{M_{\rm BH}^5}{M_{\rm NBS}^5}\left(M_{\rm NBS}\mu \right)^{10}\,.
\eeq
We have tested the above physics with a series of toy models, including the study of accretion of a massive, non self-gravitating scalar confined in a spherical cavity
with a small BH at the center (\textit{see} Appendix~\ref{app:bh_bomb}). This toy model conforms to the physics just outlined. One example, summarized in Appendix~\ref{app:string_toy}, suggests that all modes
of the NBS are excited during such an accretion process, but made quasinormal (i.e., damped) by the presence of the absorption. These are all low-frequency modes, and our argument should be valid even in such circumstance.

With $\dot{E}_{\rm abs}=\dot{M}_{\rm BH}$ and fixed NBS mass, one finds the timescale
\beq
\tau&\sim& \frac{1}{M_{\rm BH}^4M_{\rm NBS}^5\mu^{10}}\nonumber\\
&=&10^{24}\,{\rm yr}\,\frac{M_{\rm NBS}}{10^{10}M_{\odot}}\left(\frac{\chi}{10^4}\right)^4\left(\frac{0.1}{M_{\rm NBS}\mu}\right)^{10}\,,
\eeq
where $\chi\equiv M_{\rm NBS}/M_{\rm BH}$. 
In other words, the timescale for the BH to increase substantially its mass -- which we take as a conservative indicative of the lifetime of the entire NBS -- is
larger than a Hubble timescale for realistic parameters. This timescale is the result of forcing the BH with a nearly monochromatic field from the NBS. When the material of the star is nearly exhausted, a new timescale is relevant, that of the quasinormal modes of the BH surrounded by a massive scalar.
This timescale is $\tau_{\rm QNM}\sim M_{\rm BH}(M_{\rm BH}\mu)^{-6}<\tau$~\cite{Detweiler:1980uk,Brito:2015oca}, but still typically larger than a Hubble time.

When rotation is included, the entire setup may become even more stable: rotation is able to provide energy, via superradiance, to the surrounding field, and sustain
nearly stationary, but non spherically-symmetric, configurations~\cite{Herdeiro:2014goa,Brito:2015oca}. We will not discuss these effects here.

\subsection{Massive objects plunging into boson stars\label{Plunging_particle_BS}}
%
\begin{figure}[t]
\includegraphics[width=8cm,keepaspectratio]{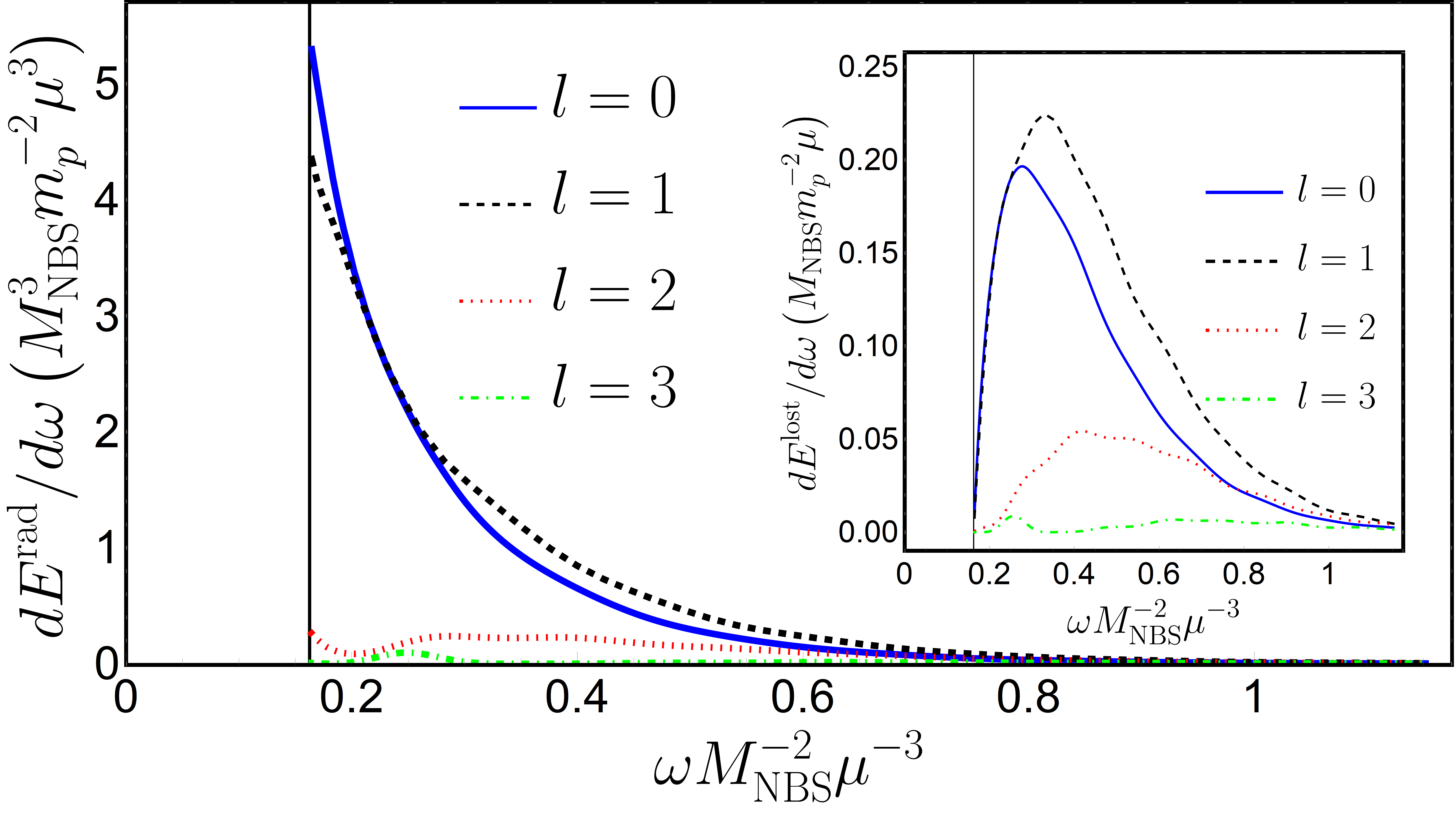} 
\includegraphics[width=8cm,keepaspectratio]{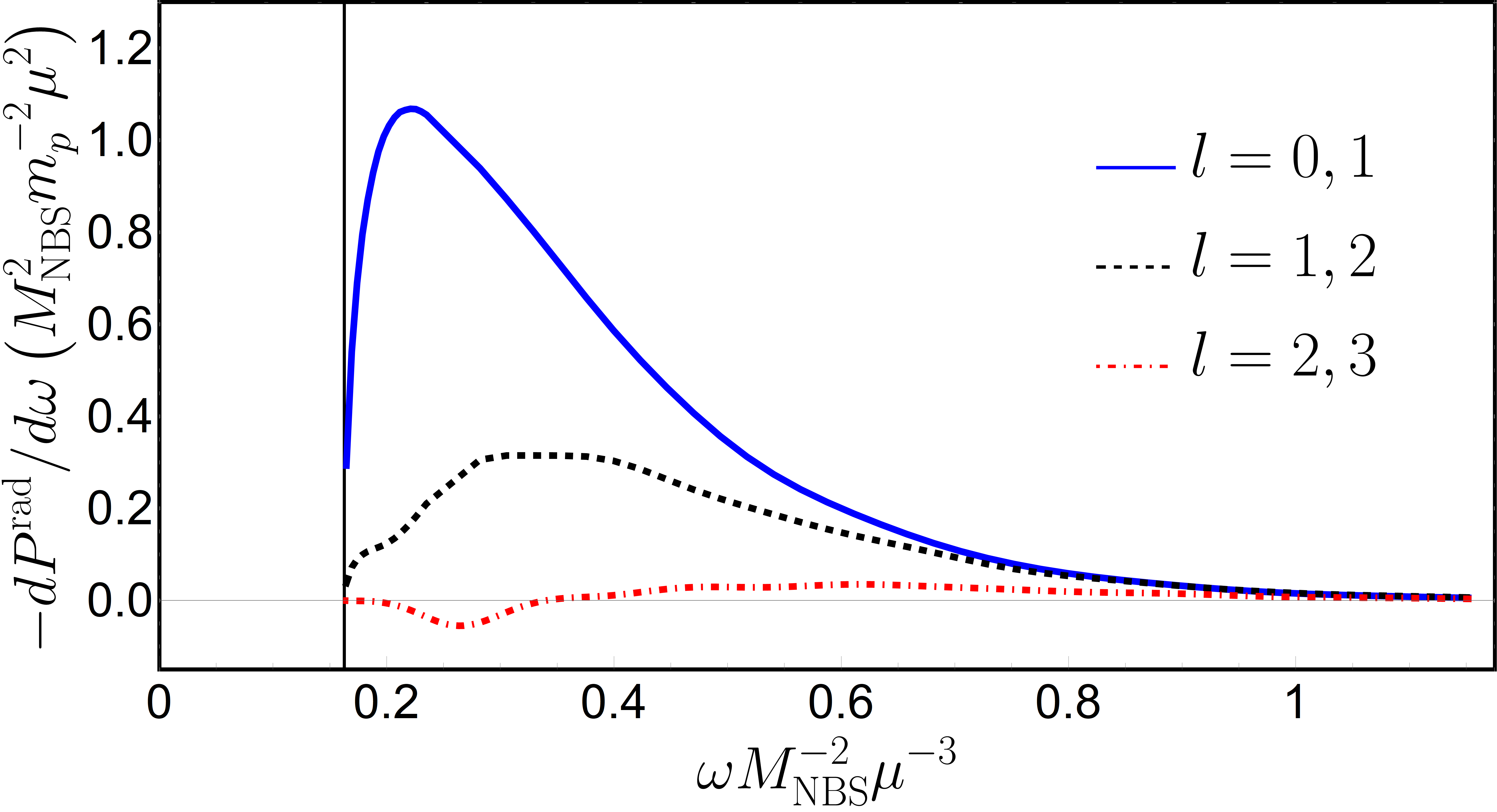}
\caption{Spectrum of radiation released when an object of mass $m_p$ plunges through an NBS with initial velocity $v_R\approx 0 $. Emission takes place for frequencies $\omega>\gamma$ (\textit{see} Eqs~\eqref{Energy_flux_BS}-\eqref{Momentum_flux_BS}). Upper panel: lowest multipole contribution $l=0,1,2,3$ to the total spectral flux of energy. Inset: multipole contributions to the radiated kinetic energy of the scalar field. Lower panel: spectral fluxes of linear momentum along $z$ associated with the lowest multipoles. The results obtained for other plunging velocities are summarized in Eqs~\eqref{eq:fit_BS_Erad_plunge_gravityon}-\eqref{eq:fit_BS_Prad_plunge_gravityon}.}
	\label{fig:PlungingSpectrabosonstar_gravityon}
\end{figure}
Consider now a massive perturber plunging, head-on, into an NBS. The perturber is assumed to have traveled from far away, but for our purposes the only relevant quantity is the perturber velocity
when it reaches the NBS surface, $\boldsymbol{v}=-v_{R}\boldsymbol{e}_z$, with $v_R\geq 0$. This setup is described in detail in Sec.~\ref{sec:External perturbers}. As we argued before (and also below), this situation could describe a massive BH ``kicked'' at formation, via GW emission, in a DM core of light fields, or, simply, stars crossing an NBS. Our framework allow us to do the first self-consistent computation of the gravitational drag acting on perturbers in such systems. Including the effect of the NBS gravitational potential on the perturber motion sets a natural critical velocity in the problem, the escape velocity $v_{\rm esc}$. For the fundamental NBS described in Fig.\ref{fig:BS}, the velocity needed to escape from the surface of the NBS is $v_{\rm esc}\sim 0.47M_{\rm NBS}\mu$.
When the velocity is smaller than this, the crossing object should be confined in the NBS with an oscillatory motion. 
For now, we study a simple one-way motion, and assume that when the particle crosses the NBS once, it simply ``disappears''. This will allow us to estimate the dynamical friction on the perturber.
This assumption is formally correct and accurate for unbound motion. For bound oscillatory motion it is not, and we work out the full case below, in Section~\ref{oscillating_particle_BS}.

Some quantities of interest are the spectral fluxes of energy and linear momentum radiated in these processes, as well as the energy lost by the perturber. These are given, respectively, by Eqs.~\eqref{Energy_flux_BS}-\eqref{Momentum_flux_BS} and \eqref{Energy_loss_BS}. The upper panel of Fig.~\ref{fig:PlungingSpectrabosonstar_gravityon} shows the contribution of the lowest multipoles to the total energy spectrum $d E^{\rm rad}/d \omega$ ($d E^{\rm lost}/d \omega$ in inset). This result was obtained through the numerical evaluation of expressions~\eqref{Energy_flux_BS}-\eqref{Energy_loss_BS} for a perturber plunging into an NBS, starting the fall from rest at~$R$. The fluxes converge exponentially with increasing values of $l$, after a sufficiently large~$l$. Our results are compatible with~$E_l^{\rm rad}\propto e^{-l}$, where~$E_l^{\rm rad}$ is the $l$-mode contribution to the energy radiated. Once the behavior of $E_l^{\rm rad}$ for large $l$ is known, one can find the total energy radiated.
For a particle plunging with zero initial velocity into an NBS we obtain $E^{\rm rad}\sim 1.28 \, m_p^2/M_{\rm NBS}$ and $E^{\rm lost}\sim 0.18 \, m_p^2 M_{\rm NBS}\mu^2$. 
Applying this procedure to other velocities, we find that the following is a good description of our results,
\be 
E^{\rm rad}=29\frac{m_p^2}{M_{\rm NBS}}\frac{e^{-3.25/X}}{X^{17/4}}\,\label{eq:fit_BS_Erad_plunge_gravityon}
\ee
\be 
E^{\rm lost}=7\,m_p^2 M_{\rm NBS}\mu^2\frac{e^{-3.54\,\left(X-0.05\right)^{-1}}}{\left(X-0.05\right)^{17/4}}\,\label{eq:fit_BS_Ekin_plunge_gravityon}
\ee
accurate to within $5\%$ of error for~$0\lesssim v_R\lesssim 2.5M_{\rm NBS}\mu$. This interval spans over non-relativistic astrophysical relevant velocities ({\it e.g.}, $0\lesssim v_R[{\rm km/s}]\lesssim 6000$ for the DM core of the Milky Way). 
Here, 
\be
X\equiv \frac{v_R}{M_{\rm NBS}\mu}+0.68\,.
\ee

The lower panel of Fig.~\ref{fig:PlungingSpectrabosonstar_gravityon} shows the multipolar contribution to the spectral flux of linear momentum along $z$. The linear momentum radiated also converges exponentially in $l$, after a sufficiently large $l$. For a perturber starting at rest, the total linear momentum radiated along $z$ in the whole process is $P^{\rm rad}\sim -0.43 \, m_p^2\mu$.
The fitting expression
\be 
P^{\rm rad}=-2.4\,m_p^2\mu\frac{e^{-2.26\,\left(X-0.27\right)^{-1}}}{\left(X-0.27\right)^{17/4}}\,,\label{eq:fit_BS_Prad_plunge_gravityon}
\ee
is a good approximation to our results (within $5\%$ of error for~$0 \lesssim v_R\lesssim 2.5M_{\rm NBS}\mu$). 
\begin{figure}
\includegraphics[width=8cm,keepaspectratio]{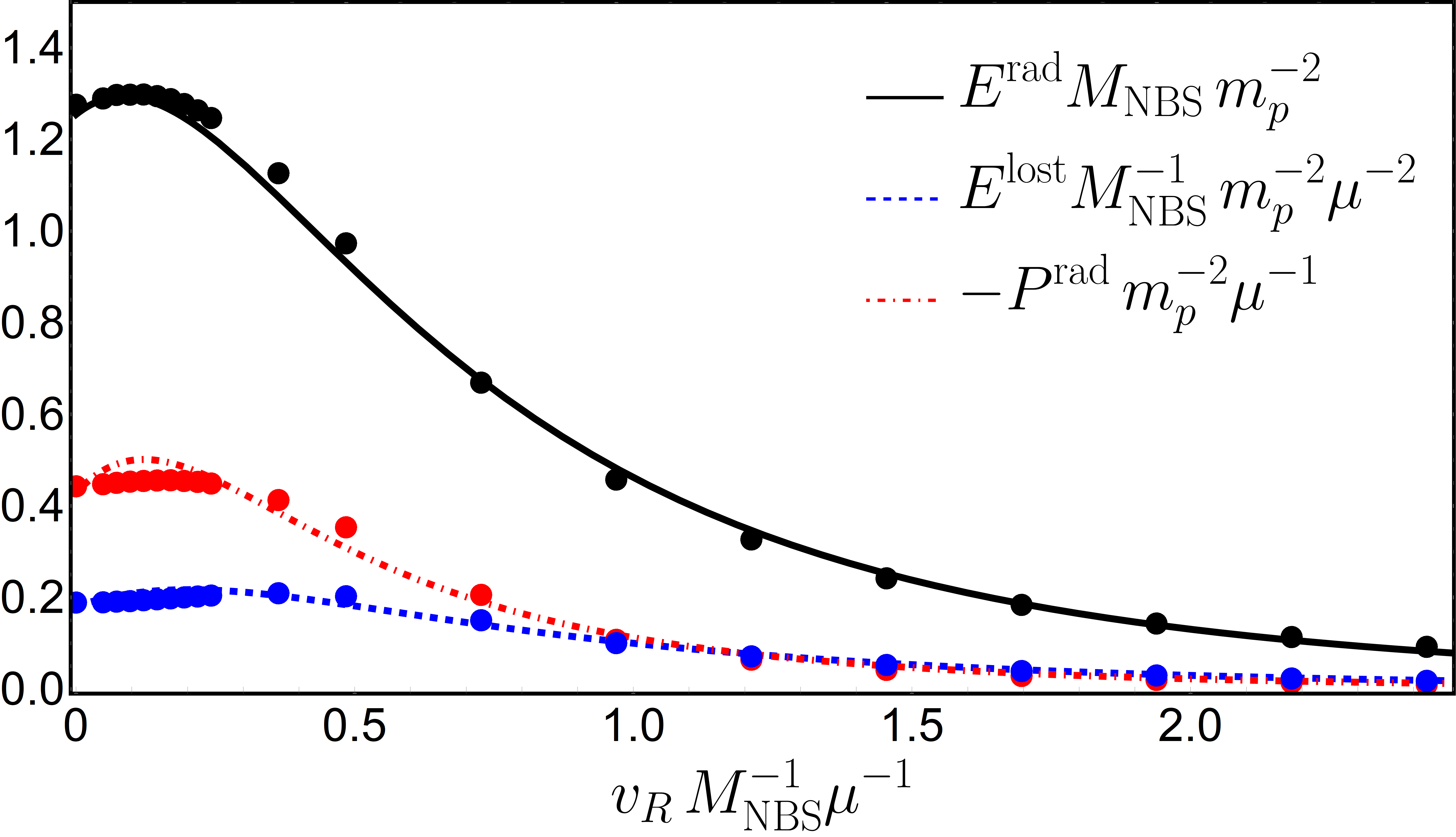} 
\caption{Total and kinetic energy, and linear momentum emitted when an object of mass $m_p$ plunges through an NBS, as a function of the initial perturber velocity. The dots correspond to the numerical data used to obtain Eqs.~\eqref{eq:fit_BS_Erad_plunge_gravityon}-\eqref{eq:fit_BS_Ekin_plunge_gravityon}-\eqref{eq:fit_BS_Prad_plunge_gravityon}.}
	\label{fig:fitsBS}
\end{figure}
Figure~\ref{fig:fitsBS} shows how the total radiated energy $E^{\rm rad}$, the total energy lost by the moving perturber $E^{\rm lost}$, and the linear momentum radiated $P^{\rm rad}$ vary with the change of initial velocity.

The momentum lost by a small plunging object ($m_p\mu \ll v_R$) is given by $P^{\rm lost}=-E^{\rm lost}/v_R$, as shown in Eq.~\eqref{EvsPloss}.
We have thus computed, in a self-consistent way, the dynamical friction acting upon a body moving within a NBS. The quantity $E^{\rm lost}$ is the actual kinetic energy lost by the perturber as it crosses the NBS. Note that, in accordance with the results for the energy lost -- in particular, its sign -- this is indeed a friction; the body will slow down. On the other hand, the results for the energy lost together the radiated momentum show that the NBS will acquire a small momentum in the direction of the moving perturber, described by Eq.~\eqref{eq:NBS_momentum}; note the two lines crossing each other close to~$v_R= M_{\rm NBS} \mu$ in Fig.~\ref{fig:fitsBS}.

Our results should be compared and contrasted with those of Ref.~\cite{Hui:2016ltb,Lancaster:2019mde}, where dynamical friction in these structures was estimated
without including self-gravity (therefore not accounting for the size of the scalar structure either). In contrast to those of Ref.~\cite{Hui:2016ltb}, our results are self-consistent, regular and finite at all velocities. In Appendix~\ref{app:drag}, we look at a simple toy model which indicates that the discrepancy between these results may be partially related with the trivial gravitational potential of the background medium. A non-trivial gravitational potential can confine small-frequency scalars, suppressing efficiently scalar emission. Nevertheless, the self-gravity of the scalar seems to help suppressing scalar emission for small velocities. Thus, the above results are the first self-consistent and accurate calculation of dynamical friction caused by a self-gravitating scalar on passing objects.

\subsection{A perturber oscillating at the center\label{oscillating_particle_BS}}
%
As a black hole forms through gravitational collapse in a DM core it can be ``kicked'', via GW emission, and left in an oscillatory motion around the center of the core. 
The reason for the kick is that collapse is, in general, an asymmetric process, and leads to emission of GWs which carry some momentum. This process is known
to lead to velocities of at most a few hundred kilometers per second~\cite{1973ApJ...183..657B}, generally smaller than the galactic escape velocity. Thus, the remnant BH
is bound to the galaxy and, in absence of dissipation, performs an oscillatory motion.

It is crucial to understand how the DM core reacts to this motion and to quantify the energy and momentum radiated and deposited in the scalar field. Similar issues were addressed in Ref.~\cite{Gualandris:2007nm}, in the context of the interaction between a kicked supermassive black hole and stars in galaxy cores.

At the center of a NBS the energy density is approximately constant~$\rho_E\simeq 4\times 10^{-3} M_{\rm NBS}^4\mu^6$. So, the motion of the perturber is
\begin{align}
&z_p(t)=	-\mathcal{A} \sin\left(\omega_{\rm osc} t\right)\,, \nonumber \\
&\mathcal{A} \equiv \sqrt{\frac{3}{4 \pi} \frac{v_0^2}{\rho_ E}}\,, \qquad \omega_{\rm osc}\equiv \sqrt{\frac{4 \pi \rho_E}{3}}\,,
\end{align}
where~$v_0$ is the velocity of the perturber at the center of the core.
The source is described by
\begin{align}
&P=m_p \frac{\delta(\varphi)}{r^2 \sin \theta}\nonumber \\
&\times\left[\delta\left(r-z_p(t)\right)\delta\left(\theta\right)+ \delta\left(r+z_p (t)\right)\delta\left(\theta-\pi\right)\right].
\end{align}
Using Eq.~\eqref{p_def} the function~$p$ reads
\begin{align}
p&=\frac{m_p}{2\sqrt{2 \pi}}\frac{|\tau'_{1,n}(r)|}{r}Y_l^0(0) \delta_m^0 
\sum_{n \in \mathbb{Z}}\Big[e^{-i \omega \tau_{1,n}}+e^{-i \omega \tau_{2,n}}\nn \\
&+(-1)^n \left(e^{i \omega \tau_{1,n}}+e^{i \omega \tau_{2,n}}\right)\Big]\,,
\end{align}
where we defined~\footnote{The functions~$\tau_{1,n}(r)$ and~$\tau_{2,n}(r)$ are the roots of~$r+z_p(\tau)=0$; the symmetric functions~$-\tau_{1,n}(r)$ and~$-\tau_{2,n}(r)$ are the roots of~$r-z_p(\tau)=0$.}
\begin{align}
\tau_{1,n}&\equiv \frac{1}{\omega_{\rm osc}}\left[\arcsin\left(\frac{r}{\mathcal{A}}\right)+2n \pi\right] \,, \nonumber \\
\tau_{2,n}&\equiv \frac{1}{\omega_{\rm osc}}\left[(2n+1) \pi-\arcsin\left(\frac{r}{\mathcal{A}}\right)\right] \,.
\end{align}
In the last expressions we are using the principal branch of the inverse sine function. It is easy to see that the function~$p$ can be put in the form
\begin{align} \label{p_osc_start}
p&=\frac{m_p}{\sqrt{2 \pi}}\frac{Y_l^0(0)}{\sqrt{\mathcal{A}^2-r^2}}\frac{\delta_m^0}{\omega_{\rm osc}}\,\Theta\left(\mathcal{A}-r\right) \nonumber \\
&\times \sum_{n \in \mathbb{Z}}\bigg[\delta_l^{\rm even} \left(\cos\left[\omega \tau_{1,n}(r)\right]+\cos\left[\omega \tau_{2,n}(r)\right]\right) \nonumber \\
&-i\, \delta_l^{\rm odd} \left(\sin\left[\omega \tau_{1,n}(r)\right]+\sin\left[\omega \tau_{2,n}(r)\right]\right)\bigg]\,,
\end{align}
Using the mathematical identities
\begin{align*}
&\sum_{n \in \mathbb{Z}} \sin \left(2 n \pi \frac{\omega}{\omega_{\rm osc}}\right)=0 \,, \\
&\sum_{n \in \mathbb{Z}} \cos \left(2 n \pi \frac{\omega}{\omega_{\rm osc}}\right)=\omega_{\rm osc} \sum_{n \in \mathbb{Z}} \delta(\omega-n \omega_{\rm osc})\,,
\end{align*}
together with some trivial trigonometric identities, one can rewrite~\eqref{p_osc_start} as
\begin{align*} 
&p=m_p\sqrt{\frac{2}{\pi}}\frac{Y_l^0(0)}{\sqrt{\mathcal{A}^2-r^2}}\delta_m^0 \,\Theta\left(\mathcal{A}-r\right) \sum_{n \in \mathbb{Z}} \delta(\omega-2n\omega_{\rm osc}) \nonumber \\
&\times \bigg[\delta_l^{\rm even} \cos\left(2 n \arcsin\frac{r}{\mathcal{A}}\right)-i\, \delta_l^{\rm odd} \sin\left(2 n \arcsin\frac{r}{\mathcal{A}}\right)\bigg]\,.
\end{align*}
With the help of the trigonometric identities
\begin{align*}
&\cos(2n x)=\sum_{k=0}^{n}(-1)^k \binom{2n}{2k}\sin^{2k} x \cos^{2(n-k)}x\,, \\
&\sin(2n x)=\sum_{k=0}^{n-1}(-1)^k \binom{2n}{2k+1}\sin^{2k+1} x \cos^{2(n-k)-1}x\,,
\end{align*}
the last expression can be written in the alternative form
\begin{align} 
&p=m_p\sqrt{\frac{2}{\pi}}Y_l^0(0)\delta_m^0 \,\Theta\left(\mathcal{A}-r\right) \sum_{n \in \mathbb{Z}} \frac{1}{\mathcal{A}^{2n}}\delta(\omega-2n\omega_{\rm osc}) \nonumber \\
&\times \bigg[-i\, \delta_l^{\rm odd} \sum_{k=0}^{n-1}(-1)^k \binom{2n}{2k+1}r^{2k+1}\left(\mathcal{A}^2-r^2\right)^{n-k-1} \nonumber \\
&+\delta_l^{\rm even}\sum_{k=0}^{n}(-1)^k \binom{2n}{2k} r^{2k}\left(\mathcal{A}^2-r^2\right)^{n-k-\frac{1}{2}}\bigg]\,.
\end{align}

We want to calculate the energy radiated through scalar waves due to the oscillatory motion of the massive object. First, note that the oscillation frequency is $\omega_{\rm osc}\sim 0.135M^2_{\rm NBS}\mu^3\lesssim \gamma$. Only the modes with $n\geq 1$ arrive at infinity; so, only these contribute to the energy radiated.
Applying the formalism described in Section~\ref{sec:SmallPert}, we obtain
\begin{align}
&Z_1^\infty=4 \pi \int_0^\mathcal{A}dr'F_{4,6}^{-1}(r')p(r')\,, \nonumber \\
&Z_2^\infty(\omega,l,0)=Z_1^\infty(-\omega,l,0)^*\,.
\end{align}
The energy radiated per unit of time is (\textit{see} Eq.~\eqref{Energy_flux_rate})
\begin{align}
&\dot{E}^{\rm rad}=\frac{2}{\pi} \sum_{l, n}\left(\mu-\gamma+2n\omega_{\rm osc}\right) \nonumber \\
&\qquad\times {\rm Re}\left[\sqrt{2 \mu (2 n \omega_{\rm osc}-\gamma)}\right]|\widetilde{Z}_1^\infty(2n \omega_{\rm osc},l,0)|^2  \nn \\
&\simeq\frac{2}{\pi}\mu \sum_{l, n}  {\rm Re}\left[\sqrt{2 \mu (2 n \omega_{\rm osc}-\gamma)}\right]|\widetilde{Z}_1^\infty(2n \omega_{\rm osc},l,0)|^2 \,,
\end{align}
where we used the low-energy limit~$\gamma \ll \mu$ and~$\omega_{\rm osc}\ll \mu$, and defined
\begin{align*}
&\widetilde{Z}_1^\infty\equiv 4 \pi \int_0^\mathcal{A}dr'F_{4,6}^{-1}(r')\widetilde{p}(r')\,, \\
&\widetilde{p}\equiv m_p\sqrt{\frac{2}{\pi}}\frac{Y_l^0(0)}{\sqrt{\mathcal{A}^2-r^2}} \sum_{n \in \mathbb{Z}}  \bigg[\delta_l^{\rm even} \cos\left(2 n \arcsin\frac{r}{\mathcal{A}}\right)\nonumber \\
&\qquad-i\, \delta_l^{\rm odd} \sin\left(2 n \arcsin\frac{r}{\mathcal{A}}\right)\bigg]\,.
\end{align*}

One can anticipate that the dominant contribution to the radiation is given by the~$n=1$ mode, which has a frequency~$\omega=2\omega_{\rm osc}$. This is the lowest frequency radiated by the perturber and, thus, we expect it to be the one carrying more energy, because the coupling between the perturber and the scalar is stronger for lower frequencies -- as will become evident in the following sections. Indeed, this is in accordance with our numerics. So, we focus on the single~$n=1$ mode. For oscillations deep inside the NBS with an amplitude~$\mathcal{A}\ll R$ -- which is where our constant density approximation holds -- we find that the following semi-analytic expression is a good description of our numerical results: 
\begin{align}
\dot{E}^{\rm rad}&=\frac{2 \sqrt{2}}{\pi} (m_p\mu)^2 \sqrt{ \frac{2\omega_{\rm osc}-\gamma}{\mu}}\sum_l c_l\, \left(\frac{\mathcal{A}}{R}\right)^{2(l+1)},
\end{align}
with the numerical constants~$c_l$. For the first multipoles we find
\begin{align*}
&c_0\simeq 0.852\,, \qquad c_1 \simeq 67.7 \,, \qquad c_2 \simeq 30.4\,, \nonumber \\
&c_3 \simeq 438\,, \qquad\;\;\; c_4 \simeq 13.6\,, \qquad c_5\simeq 3.85\,.
\end{align*}
The above expression describes our numerics with less than~$1\%$ of error for $\mathcal{A}/R \lesssim 0.09$. These amplitudes correspond to kicks of~$v_0\lesssim 0.1 M_{\rm NBS}\mu$, which contains astrophysical relevant velocities; for the Milky Way DM core our expression covers~$v\lesssim 300\, {\rm km/s}$, which contains typical recoil velocities imparted by GW emission in gravitational collapse. Larger kicks, like the ones delivered in a merger of two supermassive BHs, have larger amplitudes and are out of our approximation. However, the framework of Section~\ref{sec:SmallPert} (without the constant density approximation) can still be applied to those cases.

Using the same reasoning that we applied to the \textit{orbiting particles} to deduce Eq.~\eqref{eq:E_loss_circular}, we can estimate the perturber's energy loss per unit of time to be
\begin{align}
&\dot{E}^{\rm lost}=\frac{2}{\pi} \sum_{l, n}\left(2n\omega_{\rm osc}-\gamma\right) \nonumber \\
&\qquad\times {\rm Re}\left[\sqrt{2 \mu (2 n \omega_{\rm osc}-\gamma)}\right]|\widetilde{Z}_1^\infty(2n \omega_{\rm osc},l,0)|^2\,.
\end{align}
Considering the single (dominant)~$n=1$ mode, the numerical evaluation of the last expression is well described by the semi-analytic formula
\begin{align}
\dot{E}^{\rm lost}&=\frac{2 \sqrt{2}}{\pi} (m_p\mu)^2 \left( \frac{2\omega_{\rm osc}-\gamma}{\mu}\right)^{\frac{3}{2}}\sum_l c_l\, \left(\frac{\mathcal{A}}{R}\right)^{2(l+1)}.
\end{align}
Again, this describes our numerics with less than~$1\%$ of error for small amplitude oscillations~$\mathcal{A}/R\leq 0.09$. 

One may wonder how long it takes for a kicked BH (or star) to settle down at the center of an halo purely due to the dynamical friction caused by dark matter. When the condition
\begin{equation}\label{AdiabatCond}
\frac{\dot{E}^{\rm lost} \left(\frac{2\pi}{\omega_{\rm osc}}\right)}{\frac{1}{2}m_p \omega_{\rm osc}^2 \mathcal{A}^2} \ll 1
\end{equation}
is verified, the system is suited to an adiabatic approximation, and we can compute how the amplitude changes with time by solving
\begin{align}
m_p \omega_{\rm osc}^2 \mathcal{A} \,\dot{\mathcal{A}}=-\dot{E}^{\rm lost}\,.
\end{align}
Several astrophysical systems fall within this approximation. For example, the Milky Way dark matter core has a mass~$M_{\rm NBS} \mu \sim 10^{-2}$; so, for an object forming through gravitational collapse and receiving a kick of~$300\,{\rm km/s}$, via GW emission, the adiabatic approximation is suitable if~$m_p/M_{\rm NBS} \ll 0.1$ -- which is verified by all known objects. Using only the dominant multipole~$l=0$ (which accounts for more than~$61\%$ of the total energy loss for~$\mathcal{A}/R\leq 0.09$, and more than~$89\%$ for~$\mathcal{A}/R\leq0.04 $) we obtain
\begin{align}
\mathcal{A}=\mathcal{A}_0\, e^{-t/\tau_{\rm s}}\,,
\end{align}
with the timescale
\begin{align}
&\tau_{s}\simeq \frac{56}{m_p M_ {\rm NBS} \mu^3} \nn \\
&\sim 10^{10} {\rm yr} \left(\frac{10^{-22}\, {\rm eV}}{\mu}\right)^2\left(\frac{10^5 M_ \odot}{m_p}\right)\left(\frac{0.01}{M_ {\rm NBS}\mu}\right)\,. 
\end{align}
So, an object kicked at the center of a NBS, interacting solely with the scalar, settles down in a timescale smaller than the Hubble time if it has a mass~$m_p\gtrsim 10^5 M_\odot$; in other words, if it is a supermassive BH. 

The above timescale is in general much larger than the period of oscillation,
\begin{equation}
	 \tau_s  \sim \frac{M_{\rm NBS}}{m_p} \left(\frac{2 \pi}{\omega_{\rm osc}}\right)\,.
\end{equation}
This suggest that treating the source as eternal is indeed a good approximation to study this process.
It is interesting to compare this result with the timescale of damping due to dynamical friction caused by stars in the galactic core. In Ref.~\cite{Gualandris:2007nm} the authors estimate that timescale to be
\begin{equation}
	\tau^*\sim 0.1\, \frac{M_{\rm c}}{m_p} \left(\frac{2 \pi}{\omega_{\rm osc}}\right) \,,
\end{equation}
where~$M_{\rm c}$ is the galactic core mass.
Using~$M_{\rm c}=M_{\rm NBS}$ we see that~$\tau^* \sim 0.1\,\tau_s$, which is smaller but still comparable to~$\tau_s$. Both ours and Ref.~\cite{Gualandris:2007nm} calculations are order of magnitude estimates, but our result suggests that dark matter may exert a dynamical friction comparable to the one caused by stars for processes happening in galactic cores.

\subsection{Low-energy binaries within boson stars\label{Orbiting_particle_BS}}
%
\begin{figure}[ht]
\includegraphics[width=8.4cm,keepaspectratio]{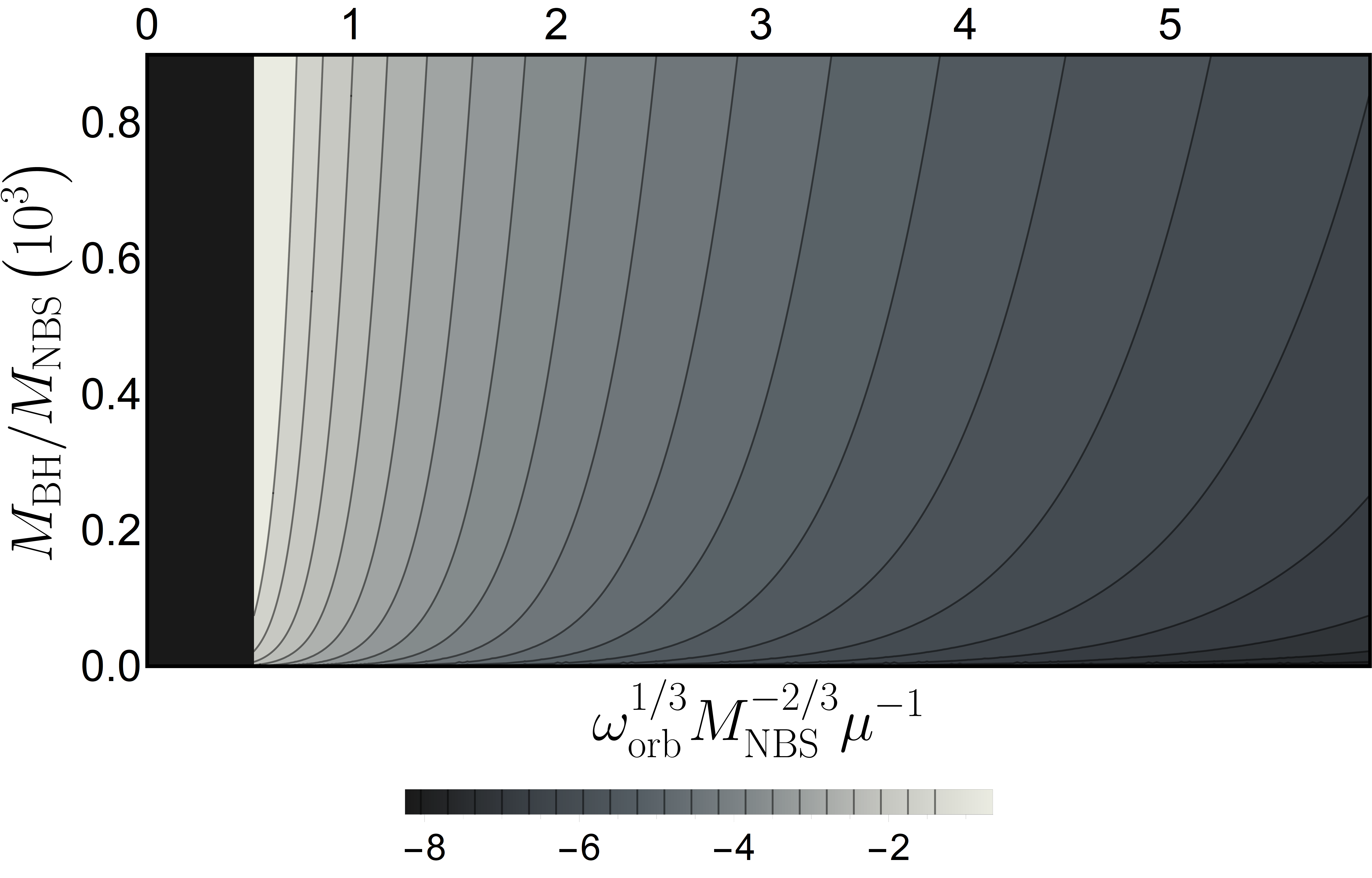} 
\caption{Logarithm of the universal rate of scalar energy radiated by an EMRI orbiting inside an NBS: $\log_{10}\left[\dot{E}_{\rm EMRI}^{\rm rad}\left(m_p^2 M_{\rm NBS}\mu^3\right)^{-1}\right]$.
The EMRI is described by a supermassive BH of mass $M_{\rm BH}$ sitting at the NBS center, and a star or stellar-mass BH in a circular orbit around it. 
Note that the maximum energy emitted is associated with the smallest frequency (largest distance). This is due to oscillating background field which imparts an energy $\mu$ to any wave. 
For a DM core with $M_{\rm NBS}\sim 10^{10} M_{\odot}$ and mass ratio $m_p/M_{\rm BH}\sim 10^{-4}$, the orbital distances corresponding to nonzero fluxes are in the range $r_{\rm orb}\lesssim 10^6 \,M_{\rm BH}$. For larger radii, the fluctuation has too low an energy and is confined to the structure. This explains the zero-flux (black) region on the left of the panel, corresponding to the suppression of perturbations with frequency $\omega \leq\gamma$.}  	
\label{fig:circularEMRI}
\end{figure}
We now focus on orbiting objects within such an NBS. These will describe binaries, either at an early or late stage in their life,
stirring the field and producing disturbances in the local DM profile. For example, looking at the matter moments in Eq.~\eqref{p_orbiting}, such systems can describe stars orbiting around the SgrA$^*$ BH at the center of the Milky Way. The supermassive BH has a mass $\sim 4\times 10^6 M_{\odot}$ with known companions. The closest known star, S2, has a pericenter distance of $\sim 2800M_{\rm BH}$ and a mass $m_p\sim 20 M_{\odot}$ with a large uncertainty~\cite{Abuter:2018drb,Abuter:2020dou}. Its orbit is, however, highly eccentric. Given the mass and sizes of the NBSs discussed here (i.e. which described the core of DM haloes) all these systems can be handled via perturbation techniques.
In addition, binaries close to supermassive BHs, and therefore to galactic centers, have been observed recently via electromagnetic counterparts to GWs~\cite{Graham:2020gwr}.
\subsubsection{Scalar emission}
Let us consider first an EMRI: a perturber of mass $m_p$ orbiting a supermassive BH, of mass $M_{\rm BH}\gg m_p$ placed at the center of a NBS. Solving the perturbation equations~\eqref{BS_Perturbation_Matrix_Sourced}, with the source defined in Eq.~\eqref{p_orbiting}, with $m_p(1+(-1)^m)\to m_p$, we find that, up to $3\%$ accuracy, the fluxes of energy (Eqs.~\eqref{eq:circular_flux_Erad}-\eqref{eq:E_loss_circular}) are described by~\footnote{Notice that in principle, the emission would starts for frequency larger than $\left(\gamma m^{-1}\right)$. However, since the emission in multipoles higher than the dipole is suppressed by roughly factor $10^3$, we consider only $l=1$ in \eqref{Erad_circular_BS}.}
\beq \nn
&&\dot{E}^{\rm rad}_{\rm EMRI}= 10^{-2}\, m_p^2 M_{\rm BH}^{2/3}M_{\rm NBS}^{4} \mu^{17/2}\omega_{\rm orb}^{-11/6}\Theta\left[\omega_{\rm orb}-\gamma\right]\times\nonumber\\
&&\Big[2.66- 0.49 \, M_{\rm NBS}^{4/3}\mu^2\omega_{\rm orb}^{-2/3} +0.054 \, M_{\rm NBS}^{8/3}\mu^4\omega_{\rm orb}^{-4/3} \Big],\label{Erad_circular_BS}\\
&&\dot{E}^{\rm lost}_{\rm EMRI}= 10^{-2}\, m_p^2 M_{\rm BH}^{2/3}M_{\rm NBS}^{4} \mu^{15/2}\omega_{\rm orb}^{-5/6}\Theta\left[\omega_{\rm orb}-\gamma\right]\times\nonumber\\
&&\Big[2.70- 0.96 \, M_{\rm NBS}^{4/3}\mu^2\omega_{\rm orb}^{-2/3} +0.043 \, M_{\rm NBS}^{8/3}\mu^4\omega_{\rm orb}^{-4/3} \Big].\label{Elost_circular_BS}
\eeq
%
Equations \eqref{Erad_circular_BS}-\eqref{Elost_circular_BS} were evaluated assuming a non-relativistic perturbation, therefore are valid for orbital periods 
$T=2\pi/\omega_{\rm orb}\gg 2\pi/\mu\sim 10^{-22}{\rm eV}/\mu\, {\rm yr}$. 
We show in Fig.~\ref{fig:circularEMRI} the flux of energy ($E^{\rm rad}$) as a function of the orbital period and of the BH-NBS mass ratio. Once the orbital frequency is fixed, our results are consistent with exponential convergence in $l$ for the flux.

The calculation above is easy to adapt to other systems. Consider an equal mass binary system ($M=2m_p$). Looking at the matter moments in Eq.~\eqref{p_orbiting}, it is clear that the first multipole moment that is going to be emitted is the quadrupole $m=2$. As a result of solving the perturbation equations, we find the following expression for the energy emitted in scalar waves, and the one lost by the orbiting particle (up to $3\%$ of accuracy)
\beq
&&\hspace{-0,65cm}\dot{E}^{\rm rad}= 10^{-2}\, M^{4/3} m_p^{2}M_{\rm NBS}^{4} \mu^{19/2}\omega_{\rm orb}^{-13/6}\Theta\left[2\omega_{\rm orb}-\gamma\right] \nn\\
&&\hspace{-0,65cm} \left[1.45- 0.16 \, M_{\rm NBS}^{4/3}\mu^2\omega_{\rm orb}^{-2/3} +0.015 \, M_{\rm NBS}^{8/3}\mu^4\omega_{\rm orb}^{-4/3} \right],\label{Erad_circular_BS_equalmass}\\
&&\hspace{-0,65cm}\dot{E}^{\rm lost}=10^{-2}\, M^{4/3} m_p^{2}M_{\rm NBS}^{4} \mu^{17/2}\omega_{\rm orb}^{-7/6}\Theta\left[2\omega_{\rm orb}-\gamma\right]  \nn\\
&&\hspace{-0,65cm} \left[2.97- 0.58 \, M_{\rm NBS}^{4/3}\mu^2\omega_{\rm orb}^{-2/3} +0.0051 \, M_{\rm NBS}^{8/3}\mu^4\omega_{\rm orb}^{-4/3} \right].\label{Elost_circular_BS_equalmass}
\eeq
The expression above is valid both for solar mass BHs as well as for BH masses of the order $\sim 10^4 M_{\odot}$.

In the limit of an high-frequency ($\omega_{\rm orb} \gg \gamma, \mu U_0$), but still non-relativistic ($\omega_{\rm orb} \ll \mu$) excitation, 
the relevant equations~\eqref{Sourced_SP_System1}-\eqref{Sourced_SP_System2} can be solved analytically in closed form, noticing that $|\Psi_0 \delta \Psi|\ll |\delta U|$. 
Equation~\eqref{Sourced_SP_System2} therefore reduces simply to 
\begin{align}
\nabla^2 \delta U=4 \pi P\,,\label{Sourced_SP_System2HF}
\end{align}
which has the solution
\begin{align}
&\delta U=\frac{2}{\sqrt{2 \pi}}\sum_{l,m} \frac{u(r)}{r} Y_l^m(\theta, 0) e^{-i m \left(\omega_{\rm orb} t -\varphi\right)} \,,
\end{align}
with
\begin{align}
&u=-\left(2 \pi\right)^{3/2} m_p \left[1+(-1)^m\right]\frac{Y_l^m\left(\frac{\pi}{2},0\right)}{2l+1} \nonumber \\
&\times  \left[\left(\frac{r}{r_{\rm orb}}\right)^{-l} \Theta(r-r_{\rm orb})+\left(\frac{r}{r_{\rm orb}}\right)^{l+1} \Theta(r_{\rm orb}-r)\right]\,.\nonumber
\end{align}
Then, using the decomposition
\begin{align}
\delta \Psi = \frac{2}{\sqrt{2 \pi}}\sum_{l,m} \frac{Z(r)}{r} Y_l^m(\theta, 0)e^{-i m\left(\omega_{\rm orb} t-\varphi\right)}\,,
\end{align}
equation~\eqref{Sourced_SP_System1} becomes
\begin{align}
\partial_r^2 Z+\left(2 \mu m\omega_{\rm orb}-\frac{l(l+1)}{r^2}\right)Z=2 \mu^2 \Psi_0  u\,.
\end{align}
Using the method of variation of parameters, one can solve the last equation imposing the Sommerfeld radiation condition at large distances and regularity at the origin. The obtained solution is, at large distances, 
\begin{align} \label{vop_hf}
Z(r \to \infty)= i \pi\mu^2Z_\infty(r\to \infty) \int_{0}^{\infty} dr' Z_0\Psi_0 u\,,
\end{align}
where $Z_0$ and $Z_\infty$ are homogeneous solutions satisfying, respectively, regularity at the origin and the Sommerfeld radiation condition at large distances, and are given by
\begin{align}
Z_0&=\sqrt{r}\, J_{l+1/2}\left(\sqrt{2 \mu m \omega_{\rm orb}}r\right)\,, \\
Z_\infty&= \sqrt{r} H^{(1)}_{l+1/2}\,\left(\sqrt{2\mu m \omega_{\rm orb}}r\right)\,,
\end{align}
with $J_\nu(x), H^{(1)}_{\nu}(x)$ Bessel and Hankel functions~\cite{Abramowitz:1970as}.
Using the asymptotic form 
\begin{equation}
Z_\infty (r\to \infty)\simeq (-i)^{l+1}\sqrt{\frac{2}{\pi}} \frac{e^{i \sqrt{2\mu m \omega_{\rm orb}}\,r }}{\left(2\mu m \omega_{\rm orb}\right)^{1/4}}\,,
\end{equation} 
and assuming that~$r_{\rm orb}\ll R$, and~$\omega_{\rm orb}/\mu\gg \left(r_{\rm orb} \mu\right)^{-2}$, the integration in~\eqref{vop_hf} converges a few wavelengths from the binary and gives
\begin{align}
&Z(r \to \infty) \simeq -(-i)^l \left(2 \pi\right)^2 \mu^{2} m_p \Psi_0(0)r_{\rm orb}^l\nonumber \\
&\times \left[1+(-1)^m\right]  \frac{2^{-\frac{l}{2}-\frac{3}{2}}\,e^{i \sqrt{2\mu m \omega_{\rm orb}}\,r }}{\left(\mu m \omega_{\rm orb}\right)^{1-\frac{l}{2}}} \frac{Y_l^m\left(\frac{\pi}{2},0\right)}{\Gamma\left(l+\frac{3}{2}\right)}\,.
\end{align}
So, the dominant $l=m$ modes give the scalar perturbation
\begin{align}
&\delta \Psi (r\to \infty) \simeq -8 \pi^{\frac{3}{2}}\mu^2 m_p \Psi_0(0) \sum_{m=1}^{+\infty}(-i)^m\left[1+(-1)^m\right] \nonumber \\
&\times  \frac{Y_l^m\left(\frac{\pi}{2},0\right)}{\Gamma\left(m+\frac{3}{2}\right)} \frac{(\mu m)^{\frac{m}{2}-1}(M \omega_{\rm orb})^{\frac{m}{3}}}{2^{2+\frac{m}{2}}\omega_{\rm orb}^{\left(1+\frac{m}{2}\right)}}e^{i \sqrt{2\mu m \omega_{\rm orb}}\,r }\,,
\end{align}
where we have used Kepler's law $r_{\rm orb}^3=M/\omega_{\rm orb}^2$.
Then, the flux of energy is given by
\begin{align}
&\dot{E}^{\rm rad}=- r^2 \lim_{r \to \infty}\int d\theta d\varphi \sin \theta \,T^S_{t r}\nonumber \\
&= 0.28\, \pi^{3} \left(\mu m_p\right)^2 \left(\mu M_{\rm NBS}\right)^4 \sum_{m=1}^{+\infty}\left[1+(-1)^m\right]^2 \nonumber\\ &\times\left(1+\frac{m \omega_ {\rm orb}}{\mu}\right) \left(\frac{Y_m^m\left(\frac{\pi}{2},0\right)}{\Gamma\left(m+\frac{3}{2}\right)} \frac{m^{\left(\frac{m}{2}-\frac{3}{4}\right)}(M\omega_{\rm orb})^{\frac{m}{3}}}{2^{\left(\frac{7}{4}+\frac{m}{2}\right)}(\omega_{\rm orb}/\mu)^{\left(\frac{3}{4}+\frac{m}{2}\right)}}\right)^2\,.
\end{align}
The last expression can be further simplified using~$\left(1+m\omega_ {\rm orb}/\mu\right)\simeq 1$, since we are considering low-energy excitations of the scalar field.
The same reasoning that we used to derive~\eqref{eq:E_loss_circular} can be applied here to find that the binary loses energy at a rate
\begin{align}
&\dot{E}^{\rm lost}\simeq 0.28\, \pi^{3} \left(\mu m_p\right)^2 \left(\mu M_{\rm NBS}\right)^4 \sum_{m=1}^{+\infty}\left[1+(-1)^m\right]^2 \nonumber\\ &\times \left(\frac{Y_m^m\left(\frac{\pi}{2},0\right)}{\Gamma\left(m+\frac{3}{2}\right)} \frac{m^{\left(\frac{m}{2}-\frac{1}{4}\right)}(M\omega_{\rm orb})^{\frac{m}{3}}}{2^{\left(\frac{7}{4}+\frac{m}{2}\right)}(\omega_{\rm orb}/\mu)^{\left(\frac{1}{4}+\frac{m}{2}\right)}}\right)^2\,.
\end{align}

These analytic results are in excellent agreement with our numerics for both EMRIs (Eq.~\eqref{Erad_circular_BS}) and equal mass binaries (Eqs.~\eqref{Erad_circular_BS_equalmass}): the leading terms agrees with the numerical within $4\%$. Such agreement is a cross-check both on our numerical routine and our simple analytical description.

%
\subsubsection{Comparison with gravitational wave emission}
In vacuum, the orbit of a binary system shrinks in time, due to the emission of GWs. At leading order, loss via GWs is described by the quadrupole formula~\cite{Peters:1963ux}-\cite{Poisson:1993vp},
\be
\dot{E}^{\rm GW}=\frac{32}{5}\eta^2\left(M\omega_{\rm orb}\right)^{10/3}\,,
\label{eq:quadrupole}
\ee
where $\eta=m_1m_2/(m_1+m_2)^2$ is the symmetric mass ratio of a binary of component masses $m_1, m_2$ and total mass $M=m_1+m_2$. To estimate the flux of energy emitted in the scalar channel, we consider the orbit to be circular, with the radius equal to the semi-major axis ($\sim 970$ au) of the S2 star. The NBS scalar provides an extra channel for energy loss. For EMRIs ($m_p=\eta M$ and $M_{\rm BH}=M$), combining together Eqs.~\eqref{Elost_circular_BS}-\eqref{eq:quadrupole} we get~\footnote{Since the total scalar field mass contained in a sphere of radius $r_{\rm orb}\ll R$ is negligible with respect to the mass of the central BH $M_{\rm NBS}(r_{\rm orb})/M\sim 10^{-10}$, we can consider that the entire GW flux emitted is due to the quadrupole moment of the binary alone, neglecting the gravitational field of the DM halo.}
\beq
&&\frac{\dot{E}^{\rm lost}}{\dot{E}^{\rm GW}}\simeq   10^{-3}\, \left[\frac{M_{\rm NBS}}{10^{10}M_{\odot}}\right]^4\left[\frac{10^{6}M_{\odot}}{M}\right]^{2/3}\left[\frac{T}{16 {\rm yr}}\right]^{31/6} \nn\\
&&\,\times  \left[\frac{\mu}{10^{-22}{\rm eV}}\right]^{17/2},
\eeq
where we normalized to the typical values for the EMRI composed by Sagittarius $\text{\rm A}^*$ and S2 star, surrounded by a DM halo.

The energy balance equation imposes that the loss in the orbital energy of the binary is due to the energy carried away by scalar and gravitational waves~\cite{1989ApJ...345..434T,Stairs:2003eg} 
\be
\frac{d E^{\rm orb}}{dt}=-\left(\dot{E}^{\rm lost}+\dot{E}^{\rm GW}\right)\,.\label{eq:energy_balance}
\ee

Thus, energy loss leads to a secular change in orbital period
\be
\dot{T}\simeq-\frac{192\pi\left(2\pi\right)^{5/3}\eta M^{5/3}}{5T^{5/3}}-\frac{5 \eta M M_{\rm NBS}^{4} T^{5/2}}{ 10^{3}\mu^{-15/2}}\,.\nonumber
\ee
%
%
%
It is amusing to estimate such secular change for astrophysical parameters similar to those of S2 star orbiting around SgrA$^*$, 
\beq
&&\dot{T}\simeq \, -\frac{2.42}{10^{15}}  \left[\frac{M}{10^{6}M_{\odot}}\right]^{2/3}\left[\frac{T}{16 {\rm yr}}\right]^{-5/3}\left[\frac{m_p}{20 M_{\odot}}\right]\nn\\
&&-\frac{4}{10^{17}}\left[\frac{M_{\rm NBS}\mu}{0.01}\right]^4\left[\frac{\mu}{10^{-22}{\rm eV}}\right]^{7/2}\left[\frac{T}{16 {\rm yr}}\right]^{5	/2}\left[\frac{m_p}{20 M_{\odot}}\right]\,,\nonumber
\eeq
which seems hopelessly small.

The period change for equal-mass binary systems follows through, and is
\be
\dot{T}=-\frac{192\pi\left(2\pi\right)^{5/3}M^{5/3}}{20T^{5/3}}-\frac{3.1 M_{\rm NBS}^{4} m_p M^{2/3}T^{17/6}}{10^{3}\mu^{-17/2}}\nonumber\,.
\ee
%

\subsubsection{Backreaction and scalar depletion}
One cause for concern is that our calculation assumes a fixed scalar field background $\Psi_0$, but as the binary evolves
scalar radiation is depleting the NBS of scalar surrounding the binary. Assume, conservatively, that the flux above is only removing scalar field within a sphere of radius $\sim 10 \,\ell$
centred at the binary, with the radiation wavelength $\ell=2\pi/\omega_{\rm orb}$. Then the timescale for total depletion of the scalar in the sphere is
\beq
&& \tau \sim \frac{\rho R^3}{\dot{E}^{\rm rad}}\sim 10^{24}\,{\rm yr}\, \left[\frac{10^{-2}}{\mu M_{\rm NBS}}\right]^{2/3}\left[\frac{10^4}{\chi}\right]^{2/3} \left[\frac{20\,M_{\odot}}{m_p}\right]^{2}\nn\\
&&\times\left[\frac{10^{-22}{\rm eV}}{\mu}\right]^{11/6}\left[\frac{T}{16 {\rm yr}}\right]^{7/6}\,,
\eeq
that is much larger than the Hubble timescale. A similar value can be found for equal mass binary systems. Thus, our results seem to indicate that the background configuration remains unaffected by the emission of scalars by low frequency binaries.
\subsection{High-energy binaries within boson stars\label{BS_binaries}}
\subsubsection{Scalar emission close to coalescence}
%
\begin{figure}[ht]
\includegraphics[width=8.3cm,keepaspectratio]{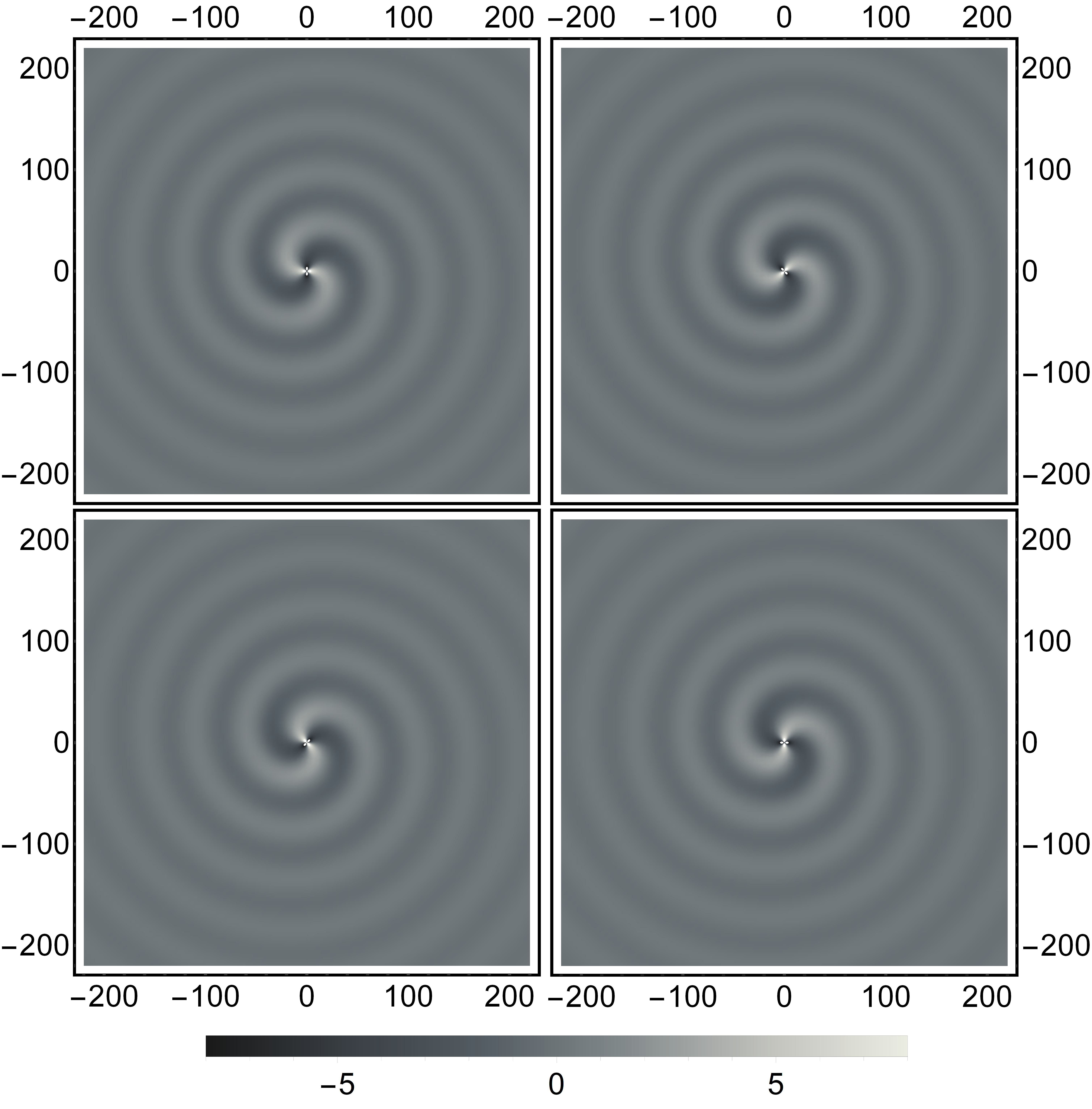}
\caption{Scalar field emission from a high energy, equal-mass binary describing a circular orbit of radius $r_{\rm orb}$, evolving inside an NBS. The axis are the normalized $x/r_{\rm orb}, y/r_{\rm orb}$ respectively, and each frame represents an equatorial slice of the scalar field perturbation $10^{17} {\rm Re}\left[\delta\Phi\right]$, induced by a binary orbiting in the equatorial plane. In the upper-left panel, particles are at $(x_1,y_1)=(r_{\rm orb},0)$, $(x_2,y_2)=(-r_{\rm orb},0)$. Moving clockwise in the panels, the system evolves for an eighth of a period between each, the binary moving anti-clockwise. The binary components have the same mass, ($m_p\sim 10^6 M_\odot$) and they are orbiting inside an NBS of mass $M_{\rm NBS}\mu\sim 0.01$ with a period of $\sim 1$ day.
}
\label{fig:highfreq_timedomain}
\end{figure}
We now wish to focus on rapidly moving binaries, such as those suitable for LIGO or LISA sources. In such a situation, the non-relativistic regime is not appropriate. Instead, one can show that the relevant description of these systems, for which the frequencies involved $\omega_{\rm orb}\gg \mu$, is accounted for by a slight modification of the previous equations, cf. Appendix~\ref{app:Newtonian} for details
\beq
&&\nabla^2\delta U= 4\pi P\,,\nonumber\\
&&-\partial^2_t \delta \Phi +\nabla^2 \delta \Phi =2 \mu^2 \Phi\, \delta U\,.\label{eq:high_binary}
\eeq

We consider two equal-mass point particles, each of mass $m_p$, on a circular motion of orbital frequency $\omega_{\rm orb}$
and radius $r_{\rm orb}$. We can solve the Poisson equation first, using a multipolar decomposition. We find
\beq
&&U=\sum_{lm}\frac{u_{lm}}{r}Y_{l}^m (\theta,0)e^{im\left(\phi-\phi_0\right)}\,,\\
&&u_{lm}=-\frac{4\pi m_p\left(1+(-1)^m\right)\,Y_{l}^m(\pi/2,0)}{2l+1} r_{\rm orb}^{-l-1}\nonumber\\
&&\times\left[r_{\rm orb}^{2l+1}r^{-l}\Theta(r-r_{\rm orb})+r^{l+1}\Theta(r_{\rm orb}-r)\right]\,.
\eeq
Here $\phi_0=\omega_{\rm orb}t$ is the azimuthal location of one particle; the other is at $\phi_0+\pi$.
If the factor $m_p\left(1+(-1)^m\right)$ is replaced by $m_p$ this same source describes a single point particle of mass $m_p$.
We now perform a Fourier transform and a multipolar decomposition of the scalar to solve Eq.~\eqref{eq:high_binary}:
\be
\delta \Phi=\frac{1}{\sqrt{2\pi}}\sum_{l,m}\int d\omega\, \frac{\delta\psi (\omega, r)}{r}e^{-i(\omega+\Omega) t} Y_{l}^m\,.
\label{eq:fourier_deltaphi}
\ee
We find the following ODE for $\delta\psi$:
\be
\delta\psi''+\left((\Omega+\omega)^2-\frac{l(l+1)}{r^2}\right)\delta\psi=\sqrt{8\pi}\mu^2\Psi_{0}\,\tilde{u}_{lm} \,,\nonumber
\ee
where $\tilde{u}_{lm}=u_{lm}\delta\left(\omega-m \omega_{\rm orb}\right)$. Here primes stand for radial derivatives. We can now solve this using variation of constants, requiring outgoing waves at large distances and regularity at
the origin. The solution is
\beq
\delta\psi&=&\delta\psi_\infty \int_0^r \frac{2\sqrt{2\pi}\mu^2\Psi_{0}\,\delta\psi_H\,\tilde{u}_{lm}}{i\omega}\nonumber\\
&+&\delta\psi_H \int_r^\infty \frac{2\sqrt{2\pi}\mu^2\Psi_{0}\,\delta\psi_\infty\,\tilde{u}_{lm}}{i\omega}\,,
\label{eq:totaldeltapsi_highfreq}
\eeq
where $\omega=m\omega_{\rm orb}$ and $\delta\psi_{H,\infty}$ are homogeneous solutions,
\beq
\delta\psi_{H}&=&\sqrt{\frac{\pi\omega r}{2}}J_{l+1/2}(\omega r)\,,\\
\delta\psi_{\infty}&=&\sqrt{\frac{\pi\omega r}{2}}\left(J_{l+1/2}(\omega r)+iY_{l+1/2}(\omega r)\right)\,.
\eeq
 The time domain response of the NBS to the perturbations induced by a binary BH system is found solving Eq.~\eqref{eq:totaldeltapsi_highfreq} and \eqref{eq:fourier_deltaphi}. Four snapshots of one period, for two equal mass BHs are shown in Fig.~\ref{fig:highfreq_timedomain}.

A binary deep inside the NBS ($r_{\rm orb}\ll R$) and with large orbital frequency~($\omega_{\rm orb}\ll 1/r_{\rm orb}$) generates a field at large distances that is independent on the size of the NBS: the integration converges a few wavelengths
away from the binary. We find the following simple result for the dominant $l=m$ modes:
\beq
&&\delta\psi(r\to \infty)=i\sqrt{2\pi} m_p\left(1+(-1)^m\right)\Psi_0 \pi^{3/2}\,2^{2-m}m^{m-2}\nonumber\\
&&\times \frac{Y_{m}^m(\pi/2,0)}{\Gamma[m+3/2]}\frac{\mu^2}{\omega_{\rm orb}^2}(M\omega_{\rm orb})^{m/3}\,e^{i\omega r}\,.
\eeq
Here, $M=2m_p$ for the equal-mass binary. If we substitute $m_p\left(1+(-1)^m\right)\to m_p$, these results also describe an EMRI, where a single particle of mass $m_p$ is revolving around a massive BH of mass $M$ (note the crucial difference that $l=m=1$ modes are radiated for EMRIs, whereas  only even modes are emitted for equal-mass binaries).
The flux is given by
\begin{align}
&\dot{E}^{\rm rad}=- r^2 \lim_{r \to \infty}\int d\theta d\varphi \sin \theta \,\delta T^S_{t r}\nonumber \\
&=128 \pi^{3}(\mu^2 m_p \Psi_0(0))^2\left(1+(-1)^m\right)^2  \nonumber\\ 
&\times\sum_{m=1}^{+\infty}\left(\frac{Y_m^m(\pi/2,0)}{\Gamma(m+3/2)} \frac{m^{m-1}(M\omega_{\rm orb})^{m/3}}{2^{m+1}\,\omega_{\rm orb}}\right)^2\,.\label{eq:energy_loss_high_binaries}
\end{align}
Since we are considering high-energy excitations of the scalar ($\omega_{\rm orb} \gg \mu$) it is easy to see that the rate of change of the NBS energy~$\dot{E}_{\rm NBS}$ is much smaller than~$\dot{E}^{\rm rad}$;~\footnote{Note that, at leading order, $$\dot{E}_ {\rm NBS}=\mu\, \dot{Q}_ {\rm NBS}=-r^2 \lim_{r \to \infty}\int d\theta d\varphi \sin \theta \,\delta j_{r}\,.$$}
so, conservation of energy (as expressed in Eq.~\eqref{LossRad}) implies that~$\dot{E}^{\rm lost}\simeq \dot{E}^{\rm rad}$.

\subsubsection{The phase dependence in vacuum and beyond}
In vacuum GR, the dynamics of a binary is governed by the energy balance equation \eqref{eq:energy_balance},
together with the quadrupole formula \eqref{eq:quadrupole}.
This implies that the orbital energy of the system $E_{\rm orb}=-M^2\eta/(2r_{\rm orb})$ must 
decrease at a rate fixed by such loss. This defines immediately the time-dependence of the GW frequency to be~$f^{-8/3}=(8\pi)^{8/3}{\cal M}^{5/3}(t_0-t)/5$, where ${\cal M}$ is the chirp mass and $f=\omega_{\rm orb}/\pi$. Once the frequency evolution is known, the GW phase simply reads
\begin{equation}
\varphi(t)=2\int^t\Omega(t')dt' \,.\label{GWphase}
\end{equation}

To take into account dissipative losses via the scalar channel, we add to the quadrupole formula the energy flux~\eqref{eq:energy_loss_high_binaries}.
In Fourier domain one can write the gauge-invariant metric fluctuations as
\begin{eqnarray}
 h_+(t)&=&A_+(t_{\rm ret})\cos\varphi(t_{\rm ret}) \,,\\
 h_\times(t)&=&A_\times(t_{\rm ret}) \sin\varphi(t_{\rm ret})\,,
\end{eqnarray}
where $t_{\rm ret}$ is the retarded time. The Fourier-transformed quantities are
\begin{equation}
 \tilde{h}_+= {\cal A}_+e^{i\Upsilon_+}\,,\qquad  \tilde{h}_\times={\cal A}_\times e^{i\Upsilon_\times}\,.
\end{equation}
Dissipative effects are included within the stationary phase approximation, where the secular time evolution is governed by 
the GW emission~\cite{Flanagan:1997sx}. In Fourier space, we decompose the phase of the GW signal 
$\tilde{h}(f)={\cal A}e^{i\Upsilon(f)}$ as:  
\begin{equation}
 \Upsilon(f) =\Upsilon_{\rm GR}^{(0)}[1+{\rm (PN\ corrections)}+\delta_{\Upsilon}]\,.
\end{equation}
where $\Upsilon_{\rm GR}^{(0)}=3/128 ({\cal M}\pi f)^{-5/3}$ represents the leading term of the 
phase's post-Newtonian expansion, and $f=\omega_{\rm orb}/\pi$. We find the following dominant correction due to the background scalar,
\be
\delta_{\Upsilon}=\frac{16\mu^4\Psi_0^2}{51\pi^3f^4}\sim 10^{-24}\left[\frac{\mu}{10^{-22}\,{\rm eV}}\right]^4\left[\frac{10^{-4}{\rm Hz}}{f}\right]^4\left[\frac{M_{\rm NBS}\mu}{0.01}\right]^4\nonumber
\ee
for equal-mass binaries. Such a correction corresponds to a $-6$ PN order correction~\cite{Yunes:2016jcc}. The smallness of the coefficient
makes it hopeless to detect with space-based detector LISA~\cite{Audley:2017drz}. Note that pulsar timing arrays operate at lower frequencies~\cite{Barack:2018yly}, and the previous Newtonian non-relativistic analysis is necessary.

\subsubsection{Backreaction and scalar depletion}
During the evolution, the binary emits scalar radiation away from the NBS. Assuming, again, that the flux above is only removing scalar field within a sphere of radius $\sim 10 \,\ell$
centred at the binary, with the radiation wavelength $\ell=2\pi/\omega_{\rm orb}$. Then the timescale for total depletion of the scalar is
\be
\tau\sim 2\times 10^{11}\,{\rm yr}\,\left(\frac{0.1}{m_p\omega_{\rm orb}}\right)^{7/3} \left(\frac{10^{-2}}{\mu M_{\rm NBS}}\right)^{2}\left(\frac{\chi}{10^4}\right)^2\nn\\
 \frac{m_p}{10^6M_{\odot}}\,,\nonumber
\ee
larger than a Hubble timescale, even for binaries close to coalescence. Thus, our results seem to describe emission of scalars during the entire lifetime of a compact binary.

\section{Scalar Q-balls}\label{ScalarQ}
We will now generalize the previous calculations to $Q$-balls, where gravity is absent but
for which self-interactions are necessary.

\subsection{Background configurations}
%
\begin{figure}	
\includegraphics[width=8.45cm,keepaspectratio]{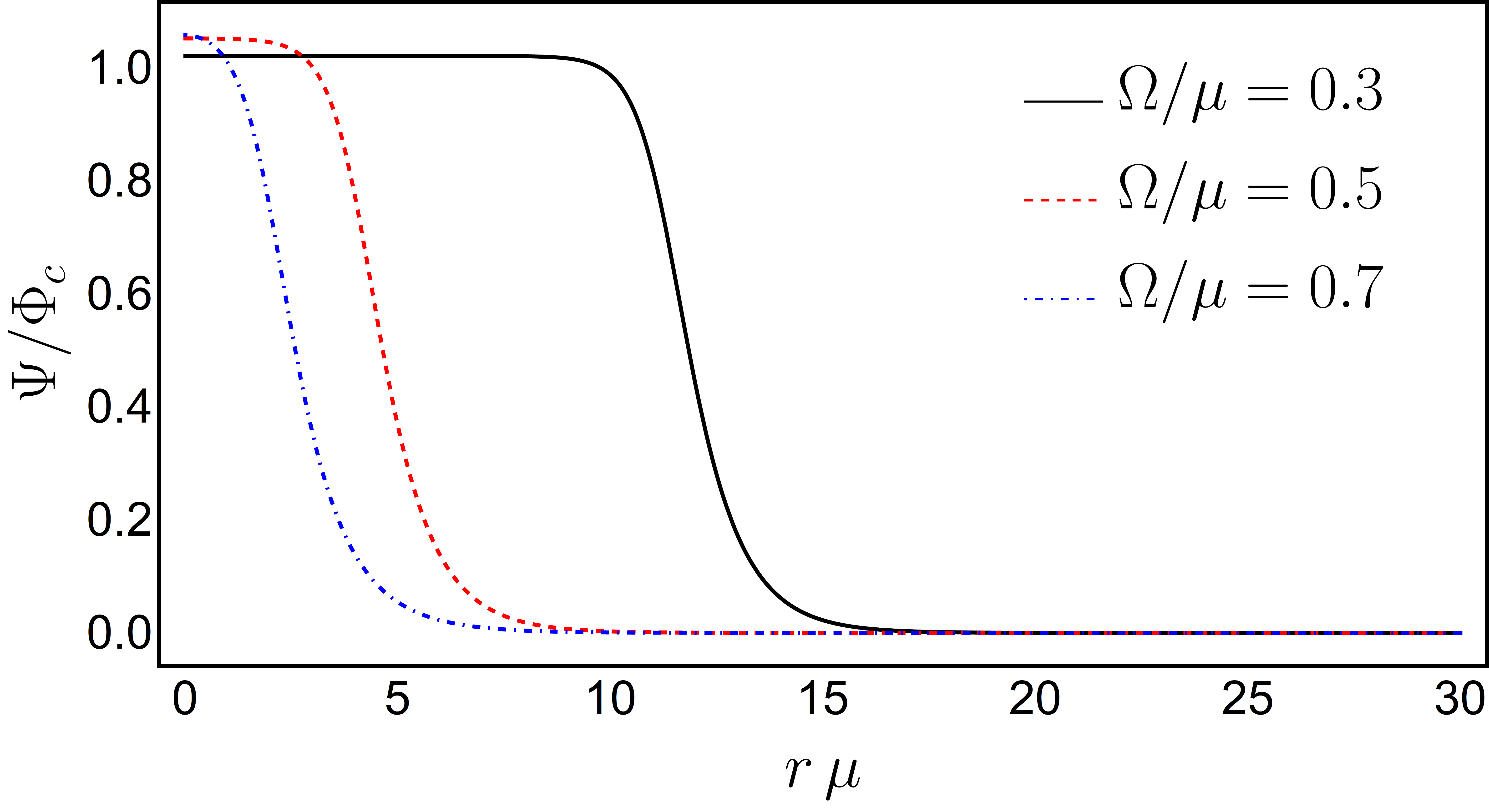} 
\caption{Three radial profiles $\Psi(r)/\Phi_c$ obtained through numerical integration of Eq.~\eqref{EOM_Qball_radial1} with appropriate boundary conditions ($\Psi(\infty)\rightarrow 0$ and $\partial_r\Psi(0)=0$). Each curve corresponds to a different Q-ball. }\label{fig:QBalls}
\end{figure}
The field equation for $\Phi$ is obtained through the variation of action~\eqref{action} with respect to $\Phi^*$ and reads
\beq
&&\nabla^\mu\partial_\mu  \Phi-2 \frac{d \mathcal{U}_{\rm Q}}{d|\Phi|^2} \Phi=0\,,\label{EOM_Qball}
\eeq
where we used $g_{\mu \nu}=\eta_{\mu \nu}$ and the potential $\mathcal{U}_{\rm Q}$ defined in Eq.\eqref{Potential_Qball}.
We now look for localized solutions of this model with the form~\eqref{BKG_ansatz} -- the so-called Q-balls. This ansatz yields the radial equation
\be 
\partial^2_r \Psi+\frac{2}{r}\partial_r \Psi+\left[\Omega^2-2 \frac{d \mathcal{U}_{\rm Q}}{d|\Phi|^2}\right]\Psi=0\,.\label{EOM_Qball_radial}
\ee
For the class of nonlinear potentials~\eqref{Potential_Qball}, the last equation becomes  
\be 
\partial^2_r \Psi+\frac{2}{r}\partial_r \Psi+\left[\Omega^2-\mu^2 \left(1-\frac{\Psi^2}{\Phi_c^2}\right)\left(1-3\frac{\Psi^2}{\Phi_c^2}\right)\right]\Psi=0\,.\label{EOM_Qball_radial1}
\ee
According to the results of Ref.~\cite{Coleman:1985ki}, there exist stable Q-ball solutions for any $0<\Omega<\mu$, independently of the free parameter $\Phi_c$. 
Additionally, it is known that, in the limit $\Omega/\mu \ll1$, the radial function $\Psi$ mimics an Heaviside step function (the so-called \textit{thin-wall} Q-ball)~\cite{Coleman:1985ki,Ioannidou,Tsumagari2008}. On the other hand, in the regime $\Omega/\mu \sim 1$, the function $\Psi$ starts to fall earlier and drops very slowly (\textit{thick-wall} Q-ball)~\cite{Ioannidou,Tsumagari2008}. In particular, using the results of Ref.~\cite{Tsumagari2008} one can show that, in the thin-wall limit,
\begin{equation}\label{Psi_thin}
\Psi(r) \simeq\Phi_c\left[1+\left(\frac{\Omega}{2 \mu}\right)^2\right]\Theta\left(\frac{\mu}{\Omega^2}-r\right)\,.
\end{equation}
Notice that the Q-ball radius is approximately given by $R_Q\simeq\mu/\Omega^2$.

\begin{figure}	
	\includegraphics[width=8.5cm,keepaspectratio]{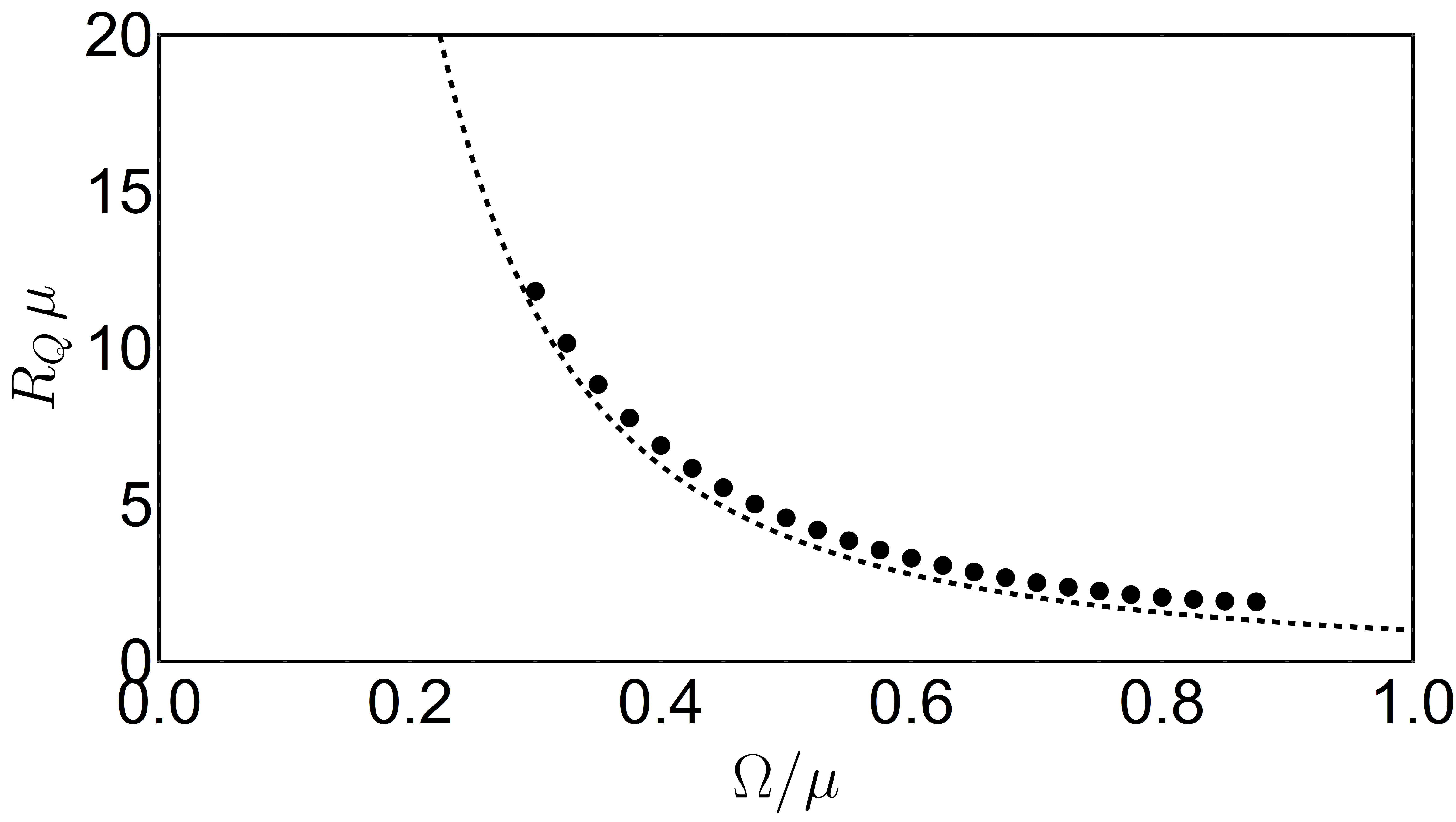} 
	\caption{Numerical results for the dependence of the Q-ball radius $R_Q \mu$ on the internal frequency $\Omega/\mu$, obtained through direct integration of Eq.~\eqref{EOM_Qball_radial1}. The dashed line is the thin-wall limit prediction, Eq.~\eqref{Psi_thin}. A fit on the numerical results gives $R_Q\sim 1.08 \mu \Omega^{-2}$, within $2\%$ of error, showing a good accordance with the predicted behaviour, Eq.~\eqref{Psi_thin}.
	} \label{fig:Radius_Omega}
\end{figure}

A few examples of radial profiles $\Psi(r)$ constructed numerically from Eq.~\eqref{EOM_Qball_radial1} are shown in Fig.~\ref{fig:QBalls}. 
From these results it is already evident that, when $\Omega/\mu \to 0$, the scalar does acquire a Heaviside-type profile. In such a limit the scalar drops to zero on the outside, on a lengthcale
$\sim 1/\mu$. These results also indicate that the radius of the Q-ball grows when $\Omega/\mu \to 0$. This is made more explicit in
Fig.~\ref{fig:Radius_Omega}, showing the numerical results for the dependence of the Q-ball radius $R_Q$~\footnote{We define the Q-ball radius $R_Q$ to be such that $\dfrac{\Psi(R_Q)}{\Psi(0)}=1/2$.}  on the frequency $\Omega$. The dashed line, corresponding to the thin-wall limit \eqref{Psi_thin}, agrees remarkably well with the numerics. 

The Q-ball charge $Q$ and mass $M_Q$ are obtained through~\eqref{NoetherCharge} and \eqref{Energy}, respectively, and read  
\beq
&&\hspace{-0.25cm} Q=4\pi \Omega \int dr\, r^2 \Psi^2(r)\,,\\
&&\hspace{-0.25cm} M_Q=\frac{1}{2} Q \Omega+4\pi \int dr\,r^2\left(\frac{(\partial_r \Psi)^2}{2}+\mathcal{U}(\Psi^2)\right)\,.
\eeq
For thin-wall Q-balls these become
\beq
Q&=&\frac{4 \pi}{3} \frac{\Omega^4}{\mu^6} \Psi_c^2\,,\\
M_Q&=&\frac{2 \pi}{3} \frac{\Omega^5}{\mu^6} \Psi_c^2\,.
\eeq
We are using a flat background spacetime, which requires that $M_Q/R_Q \ll 1$. In the thin-wall limit, this corresponds to
\be
\Omega/\mu \ll \Psi_c^{-2/7}\,.
\ee
\subsection{Small perturbations}

We now wish to understand the effect of a small perturbation on such Q-ball configurations. They can be considered either as sourceless small deformations of the background, or sourced by an external particle. Such perturber could be another Q-ball or simply some charge, piercing the Q-ball or orbiting around it. In the following, the external probe is modelled as pointlike, which means that our results are valid only for objects whose spatial extent are $\ll R_{\rm Q}$.
We consider an interaction between the perturber and the Q-ball described by the action
\be
\mathcal{S}_\text{int}\equiv -\int d^4x \sqrt{-g} \,{\rm Re}\left(\Phi\right) T_p\,,\label{coupling_Qball}
\ee
with $T_p\equiv g_{\mu \nu} T_p^{\mu \nu}$ being the trace of the particle's stress-energy tensor defined in Eq.~\eqref{Stress_energy_particle}. This coupling
allows for equations of motion that are simultaneously simple enough to be handled via our perturbation scheme, described in Sec.~\ref{Setup}, and it shows interesting dynamical features, as we shall see later. In the present analysis, we neglect the backreaction on the particle motion, therefore, the particle's world line $x_p^\mu(\tau)$ is considered to be known. 

An external particle sources a scalar field fluctuation of the form~\eqref{Perturbation} in the Q-ball background, which satisfies the linearized equation
\beq
&&-\partial_t^2\delta\Psi+\nabla^2 \delta\Psi +\left[\Omega^2-\mu^2\left(1- 8\frac{\Psi^2}{\Phi_c^2}+9 \frac{\Psi^4}{\Phi_c^4}\right)\right]\delta\Psi\nonumber\\
&&+2 i \Omega \partial_t \delta\Psi+2\mu^2 \frac{\Psi^2}{\Phi_c^2} \left(2-3 \frac{\Psi^2}{\Phi_c^2}\right)\delta\Psi^*=T_p e^{i \Omega t}\,,
\label{eq:Qball_pert_eq}
\eeq
and its complex conjugate. The sourceless case, corresponding to small Q-ball deformations, is simply recovered by setting $T_p=0$.
Decomposing the particle stress-energy trace as
\beq
T_p e^{i \Omega t}&=&\sum_{l,m} \int \frac{d \omega}{\sqrt{2 \pi} r}\nonumber\\
&\times&\Big[T_1^{\omega l m} Y_l^m e^{-i \omega t}+\left(T_2^{\omega l m }\right)^*\left(Y_l^m\right)^* e^{i \omega t}\Big]\,,\label{MatterDecomposition}
\eeq
where $T_1^{\omega l m}$ and $T_2^{\omega l m}$ are radial complex-functions defined by ,
\begin{eqnarray}
&&T_1^{\omega l m}\equiv \frac{r}{2 \sqrt{2 \pi}}\int dt d\theta d\varphi \sin \theta\, T_p e^{i (\omega+ \Omega)t} \left(Y_l^m\right)^*\,, \hspace{0.5cm} \label{eq:source_decomp_T1}\\
&&T_2^{\omega l m}\equiv \frac{r}{2 \sqrt{2 \pi}}\int dt d\theta d\varphi \sin \theta\, T_p e^{i (\omega- \Omega)t} \left(Y_l^m\right)^*\,.\hspace{0.5cm}\label{eq:source_decomp_T2}
\end{eqnarray}
Plugging the decompositions~\eqref{Decomposition} and~\eqref{MatterDecomposition} in Eq.~\eqref{eq:Qball_pert_eq}, one obtains the matrix equation~\footnote{The symmetry of this system implies that the radial functions satisfy $Z_2(\omega,l;r)=Z_1(-\omega,l;r)^*$. The functions $Z_1$ and $Z_2$ are clearly independent of the azimuthal number $m$.}
\begin{equation} 
\partial_r \boldsymbol{Z} -V_Q(r) \boldsymbol{Z}=\boldsymbol{T}\,,\label{Qball_Perturbation_Matrix_Sourced}
\end{equation}
where the vector $\boldsymbol{Z}\equiv (Z_1, Z_2, \partial_r Z_1, \partial_r Z_2)^T$, the matrix $V_Q$ is given by
\begin{equation*}
	V_Q\equiv
	\begin{pmatrix} 
	0 & 0 & 1 & 0 \\
	0 & 0 & 0 & 1 \\
	V_s-(\omega+\Omega)^2 & V_c & 0 & 0  \\
	V_c & V_s-(\omega-\Omega)^2 & 0 & 0  
	\end{pmatrix}\,,
\end{equation*}
and we defined the radial potentials
\beq
V_s(r)&\equiv& \frac{l(l+1)}{r^2}+\mu^2\left(1- 8\frac{\Psi_0^2}{\Phi_c^2}+9 \frac{\Psi_0^4}{\Phi_c^4}\right)\,, \\
V_c(r)&\equiv& - 2\mu^2 \frac{\Psi_0^2}{\Phi_c^2} \left(2-3 \frac{\Psi_0^2}{\Phi_c^2}\right)\,.
\eeq
and the source term\footnote{Again, to simplify the notation, we omit the labels $\omega$, $l$ and $m$ in the functions $T_1^{\omega l m}$ and $T_2^{\omega l m}$.}
\be
\boldsymbol{T}(r)\equiv \big(0,0,T_1,T_2\big)^T\,.\label{stress_energy_decomposition_qball}
\ee
To solve the small perturbations problem, either in the sourced or sourceless case, we need to establish suitable boundary conditions. We require regular solutions at the origin,
\begin{equation}
\boldsymbol{Z}(r \to 0)\sim \left(a r^{l+1},b r^{l+1},a (l+1)r^l,b (l+1)r^l\right)^T\,,\nonumber
\end{equation} 
with (complex) constants $a$ and $b$, and satisfying the Sommerfeld radiation condition at infinity
\begin{eqnarray}
\label{Qball_FO_Inf}
&&\boldsymbol{Z}(r \to \infty)\sim \nn\\
&&\left(Z_1^\infty e^{i k_1 r},
Z_2^\infty e^{i k_2 r},
i k_1 Z_1^\infty e^{i k_1r},
i k_2 Z_2^\infty e^{i k_2r}\right)\,,
\end{eqnarray} 
with 
\beq
k_1&\equiv& \epsilon_1 \sqrt{\left(\omega+\Omega\right)^2-\mu^2}\,, \\ 
k_2&\equiv& \epsilon_2 \left(\sqrt{\left(\omega-\Omega\right)^2-\mu^2}\right)^*\,.
\eeq
where we are using the principal complex square root.

Consider then the set of independent solutions $\{\boldsymbol{Z_{(1)}},\boldsymbol{Z_{(2)}},\boldsymbol{Z_{(3)}},\boldsymbol{Z_{(4)}}\}$ uniquely determined by
\beq
\boldsymbol{Z_{(1)}}(r \to 0)&\sim& \Big(r^{l+1},0,(l+1)r^l,0\Big)^T\,,\nonumber \\
\boldsymbol{Z_{(2)}}(r \to 0)&\sim& \Big(0,r^{l+1},0,(l+1)r^l\Big)^T\,,\nonumber \\
\boldsymbol{Z_{(3)}}(r \to \infty)&\sim& \Big(e^{i k_1 r},0,i k_1 e^{i k_1 r},0\Big)^T\,,\nonumber \\
\boldsymbol{Z_{(4)}}(r \to \infty)&\sim& \Big(0,e^{i k_2 r},0,i k_2 e^{i k_2 r}\Big)^T\,.
\eeq
The $4 \times 4$ matrix $F(r)\equiv\big(\boldsymbol{Z_{(1)}},\boldsymbol{Z_{(2)}},\boldsymbol{Z_{(3)}},\boldsymbol{Z_{(4)}}\big)$ is the fundamental matrix of the system~\eqref{Qball_Perturbation_Matrix_Sourced}. As shown in Appendix~\ref{app:detF}, for a system of the form~\eqref{Qball_Perturbation_Matrix_Sourced}, the determinant ${\rm det}(F)$ is independent of $r$.
\subsubsection{Sourceless perturbations}
Free oscillations of Q-ball configurations are regular scalar fluctuations satisfying the Sommerfeld radiation condition at infinity. They correspond to scalar perturbations of the form 
\begin{equation}
\delta \Psi=\frac{1}{\sqrt{2 \pi} r} \left[Z_1 Y_l^m e^{-i\omega t}+Z_2^*\left(Y_l^m\right)^* e^{i\omega^* t}\right]\,,\label{field_decomposition_QNM_qball}
\end{equation}
where $Z_1$ and $Z_2$ are solutions of system~\eqref{Qball_Perturbation_Matrix_Sourced}, with $\boldsymbol{T}=0$. For complex-valued $\omega$, the free oscillations are QNMs. For a real $\omega$, these are termed normal modes. Notice that for the discrete set $\{\omega_{\rm QNM}\}$ of QNM frequencies, the solutions $\{\boldsymbol{Z_{(1)}},\boldsymbol{Z_{(2)}},\boldsymbol{Z_{(3)}},\boldsymbol{Z_{(4)}}\}$ are not linearly independent. In fact, it is easy to see that the condition ${\rm det}(F)=0$ holds if and only if $\omega$ is a QNM frequency (\textit{i.e.,} $\omega \in \{\omega_{\rm QNM}\}$).

\subsubsection{External perturbers}
Let us turn now to the perturbations induced by an external particle, whose interacting with the background scalar field. How is such a body exciting the Q-ball, how much radiation does the interaction give rise to, what backreaction does the Q-ball
exert on the perturber? These are all questions that can be raised in this context, and that we wish to answer. As an interesting toy model, in Appendix.~\ref{app:internal_modes}, we study how the scalar field inside a spherical box is excited by a particle in circular orbital motion. This simple setup, akin to resonant harmonic oscillators, illustrates how a charged particle in circular motion can excite the proper modes of oscillation of a box filled with scalar field.

To obtain physical observable quantities one needs to find the solutions of system \eqref{Qball_Perturbation_Matrix_Sourced} that are regular at the origin and satisfy the Sommerfeld condition at infinity. These can be obtained through the method of variation of parameters 
\beq
Z_1(r)&=&\sum_{k=3}^4 \Bigg[\sum_{n=1}^2 F_{1,n}(r) \int_\infty^r dr' F^{-1}_{n,k}(r') \boldsymbol{T}_k(r')\nonumber \\ 
&+&\sum_{n=3}^4 F_{1,n}(r) \int_0^r dr' F^{-1}_{n,k}(r') \boldsymbol{T}_k(r') \Bigg]\,,\\
Z_2(r)&=&\sum_{k=3}^4 \Bigg[\sum_{n=1}^2 F_{2,n}(r) \int_\infty^r dr' F^{-1}_{n,k}(r') \boldsymbol{T}_k(r')\nonumber \\
&+&\sum_{n=3}^4 F_{2,n}(r) \int_0^r dr' F^{-1}_{n,k}(r') \boldsymbol{T}_k(r') \Bigg]\,.
\eeq
The total energy, linear and angular momenta radiated during a given process can be found using solely the amplitudes $Z_1^\infty$ and $Z_2^\infty$. These are given by
\beq
Z_1^\infty&=&\sum_{k=3}^4 \int_{0}^{\infty} dr' F^{-1}_{3,k}(r')\boldsymbol{T}_k(r') \,,\label{Z1inf} \\
Z_2^\infty&=&\sum_{k=3}^4 \int_{0}^{\infty} dr' F^{-1}_{4,k}(r')\boldsymbol{T}_k(r') \,. 
\eeq 
Let us now apply our framework to two physically relevant setups: a particle plunging into a Q-ball configuration, and a particle in a circular orbit around the Q-ball.
\paragraph*{\underline{Plunging particle.}}
Consider a particle moving at a constant velocity $\boldsymbol{v}=-v\boldsymbol{e}_z$ (with $v>0$), plunging into a Q-ball, and crossing its center at $t=0$. 
In this case, the trace of the particle's stress-energy tensor reads
\beq
T_p=&&-\left[\delta\left(r+v t\right)\delta\left(\theta\right) \Theta(-t)+ \delta\left(r-v t\right)\delta\left(\theta-\pi\right) \Theta(t)\right]\nonumber \\
&&\times m_p \,\delta(\varphi)\sqrt{1-v^2}/(r^2 \sin \theta)\,.
\label{T_p_plunging}
\eeq
Therefore, the source decompositions in Eqs.~\eqref{eq:source_decomp_T1}-\eqref{eq:source_decomp_T2} read as
\beq
T_1=&&-\left[\cos\left((\omega+\Omega)r/v\right) \delta_l^\text{even}-i \sin\left((\omega+\Omega)r/v\right) \delta_l^\text{odd}\right]\nonumber \\
&&\times m_p \, Y_l^0(0,0) \delta_m^0\sqrt{1-v^2}/(\sqrt{2\pi}r v)\,,	\\
T_2=&&-\left[\cos\left((\omega-\Omega)r/v\right) \delta_l^\text{even}-i \sin\left((\omega-\Omega)r/v\right) \delta_l^\text{odd}\right]\nonumber \\
&&\times m_p \, Y_l^0(0,0) \delta_m^0\sqrt{1-v^2}/(\sqrt{2\pi}r v)\,.
\eeq
These satisfy the property 
\be
T_2(\omega,l,0;r)= T_1(-\omega,l,0;r)^*\,.
\ee
Thus, due to the form of the system~\eqref{Qball_Perturbation_Matrix_Sourced}, one has
\beq
Z_2(\omega,l,0;r)&=&Z_1(-\omega,l,0;r)^*\,,\\
Z_2^\infty(\omega,l,0)&=&Z_1^\infty(-\omega,l,0)^*\,.\label{property_Z1Z2_plunge_qball}
\eeq
Finally, the spectral fluxes~\eqref{Energy_flux}, \eqref{Momentum_flux} and \eqref{AngularMomentum_flux} become, respectively,
\beq
\frac{d E^{\rm rad}}{d \omega}&=&4 \left|\omega+\Omega\right|\nonumber \\
&\times& {\rm Re}\left[\sqrt{(\omega+\Omega)^2-\mu^2}\right] \sum_l \left|Z_1^\infty(\omega,l,0)\right|^2\,, \label{Energy_flux_qball}\\
\frac{d P_z^{\rm rad}}{d \omega}&=&\sum_{l}\frac{8(l+1) \Theta\left[\left(\omega+\Omega\right)^2-\mu^2\right]\left|(\omega+\Omega)^2-\mu^2\right|}{\sqrt{(2l+1)(2l+3)}} \nonumber \\
&\times&\text{Re}\left[Z_1^\infty(\omega,l,0) Z_1^\infty(\omega,l+1,0)^*\right]\,,\label{Momentum_flux_qball}\\
\frac{d L_z^{\rm rad}}{d \omega}&=&0\,.
\eeq
%
\paragraph*{\underline{Orbiting particle}}
The next setup is composed by a particle describing a circular orbit of radius $r_{\rm orb}$ and angular frequency $\omega_{\rm orb}$ inside a Q-ball and in its equatorial plane. The trace of the particle's stress-energy tensor is
\beq
T_p&=&-\frac{m_p}{r_{\rm orb}^2}\sqrt{1-\left(\omega_{\rm orb} r_{\rm orb}\right)^2} \nonumber\\
&\times&\delta(r-r_{\rm orb})\delta\left(\theta-\frac{\pi}{2}\right)\delta(\varphi-\omega_{\rm orb} t)\,, \label{T_p_orbiting}
\eeq
which implies
\beq
T_{1,2}&=&-m_p \sqrt{\pi/2}\, Y_l^m\left(\pi/2,0\right) \sqrt{1-\left(\omega_{\rm orb} r_{\rm orb}\right)^2}/r_{\rm orb}\  \nonumber \\
&\times&\delta\left(r-r_{\rm orb}\right)\delta\left(\omega \pm\Omega-m \omega_{\rm orb}\right)\,. \label{T12_orbiting}
\eeq
%
%
Notice that $T_2(\omega,l,m)=(-1)^m T_1(-\omega,l,-m)$, hence due to the form of system~\eqref{Qball_Perturbation_Matrix_Sourced}, we have
\beq
Z_2(\omega,l,m;r)&=&(-1)^m Z_1(-\omega,l,-m;r)^*\,,\\
Z_2^\infty(\omega,l,m)&=&(-1)^m Z_1^\infty(-\omega,l,-m)^*\,.
\eeq
Then, the emission rate expressions~\eqref{Energy_flux_rate} and~\eqref{AngularMomentum_flux_rate} imply, omitting the arguments $(\omega,l,m)$, 
\beq
&&\dot{E}^{\rm rad}= \frac{2}{\pi}\int d\omega \left|\omega+\Omega\right| {\rm Re}\left[\sqrt{(\omega+\Omega)^2-\mu^2}\right]\sum_{l,m} \left|Z_1^\infty\right|^2  \,, \nn\\
&&\dot{L}_z^{\rm rad}=\frac{2}{\pi}\int d\omega \,\epsilon_1(\omega){\rm Re}\left[\sqrt{(\omega+\Omega)^2-\mu^2}\right]\sum_{l,m} m \left|Z_1^\infty\right|^2. \nn
\eeq
where we remind that $\epsilon_1\equiv \text{sign}(\omega+\Omega+ \mu)$. Re-writing expression~\eqref{T12_orbiting} in the form
\be
T_{1,2}=\widetilde{T}(\omega_{\rm orb},r_{\rm orb}) \,\delta\left(r-r_{\rm orb}\right) \delta\left(\omega\pm\Omega-m \omega_{\rm orb}\right)\,,
\ee
the previous expressions for the rate of emission read
\beq
\dot{E}^{\rm rad}&=&\frac{2}{\pi}  \sum_{l,m}\widetilde{T}^2\Big[a_1\left|F_{3,3}^{-1}\left(m\omega_{\rm orb}-\Omega;\,r_{\rm orb}\right)\right|^2\nn\\
&&+a_2\left|F_{3,4}^{-1}\left(m\omega_{\rm orb}+\Omega;\,r_{\rm orb}\right)\right|^2\Big]\,,\label{Energy_flux_orbiting}
\eeq
\beq
\dot{L}_z^{\rm rad}&=&\frac{2}{\pi}  \sum_{l,m}m \widetilde{T}^2\Big[\epsilon_1 a_1 \left|F_{3,3}^{-1}\left(m\omega_{\rm orb}-\Omega;\,r_{\rm orb}\right)\right|^2\nn\\
&&+\epsilon_1 a_2\left|F_{3,4}^{-1}\left(m\omega_{\rm orb}+\Omega;\,r_{\rm orb}\right)\right|^2\Big]\,.\label{AngularMomentum_flux_orbiting}
\eeq
where 
\beq
&&\hspace{-0.5cm}a_1=|m \omega_{\rm orb}|{\rm Re}\left[\sqrt{\left(m \omega_{\rm orb}\right)^2- \mu^2}\right]\,,\nn \\
&&\hspace{-0.5cm} a_2=\left|m \omega_{\rm orb}+2\Omega\right|{\rm Re}\left[\sqrt{\left(m \omega_{\rm orb}+2\Omega\right)^2- \mu^2}\right].
\eeq
\subsection{Free oscillations}
%
%
\begin{table*}[th] 
	\begin{tabular}{ccccc}
		\hline
		\hline
		$l$ &  \multicolumn{4}{c}{$\omega_{\rm QNM}/\mu$} \\ 
		\hline
		\hline
        0 &     $0.439$ & $0.689$ & $0.931 - 1.2 \times 10^{-4} i$& $1.153 - 1.6\times 10^{-2} i$ \\
		1 & $0.300$  &  $0.555$  &   $0.806 - 9.8 \times 10^{-4} i$ &  $1.04 - 3.3\times 10^{-3}i$\\
		\hline
		\hline
	\end{tabular} 
	\caption{Some QNM frequencies of a Q-ball configuration with $\Omega/\mu=0.3$, for $l=\{0,1,2\}$. Note that the first column corresponds to normal modes, with $\omega<\mu$, hence screened from distant observers: they are confined to a spatial extent $\sim R_{\rm Q}$, the radius of the Q-ball (these modes are the counterpart of the NBS modes in Table \ref{table:QNM_BS_invariant}). There is an infinity of QNM frequencies, parametrized by an integer overtone index $n$. At large $n$, ${\rm Re}\left(\omega_{\rm QNM}\right)\sim 0.22n\sim \pi n/R_{\rm Q}$, as might be anticipated by a WKB analysis. Our results for the imaginary part of $\omega_{\rm QNM}$ carry a large uncertainty, and should be taken as order of magnitude estimate only.
	}
	\label{table:QNM_Qball}
\end{table*}
The numerical search for QNM frequencies for Q-balls is summarized in Table~\ref{table:QNM_Qball}, for the particular configuration with $\Omega=0.3\mu$. Whenever $\omega_{\rm QNM}$ are pure real numbers, they refer to normal modes of the object. For a mode to be normal, it must not be dissipated to infinity, hence the condition $\omega<\mu-\Omega$ is necessary, which also implies that such modes are screened from far-away observers, by the Q-ball background itself. This means that perturbations associated with the real-valued frequencies in Table~\ref{table:QNM_Qball} do not reach spatial infinity. Such modes are the analogs of the NBS modes, which were {\it all} normal (cf. Table~\ref{table:QNM_BS_invariant}). Q-balls, in addition to such modes also have quasinormal modes, which decay in time since they are sufficiently large energy to propagate at large distances.

\subsection{Particles plunging into Q-balls\label{sec:Plunging_particle}}
%
\begin{figure}[ht]
\includegraphics[width=8.5cm,keepaspectratio]{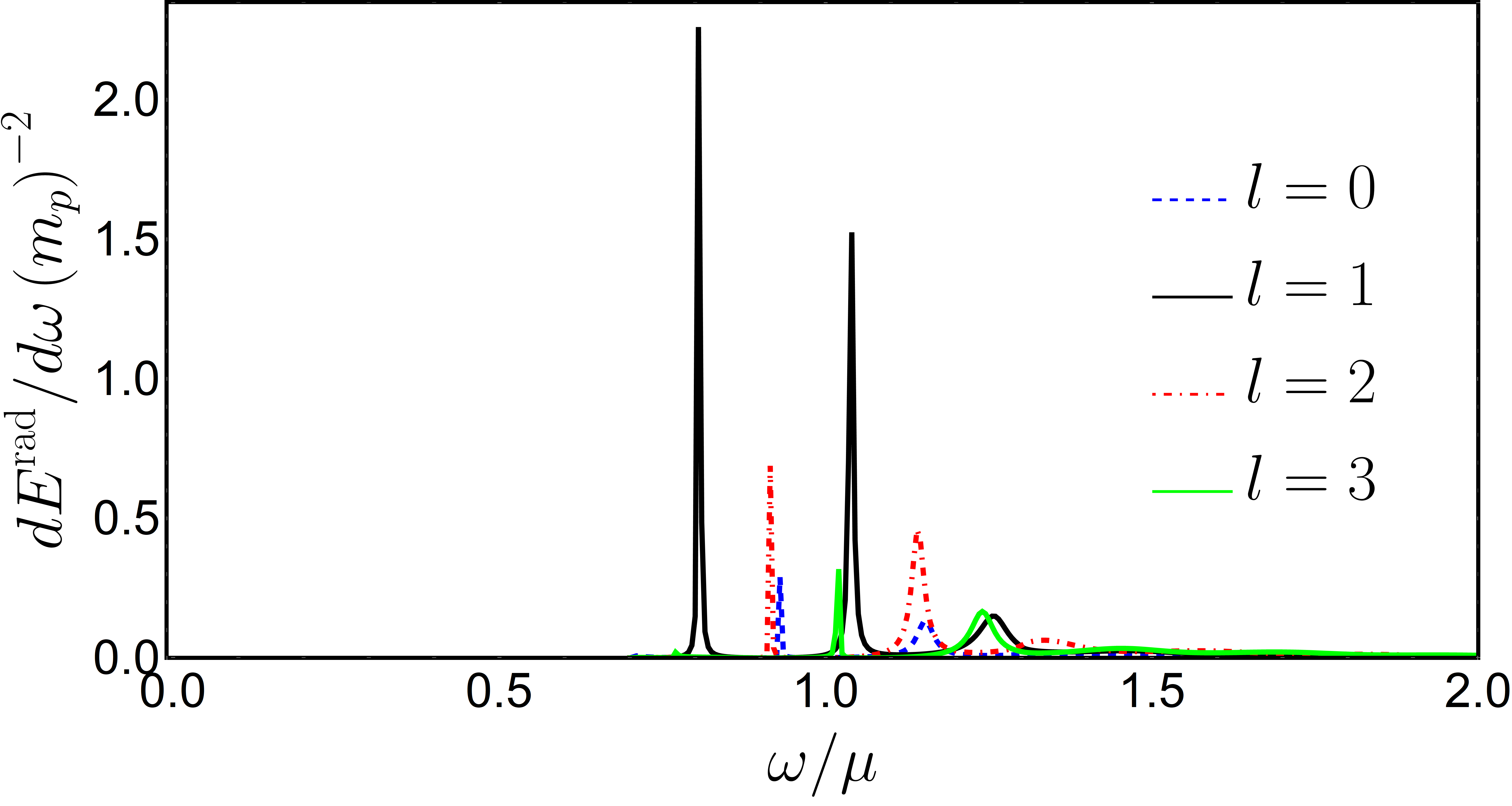} 
\caption{Energy spectra of scalar radiation emitted when a particle of rest-mass $m_p$
plunges through a Q-ball with $\Omega=0.3\mu$ with a large velocity $v=0.8c$.
The spectrum was decomposed into multipoles (cf. Eq.~\eqref{Energy_flux_qball}). 
The sharp peaks correspond to the excitation of QNM frequencies $\omega_{\rm QNM}$ (see~Tab.~\ref{table:QNM_Qball}).
}
\label{fig:Plunging_Spectra_Qball}
\end{figure}
\begin{figure}[ht]
\includegraphics[width=8.5cm,keepaspectratio]{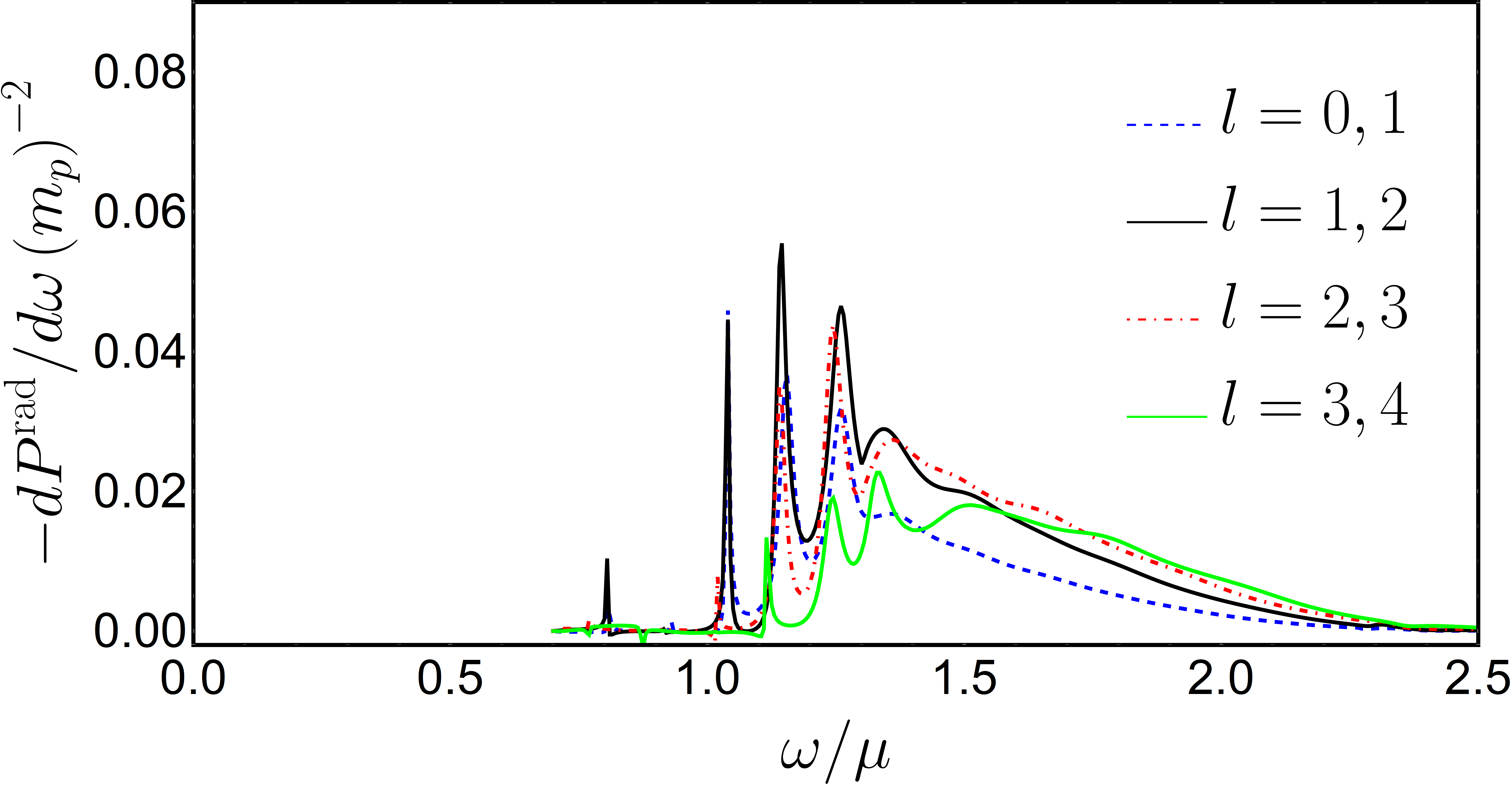}
\caption{Linear momentum radiated when a particle plunges through a Q-ball (described by $\Omega=0.3\mu$) with a velocity $v=0.8$. 
Different lines correspond to the different multipolar cross terms in Eq.~\eqref{Momentum_flux_qball}.}
\label{fig:Plunging_momentum_Qball}
\end{figure}

For concreteness, here we restrict the discussion to a large-velocity plunge $v=0.8c$.
The multipolar energy spectrum $d E^{\rm rad}_l/d \omega$ of radiation released during such process is shown in Fig.~\ref{fig:Plunging_Spectra_Qball} for the first lowest multipoles, obtained through numerical evaluation of Eq.~\eqref{Energy_flux_qball}. Just like a hammer hitting a bell excites its characteristic vibration modes, the effect of a plunging particle is to excite the QNMs of a Q-ball.
Figure~\ref{fig:Plunging_Spectra_Qball} illustrates this feature very clearly, the peaks in the energy spectrum are all coincident with the QNMs, some of them identified in Table~\ref{table:QNM_Qball}. This feature was absent in the dynamics of NBS, simply because the modes of NBS (Table \ref{table:QNM_BS_invariant}) are all normal and confined to the NBS itself: they do not propagate to large distances. Most of the radiation is dipolar, also apparent in Fig.~\ref{fig:Plunging_Spectra_Qball}, but a substantial amount is emitted in other multipoles as well.
For example, the $l=4$ mode still carries roughly $10\%$ of the total radiated energy. Our results are compatible with an exponential suppression at large $l$, of the form
$E^{\rm rad}_l\sim 0.085 e^{-0.39 l}$. We can use this to sum over multipoles, and find the total energy radiated, 
\be
E^{\rm rad}\sim 0.188 \, m_p^2 \,\mu\,.
\ee

The emitted radiation carries momentum, which is caused by an interference term between multipoles (cf. Eq.~\eqref{Momentum_flux_qball}). For radiation entirely emitted in one single direction, the linear momentum $P^{\rm rad}=E^{\rm rad}/c$. However, this is in general only a (poor) upper bound on the radiated linear momentum, as a number of multipoles are involved in the process.
Figure~\ref{fig:Plunging_momentum_Qball} shows the contribution of the modes $l\leq 4$ to the spectral flux of linear momentum $d P_z^{\rm rad}/d \omega$, obtained through numerical evaluation of~\eqref{Momentum_flux_qball}. Again, most of the contribution comes from the excitation of the Q-ball's QNMs. Note the interesting aspect that in some frequency ranges and for some interference terms, the momentum is positive, i.e., along the direction of the motion. 
We observed numerically that the total flux of linear momentum $P_z^{\rm rad}$ converge exponentially in $l$, for sufficiently large $l$. The total radiated momentum is negative, and thus represents a slowing-down of the moving point particle. Using a similar fitting procedure to sum over multipoles, we find for this particular configuration,
\be
P^{\rm rad}\sim-0.088 \, m_p^2\, \mu\,.
\ee
%

\subsection{Orbiting particles\label{sec:Orbiting_particle}}
%
\begin{figure*}[ht]
\begin{tabular}{c}
\includegraphics[width=8.4cm,keepaspectratio]{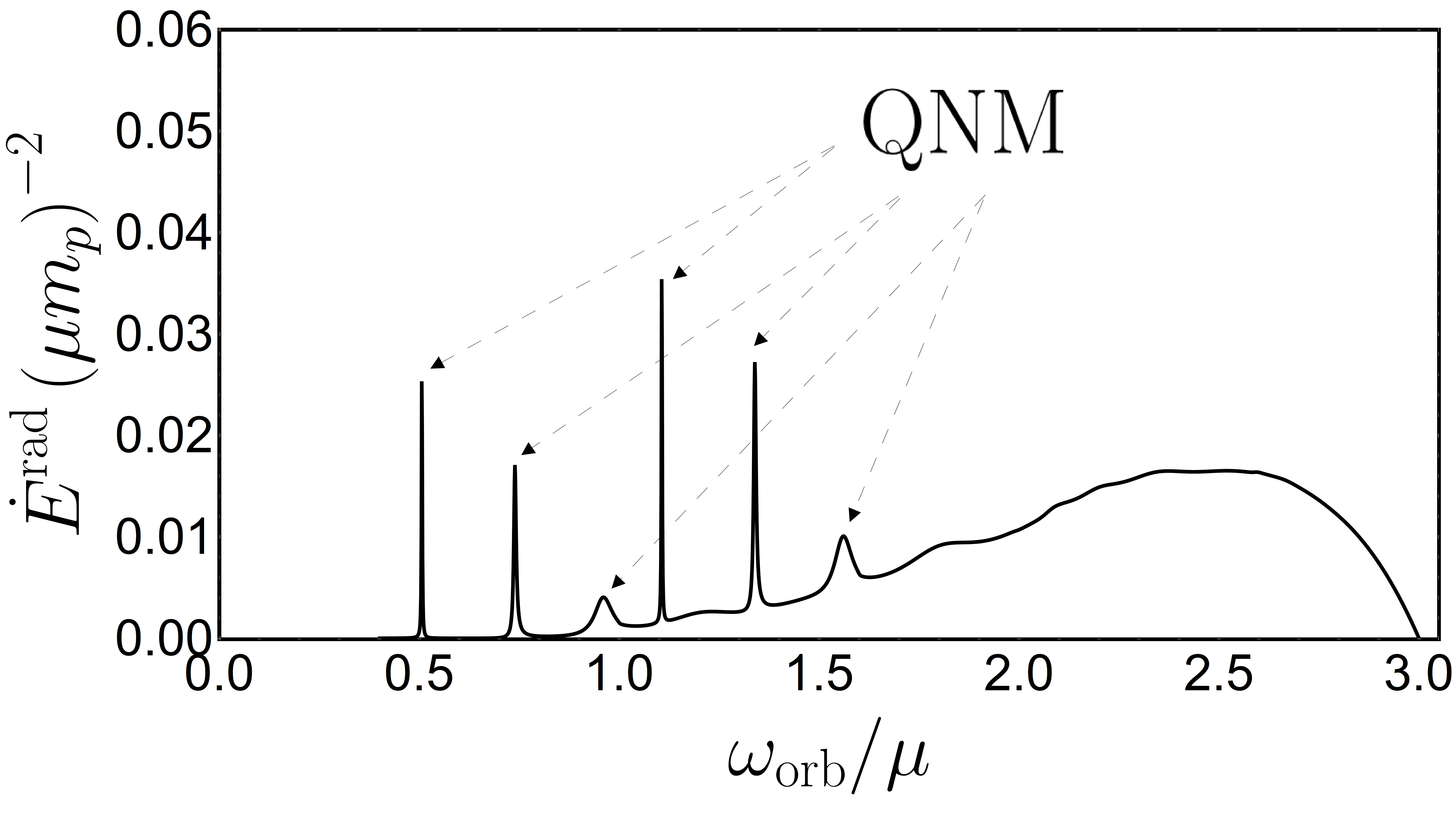} \includegraphics[width=8.4cm,keepaspectratio]{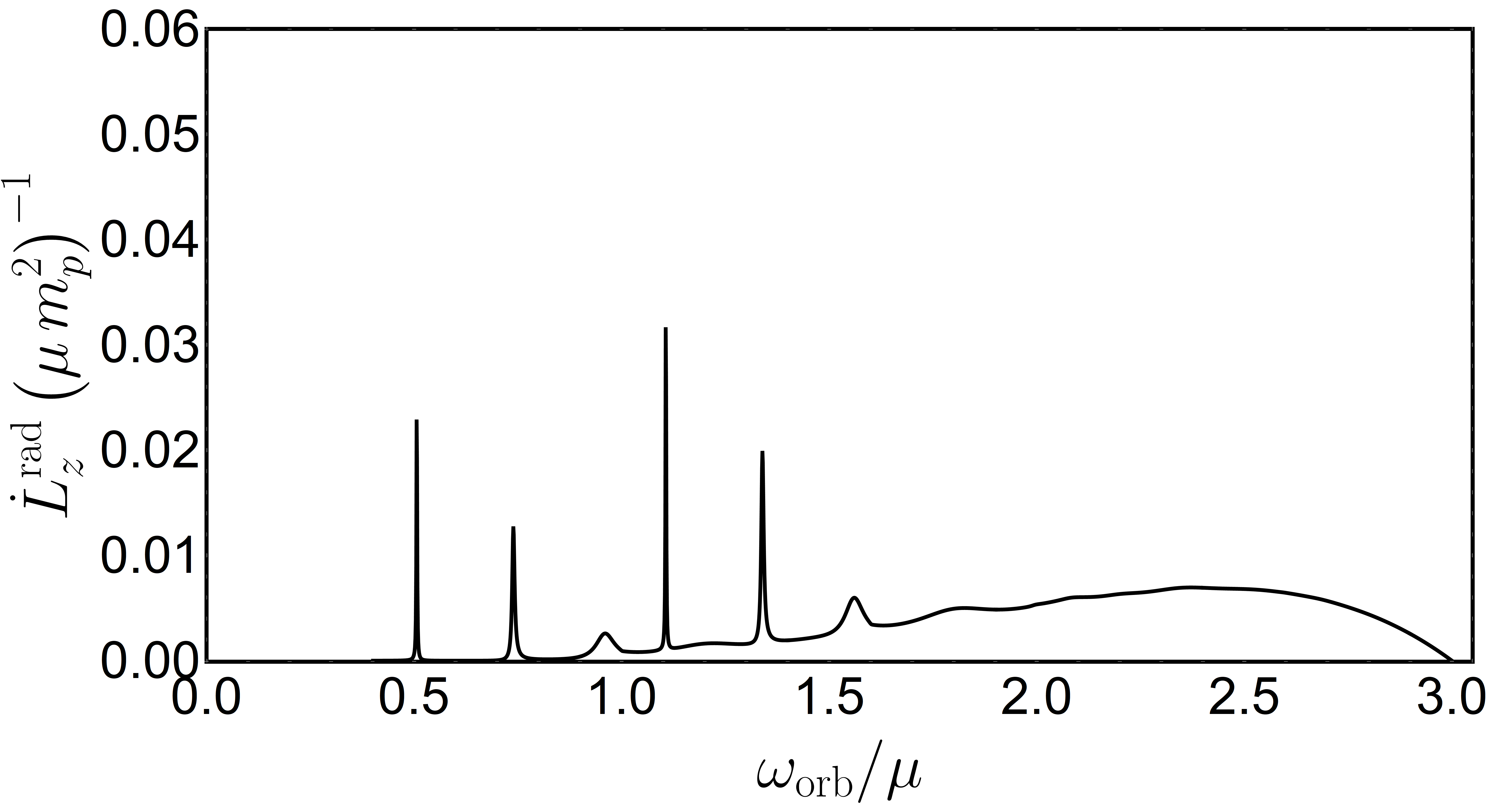}
\end{tabular}
\caption{Average dipolar ($l=1$, including $m=\pm1$) rate of energy (left), and angular momentum (right) radiated by a particle describing a circular orbit around a Q-ball with $\Omega=0.3\mu$, at radius $r_{\rm orb}\mu=1/3$ and with orbital frequency $\omega_{\rm orb}$. The peaks are associated with the excitation of QNM frequencies $\omega_{\rm QNM}$ for $\omega_{\rm orb}=\text{Re}\left(\omega_{\rm QNM}\right)\pm \Omega$ -- each QNM frequency is excited by two different $\omega_{\rm orb}$ spaced by $2\Omega$. The excitation of the QNM frequencies with $\text{Re}(\omega_{\rm QNM})=\{0.806, \,1.04$ (in Tab.~\ref{table:QNM_Qball})$,\,1.298\}\mu$ is clearly seen from these plots. However, not all the QNM frequencies can be efficiently excited: $\text{Re}(\omega_{\rm QNM})/\mu=2.30$ (in Tab.~\ref{table:QNM_Qball}) is an example.}
	\label{fig:OrbitingFluxesQball}
\end{figure*}

\begin{figure}[ht]
\includegraphics[width=8.4cm,keepaspectratio]{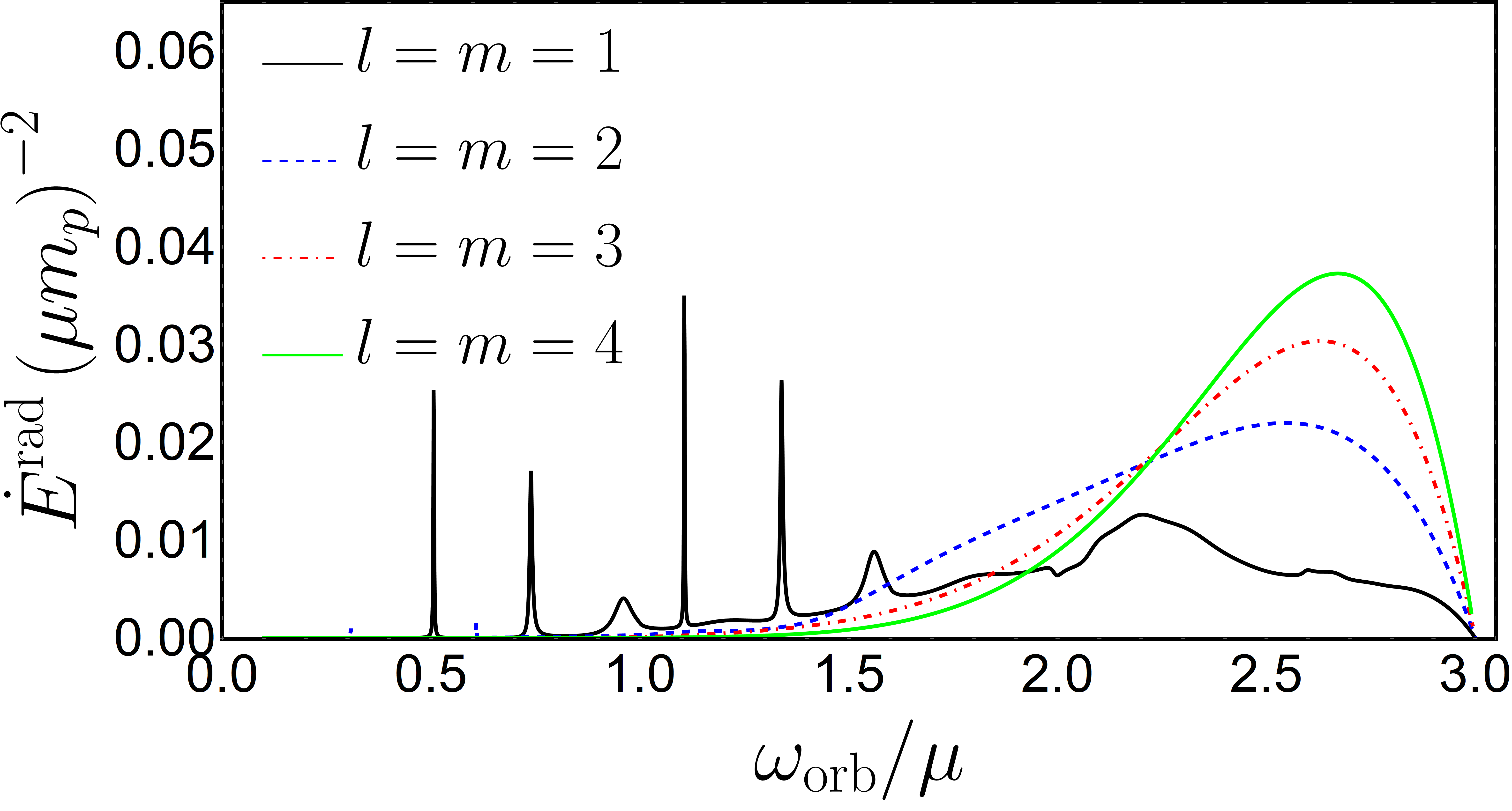} \caption{Average rate of energy radiated by a particle describing a circular orbit around a Q-ball with $\Omega=0.3\mu$, at a radius $r_{\rm orb}\mu=1/3$ and with orbital frequency $\omega_{\rm orb}$ for different values of $l=m$. At low frequencies the radiation is mostly dipolar. At large orbital frequencies the radiation is synchrotron-like and peaked at large $l=m$. In the high-frequency regime, there is a critical multipole $m$ beyond which the energy radiated decreases exponentially
(see main text for further details). 
There are QNM peaks for all multipoles, but they are visible only for the dipolar and quadrupolar.} 	
\label{fig:OrbitingFluxesQballMorels}
\end{figure}
The average dipolar flux of energy and angular momentum, emitted by a particle in circular orbit inside a Q-ball ($\Omega=0.3\mu$), at an orbital distance $r_{\rm orb}\mu=1/3$, 
are shown in Figs.~\ref{fig:OrbitingFluxesQball}. The pointlike source is assumed to be orbiting due to some external force, and its orbital frequency is varied, scanning possible resonant behavior of the Q-ball. As expected, and verified numerically, the quantity $\dot{E}^{\rm rad}$ is an even function of $\omega_{\rm orb}$, whereas $\dot{L}_z^{\rm rad}$ is an odd one. A few features are apparent in the results above (obtained evaluating Eqs.~\eqref{Energy_flux_orbiting}-\eqref{AngularMomentum_flux_orbiting}).
The fluxes have clear peaks, which correspond to the resonant excitation of the QNMs of the Q-ball. It's worth to note that for each QNM frequency listed in Tab.~\ref{table:QNM_Qball} there are two peaks associated with different orbital frequencies separated by a distance $2\Omega$: the resonances now occur at $\omega_{\rm orb}=\Omega\pm\omega_{\rm QNM}$. This is directly due to the decomposition in Eq.~\eqref{MatterDecomposition}.

In flat space, a scalar charge on a circular orbit also emits radiation~\cite{Cardoso:2007uy,Cardoso:2011xi}. For small orbital frequencies and massless fields, the flux is dipolar and of order $\dot{E}\sim q^2r_{\rm orb}^2\omega_{\rm orb}^4/(12\pi)$~\cite{Cardoso:2007uy,Cardoso:2011xi} (given the interaction \eqref{coupling_Qball}, the scalar charge $q=m_p$).
This explains the rise of the dipolar flux when the orbital frequency increases. However, at large frequencies, the radiation becomes of synchrotron type, and the radiation is emitted preferentially in higher multipoles~\cite{Misner:1972jf,Breuer}. This is apparent in Fig.~\ref{fig:OrbitingFluxesQballMorels} where we show the contribution of higher multipoles to the flux. 
Note that all other multipoles also have resonant peaks, but these are less pronounced than the dipolar. At large Lorentz factors $\gamma$, there is a critical $m$ mode after which the fluxes 
becomes exponentially suppressed. The critical multipole is of order $m_{\rm crit}\propto \gamma^2$~\cite{Misner:1972jf,Breuer}. Thus an evaluation of a large number of multipoles is necessary to have an accurate estimate of fluxes at large velocities. Our results are consistent with such a prediction. We find that as $\omega_{\rm orb}$ increases, the flux peaks at higher and higher $m$, but there's always a threshold $m$ beyond which the radiation output is exponentially suppressed. Finally, since this process is not axially symmetric, one cannot use expression~\eqref{Momentum_flux} to compute the flux of linear momentum along $z$. Nevertheless, it is straightforward to show that the average rate of linear momentum radiated $\dot{P}_z^{\rm rad}$ vanishes.

\begin{figure}[ht]
\includegraphics[width=8.3cm,keepaspectratio]{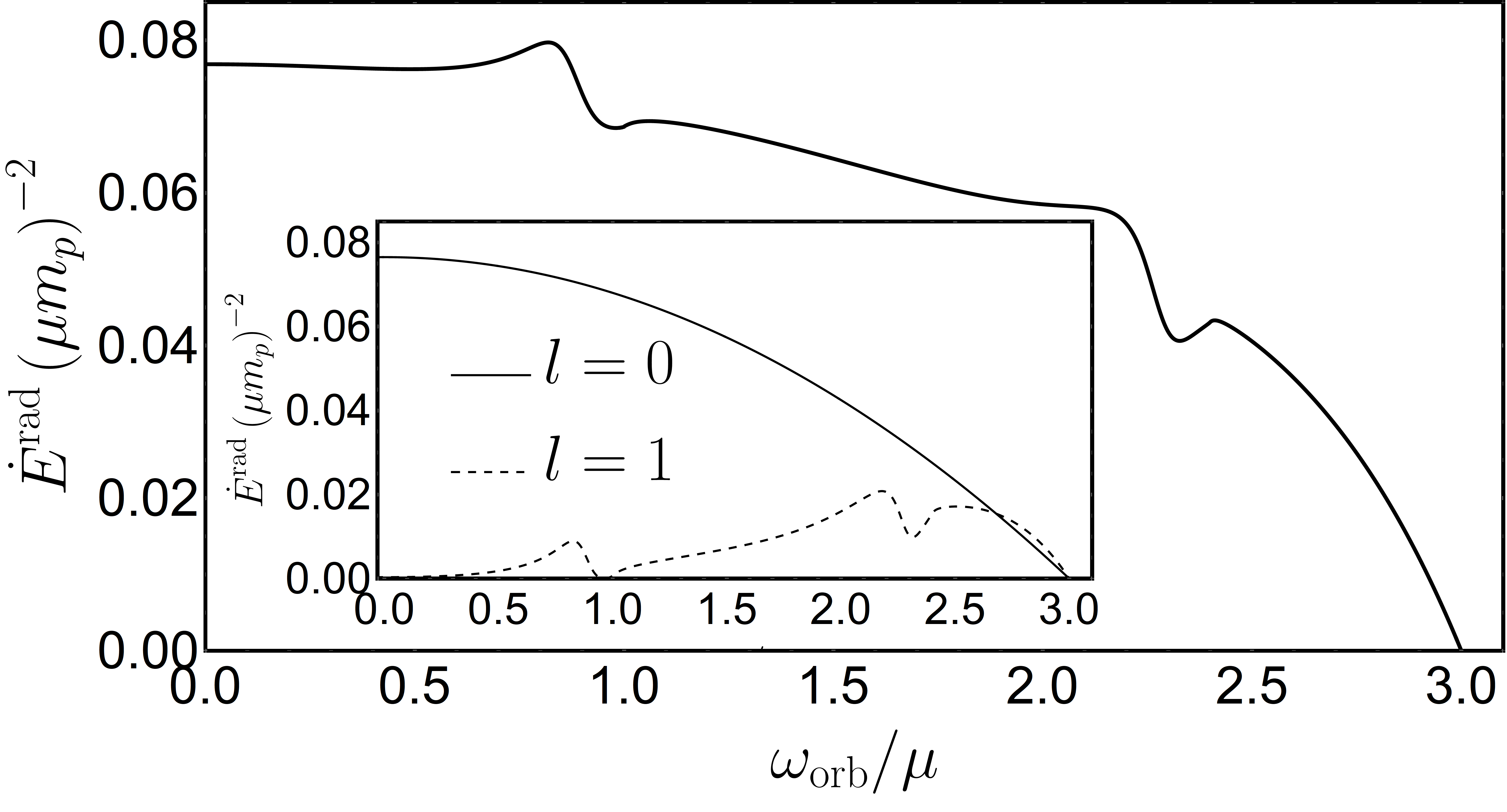} \caption{Average rate of energy radiated by a particle describing a circular orbit around a Q-ball with $\Omega=0.7\mu$, at radius $r_{\rm orb}\mu=1/3$ and with orbital frequency $\omega_{\rm orb}$. For such a scalar configuration there is radiation emitted also in the monopole mode, and it dominates the emission, as seen in the inset.} 	\label{fig:OrbitingFluxesQballOmega0p7}
\end{figure}
\begin{figure}[ht]
\includegraphics[width=8.4cm,keepaspectratio]{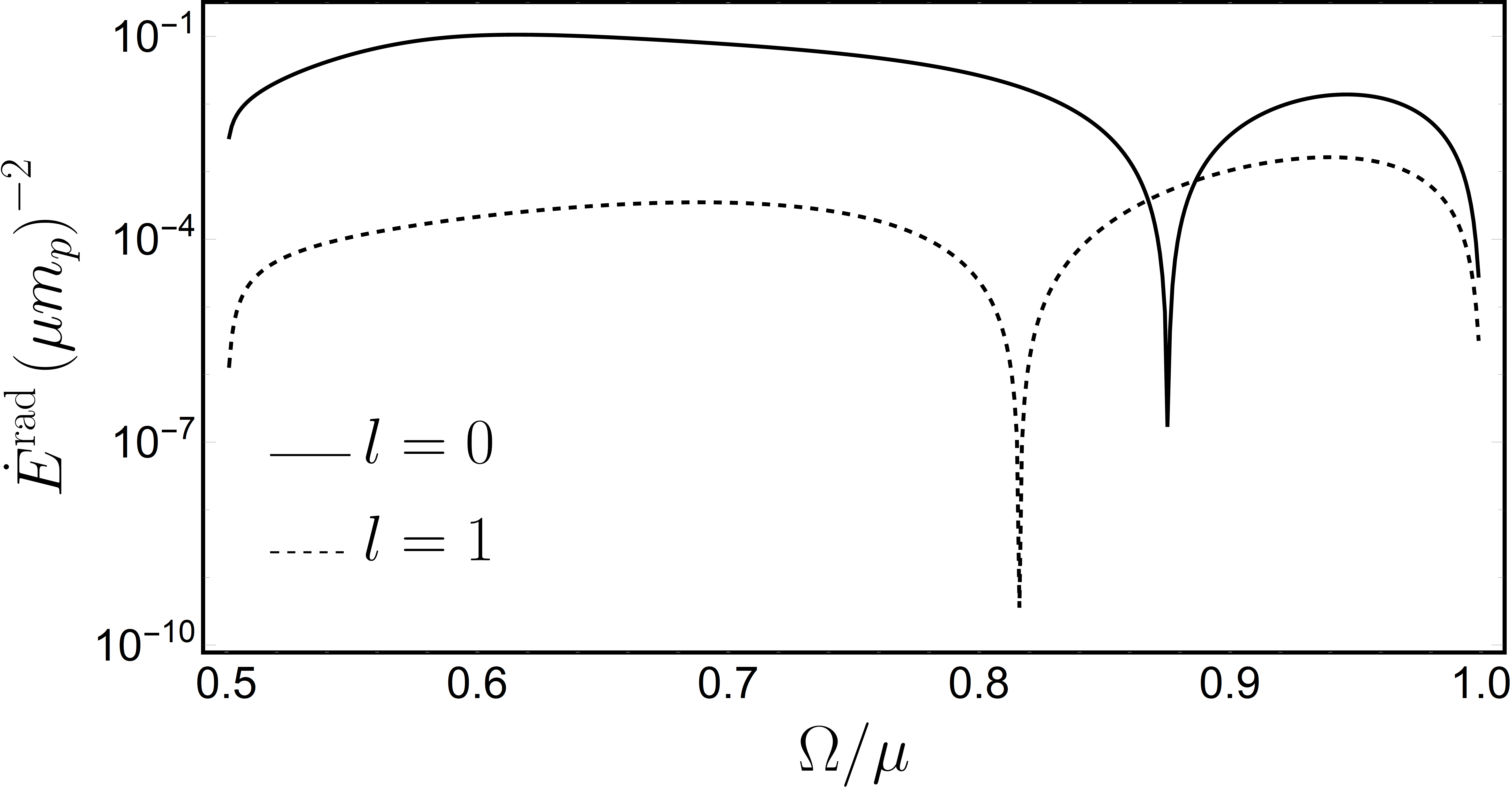} \caption{Average rate of energy radiated in the case of a particle standing at a fixed radius $r_{\rm orb}\mu=1/3$ as function of $\Omega/\mu$. It is shown the dominant contributions from the modes $l=0$ and $l=1$. The average rate of angular momentum radiated in this case vanishes.} 	\label{fig:NotOrbitingFluxesQball}
\end{figure}
One interesting aspect, not seen in the study of NBSs, concerns {\it monopolar} emission and emission from particles {\it at rest}. Both features are usually absent.
It follows from Eq.~\eqref{Energy_flux_orbiting}, that for Q-ball configurations with $\Omega\leq \mu/2$ there is no emission of $l=0$, and the first mode contributing to the radiation is $l=1$. For these objects there is no radiation emitted if the particle is at rest, with $\omega_{\rm orb}=0$. However, for Q-balls
with $\Omega/\mu>1/2$ there is indeed emission of $l=0$ modes, contributing more than (or, at least as much as) the $l=1$ modes to the radiation (\textit{see} Fig.~\ref{fig:OrbitingFluxesQballOmega0p7}). Interestingly, for these Q-balls there is also radiation emitted even when the particle is at rest (\textit{see} Fig.~\ref{fig:NotOrbitingFluxesQball}). This type of behavior is due to the coupling~\eqref{coupling_Qball} between two dynamical entities: the external perturber (through $T_p$) and the Q-ball configuration (through $\Phi$). The different coupling considered for NBSs, led to the absence of these features. 

\section{Conclusions and outlook}\label{Conclusions}
This work shows how self-gravitating NBSs and Q-balls respond to time-varying, localized matter fluctuations.
These are structures that behave classically: they are composed of $N ~\sim 10^{100}\left(10^{-22}{\rm eV}/\mu\right)^2$ particles; a binary of two supermassive BHs in the late stages of coalescence emits more than $10^{60}$ particles. Our results show unique features of bosonic ultralight structures. For example, they are not easily depleted by binaries. Even a supermassive BH binary close to coalescence would need a Hubble time or more to completely deplete the scalar in a sphere of ten-wavelength radius around the binary. In other words, the perturbative framework is consistent and robust. We have shown how a consistent, self-gravitating NBS background leads to regular, finite dynamical friction acting on passing bodies, contrasting with previous calculations using 
infinite non self-gravitating distributions~\cite{Hui:2016ltb}.

Clearly, our results can and should be extended to eccentric motion, or to self-gravitating vectorial configurations or even other nonlinearly interacting scalars~\cite{Coleman:1985ki}.
Our results should also be a useful benchmark for numerical relativity simulations involving boson stars in the extreme mass ratio regime, when and if the field is able to accommodate such challenging setups. We have considered Newtonian boson stars. Extension of our results to relativistic boson stars is nontrivial, but would provide a full knowledge of the spectrum of boson stars and of their response to eternal agents. Although we studied NBSs only, our methods can be extended to clouds arising from superradiant instabilities of spinning BHs~\cite{Brito:2015oca}. We don't expect qualitatively new aspects when the spatial extent of those clouds is large.

\noindent{\bf{\em Acknowledgements.}}
%
We are indebted to the Theory Institute at CERN and to Waseda University for warm hospitality while this work was being completed.
We are indebted to Ana Sousa Carvalho for advice and for producing one of the figures, and to Emanuele Berti
for a careful reading, and for many useful suggestions and comments. We thank Katy Clough for comments on an earlier draft.
V.~C.\ acknowledges financial support provided under the European Union's H2020 ERC 
Consolidator Grant ``Matter and strong-field gravity: New frontiers in Einstein's 
theory'' grant agreement no. MaGRaTh--646597.
This project has received funding from the European Union's Horizon 2020 research and innovation 
programme under the Marie Sklodowska-Curie grant agreement No 690904.
We thank FCT for financial support through Project~No.~UIDB/00099/2020.
We acknowledge financial support provided by FCT/Portugal through grant PTDC/MAT-APL/30043/2017.
The authors would like to acknowledge networking support by the GWverse COST Action 
CA16104, ``Black holes, gravitational waves and fundamental physics.''
L.A. acknowledges financial support provided by Funda\c{c}ao para a Ci\^{e}ncia e a Tecnologia Grant number PD/BD/128232/2016 awarded in the framework of the Doctoral Programme IDPASC-Portugal.
R.V.\ was~ supported by the FCT PhD scholarship SFRH/BD/128834/2017.
%


\appendix

\section{First post-Newtonian order expansion of the Einstein-Klein-Gordon system} \label{app:Newtonian}
Here we show that the Einstein-Klein-Gordon system reduces to the Schr\"{o}dinger-Poisson system in the Newtonian limit. Then, we obtain the equations describing a perturbation to the Newtonian fields up to first post-Newtonian corrections. Finally, we consider perturbations caused by a point-like particle. In this section we follow the treatment in Chapter 8.2 of Ref.~\cite{Poisson_will_2014}.  

The Einstein-Klein-Gordon system is the set of field equations for $\Phi$ and $g_{\mu \nu}$, which is obtained through the variation of the action~\eqref{action} with respect to $\Phi^*$ and $g_{\mu \nu}$, and reads
\beq 
&&\frac{1}{\sqrt{-g}}\partial_{\mu}\left(\sqrt{-g}g^{\mu\nu}\partial_{\nu}\Phi\right)= \mu^2\Phi\,, \nn \\
&&R_{\mu\nu}= 8\pi \widetilde{T}_{\mu\nu}^S\,,\label{EOM_BosonSapp}
\eeq
where the Einstein equations are written in an alternative form using the trace-reversed stress-energy tensor of the scalar field
\be
\widetilde{T}^S_{\mu \nu }\equiv T^S_{\mu \nu}-\frac{1}{2}T^S g_{\mu \nu}=\partial_{(\mu}\Phi^* \partial_{\nu)}\Phi+\frac{1}{2}g_{\mu \nu}\mu^2|\Phi|^2  \,.\nonumber
\ee
In the last equations we used $\mathcal{U}\sim \mu^2|\Phi|^2/2$, since we want to consider a (Newtonian) weak scalar field $|\Phi|\ll1$.
More precisely, in our perturbation scheme we consider that $\Phi \sim \mathcal{O}(\epsilon)$, with $\epsilon\ll1$. Moreover, in the Newtonian limit, we consider the spacetime metric ansatz
\beq
	g_{tt}&=&-1-2U +\mathcal{O}(\epsilon^4)\,,\nonumber \\
	g_{tj}&=&\mathcal{O}(\epsilon^3)\,,\quad	g_{jk}=\mathcal{O}(\epsilon^2)\,,
\eeq
with $j,k=\{x,y,z\}$ and where $U(t,x,y,z)\sim \mathcal{O}(\epsilon^2)$. This gives the Ricci tensor components
\beq
	R_{tt}&=& \nabla^2U+ \mathcal{O}(\epsilon^4)\,, \nonumber \\
	R_{tj}&=&\mathcal{O}(\epsilon^3)\,,\quad R_{jk}= \mathcal{O}(\epsilon^2)\,,
\eeq
where we are considering that
\be
\partial_{t}U\sim \mathcal{O}(\epsilon^3)\,, \quad \partial_{t}^2U\sim \mathcal{O}(\epsilon^4)\,.
\ee
%
%
The non-relativistic limit of the scalar field $\Phi$ is incorporated in our perturbation scheme by considering that~\footnote{This can be shown rigorously by doing an expansion in powers of ($1/c$). It corresponds to the assertion that, in the non-relativistic limit, the energy-momentum relation is $E\sim \mu+\frac{1}{2 \mu}p^2+\mu U$, with $p^2 \ll \mu^2$ and $|U|\ll 1$.}
\be
	\partial_j \Phi \sim \mathcal{O}(\epsilon^2)\,, \quad \partial_t \widetilde{\Phi} \sim \mathcal{O}(\epsilon^3)\,,
\ee
where we introduced an auxiliary scalar field $\widetilde{\Phi}$ such that
\begin{align}
\Phi= \frac{1}{\sqrt{\mu}} e^{-i\mu t}\widetilde{\Phi}\,.
\end{align}
Then, the components of the trace-reversed stress-energy tensor of the scalar field are
\beq
	\widetilde{T}_{tt}^S&=&\frac{1}{2} \mu |\widetilde{\Phi}|^2+\mathcal{O}(\epsilon^4)\,,\nonumber \\
	\widetilde{T}_{tj}^S&=&\mathcal{O}(\epsilon^3)\,, \quad\widetilde{T}_{jk}^S= \mathcal{O}(\epsilon^2)\,.
\eeq
Therefore, at Newtonian order, the Einstein equations reduce to the Poisson equation
\begin{equation}
	\nabla^2U=4\pi \mu |\widetilde{\Phi}|^2\,.
\end{equation}
On the other hand, it is easy to show that, at leading order $\mathcal{O}(\epsilon^3)$, the Klein-Gordon equation reduces to the Schr\"{o}dinger equation
\begin{equation}
	i \partial_t \widetilde{\Phi}=-\frac{1}{2 \mu} \nabla^2\widetilde{\Phi}+\mu U \widetilde{\Phi}\,.
\end{equation}
So, we have showed that, in the Newtonian limit, the Einstein-Klein-Gordon system for $\Phi$ and $g_{\mu \nu}$ reduces to the Schr\"{o}dinger-Poisson system for $\widetilde{\Phi}$ and $U$.

Let us now extend our perturbation scheme to first post-Newtonian order. We start by considering the spacetime metric ansatz
\begin{align}
g_{tt}&=-1-2U-2\delta U -2\left(\psi+ U^2\right)+\mathcal{O}(\epsilon^6)\,,\nonumber \\
g_{tj}&=-4 U_j+\mathcal{O}(\epsilon^5)\,,\nonumber \\
g_{jk}&=\left(1-2U\right) \delta_{jk}+\mathcal{O}(\epsilon^4)\,,
\end{align}
with the post-Newtonian terms $U_j(t,x,y,z)\sim \mathcal{O}(\epsilon^3)$, $\psi(t,x,y,z)\sim \mathcal{O}(\epsilon^4)$ and the perturbation $\delta U(t,x,y,z)\sim \mathcal{O}(\xi)$, where $\mathcal{O}(\epsilon^6)<\mathcal{O}(\xi)<\mathcal{O}(\epsilon^2)$. This results in the Ricci tensor components
\begin{align}
R_{tt}=& \nabla^2U+\nabla^2 \delta U+ 3 \partial_t^2 U+4 U \nabla^2 U+ \nabla^2 \psi+ \mathcal{O}(\epsilon^6)\,, \nonumber \\
R_{tj}=&2 \nabla^2U_j+ \mathcal{O}(\epsilon^5)\,, \nonumber \\
R_{jk}=&\nabla^2U\delta_{jk}+\mathcal{O}(\epsilon^4)\,,
\end{align}
where we imposed the harmonic coordinate condition, which results in
\begin{equation}
	\partial_tU+\partial_j U^j=0\,.
\end{equation}
Now, we introduce a perturbation $\delta \Phi$ to the Newtonian scalar field, such that
\begin{equation}
	\delta \Phi=\frac{1}{\sqrt{\mu}}e^{-i \mu t}\delta \widetilde{\Phi}\,,
\end{equation}
treated in our perturbation scheme with
\be
\delta \Phi\sim \mathcal{O}(\xi/\epsilon)\,, \quad	\partial_j\delta \Phi \sim \mathcal{O}(\xi)\,, \quad\partial_t \delta \widetilde{\Phi} \sim \mathcal{O}(\xi \,\epsilon)\,.
\ee
Then, the components of the trace-reversed stress-energy tensor of the scalar field are
\begin{align}
	\widetilde{T}_{tt}^S&=\frac{1}{2} \mu |\widetilde{\Phi}|^2+{\rm Im}\left(\widetilde{\Phi}\,\partial_t \widetilde{\Phi}^* \right)\nonumber \\
	&-\mu U|\widetilde{\Phi}|^2+\mu {\rm Re}\left(\widetilde{\Phi}^* \delta \widetilde{\Phi}\right)+\mathcal{O}(\epsilon^6)\,,\nonumber \\
	\widetilde{T}_{tj}^S&={\rm Im}\left(\widetilde{\Phi}\,\partial_j \widetilde{\Phi}^*\right)+\mathcal{O}(\epsilon^5)\,, \nonumber \\
	\widetilde{T}_{jk}^S&=\frac{1}{2} \mu |\widetilde{\Phi}|^2+\mathcal{O}(\epsilon^4)\,.
\end{align}
Thus, it is possible to show that, at first post-Newtonian order, the Einstein equations reduce to
\beq
\nabla^2 \psi&=&8\pi \Big[{\rm Im}\left(\widetilde{\Phi}\, \partial_t \widetilde{\Phi}^*\right)-3\mu U |\widetilde{\Phi}|^2\Big]\,,\nonumber \\
\nabla^2U_j&=&4 \pi \,{\rm Im}\left(\widetilde{\Phi}\,\partial_j \widetilde{\Phi}^*\right)\,, \nonumber\\
\nabla^2 \delta U&=&8 \pi \mu\, {\rm Re} \left(\widetilde{\Phi}^* \delta \widetilde{\Phi}\right)\,,\label{PN_Einstein}
\eeq	
where we used the equations that are satisfied at Newtonian order and we assumed $\partial^2_t U=0$, since this happens to be always the case in this work.
On the other hand, until order $\mathcal{O}(\epsilon^5)$, the Klein-Gordon equation reduces to
\beq
&&i \partial_t \delta \widetilde{\Phi}=-\frac{1}{2 \mu} \nabla^2 \delta \widetilde{\Phi}+\mu U \delta \widetilde{\Phi}+\mu \widetilde{\Phi} \,\delta U +\frac{1}{2 \mu} \partial_t^2 \widetilde{\Phi}\nonumber\\
&&+ i U \partial_t \widetilde{\Phi}+\mu \psi \widetilde{\Phi} -\frac{1}{\mu} U \nabla^2 \widetilde{\Phi}- 4i\, U^j \partial_j \widetilde{\Phi}\,.\label{KG_all}
\eeq	
Finally, note that, in the case $\mathcal{O}(\epsilon^4)<\mathcal{O}(\xi) <\mathcal{O}(\epsilon^2)$, the last equation becomes simply 
\begin{equation}
	i \partial_t \delta \widetilde{\Phi}=-\frac{1}{2 \mu} \nabla^2 \delta \widetilde{\Phi}+\mu U \delta \widetilde{\Phi}+\mu \widetilde{\Phi} \,\delta U\,.
\end{equation}

In the case of a perturbation caused by a point-like particle, one just needs to include the trace-reversed stress energy tensor of the point-like particle, Eq.~\eqref{Stress_energy_particle}, in the Einstein equation~\eqref{EOM_BosonSapp}. This is given by
\beq
&&\widetilde{T}^p_{\mu \nu}\equiv T^p_{\mu \nu}-\frac{1}{2}T^p g_{\mu \nu}\nonumber \\
&&=\frac{m_p}{2 u^0} \left(2 u_\mu u_\nu+g_{\mu \nu}\right) \frac{\delta (r-r_p)}{r^2} \frac{\delta(\theta-\theta_p)}{\sin \theta}\delta(\varphi-\varphi_p)\,,\nonumber
\eeq
with the particle's 4-velocity $u^\mu\equiv d x^\mu/d \tau$.
We consider that $m_p \sim \mathcal{O}(\xi)$ and that the particle is non-relativistic, so that $u^i\sim \mathcal{O}(\epsilon)$ in our perturbation scheme.
Then, the components of the trace-reversed stress-energy tensor of the particle are
\beq
	\widetilde{T}_{tt}^p&=&\frac{m_p}{2} \frac{\delta (r-r_p)}{r^2} \frac{\delta(\theta-\theta_p)}{\sin \theta}\delta(\varphi-\varphi_p)+\mathcal{O}(\epsilon^4)\,,\nonumber \\
    \widetilde{T}_{tj}^p&=&\mathcal{O}(\epsilon^3)\,, \quad\widetilde{T}_{jk}^p=\mathcal{O}(\epsilon^4)\,.
\eeq
Thus, we conclude that we just need to add an extra term to the last equation in~\eqref{PN_Einstein}, which becomes
\begin{equation}
	\nabla^2 \delta U=4\pi \left[2\mu\, {\rm Re}\left(\widetilde{\Phi}^* \delta \widetilde{\Phi}\right)+P\right]\,,
\end{equation}
with
\begin{equation}
	P(t,r,\theta,\varphi)\equiv m_p \frac{\delta (r-r_p)}{r^2} \frac{\delta(\theta-\theta_p)}{\sin \theta}\delta(\varphi-\varphi_p)\,.\nonumber
\end{equation}

Let us now consider the case of a non-relativistic point-like particle sourcing ultra-relativistic scalar perturbations to the Newtonian background. In our perturbation scheme, we consider~\footnote{Note that, in the ultra-relativistic limit, the energy-momentum relation becomes $E\sim p$, with $E\gg \mu $.}
\be
\delta \Phi\sim \mathcal{O}(\xi \epsilon^3)\,, \quad\partial_j\delta \Phi \sim \mathcal{O}(\xi \epsilon^2)\,, \quad \partial_t \delta \Phi \sim \mathcal{O}(\xi \epsilon^2)\,.\nonumber
\ee
with $\mathcal{O}(\epsilon^4)<\mathcal{O}(\xi)<\mathcal{O}(\epsilon^2)$. 
So, at Newtonian order, the perturbation in the scalar field does not enter in the Einstein equations, since we have
\be
\widetilde{T}_{tt}^S=\mathcal{O}\left(\epsilon^4\right)\,,\quad\widetilde{T}_{tj}^S=\mathcal{O}(\epsilon^3)\,, \quad \widetilde{T}_{jk}^S=\mathcal{O}(\epsilon^2)\,.\nonumber
\ee
In the case of a non-relativistic point-like particle, at Newtonian order, the Einstein equations describing the perturbation reduce to the Poisson equation~\footnote{The assumption of a non-relativistic perturber sourcing an ultra-relativistic scalar perturbation is consistent as long as the scalar is sufficiently light.}
\begin{equation}
	\nabla^2\delta U= 4\pi P\,. \label{eq_UR_1b}
\end{equation} 
Finally, at leading order, the Klein-Gordon reduces to
\begin{align}
-\partial^2_t \delta \Phi +\nabla^2 \delta \Phi =2 \mu^2 \Phi\, \delta U\,.\label{eq_UR_2b}
\end{align}
%

\section{The constancy of the fundamental matrix determinant} \label{app:detF}

Consider a first-order matrix ordinary differential equation
\begin{equation}\label{matrixsystem}
	\frac{d \boldsymbol{X}(r)}{dr} -V(r)\boldsymbol{X}(r)=0\,,
\end{equation}
with $\boldsymbol{X}$ a $N$-dimensional vector and $V$ a $N\times N$ matrix.
A fundamental matrix of this system is a matrix of the form $F(r)\equiv \big(\boldsymbol{X_{(1)}},...,\boldsymbol{X_{(N)}}\big)$, where $\{\boldsymbol{X_{(1)}},...,\boldsymbol{X_{(N)}}\}$ is a set of $N$ independent solutions of Eq.~\eqref{matrixsystem}. The determinant of this $N\times N$ matrix can be written as
\begin{equation*}
	\det F(r)=\epsilon^{i_1\,...\,i_N} X_{(i_1)}^1\,...\,X_{(i_N)}^N\,,
\end{equation*}
where $\epsilon$ is the Levi-Civita symbol, and $X_{(k)}^j$ is the $j$-th component of the vector $\boldsymbol{X_{(k)}}$. Using Eq.~\eqref{matrixsystem} it is easy to see that
\begin{equation}
	\frac{d }{dr}\det F=\sum_{k=1}^N \epsilon^{i_1\,...\,i_N}V^k_{\;\;\; j} \,X_{(i_1)}^1\,...\,X_{(i_k)}^j\,...\,X_{(i_N)}^N\,.
\end{equation}
Using the relation 
\begin{equation}
	\epsilon^{i_1\,...\,i_N} \,X_{(i_1)}^1\,...\,X_{(i_k)}^j\,...\,X_{(i_N)}^N=\delta_k^j \det F\,,
\end{equation}
one gets
\begin{equation}
	\frac{d }{dr}\det F= {\rm Tr}(V) \det F\,.
\end{equation}
If the trace ${\rm Tr}(V)\equiv V^k_{\;\;\; k}$ is identically zero (which is always the case in this work), the determinant of the fundamental matrix is constant.
\section{Incoming flux of energy at the center of an NBS} \label{app:incoming_flux}
Here, we compute the incoming flux of energy over a tiny spherical surface at the center of a fundamental NBS.
Consider a stationary NBS of the form
\begin{equation}
\Phi= \Psi(r)e^{-i\left(\mu- \gamma\right) t}\,,
\end{equation}
where $\Psi$ is a solution of system~\eqref{EOM_BS_radial}.
This stationary field can be written as a sum of incoming and outgoing parts $\Phi=\Phi_{\rm in}+\Phi_{\rm out}$ where
\begin{align}
	\Phi_{\rm in}&\equiv e^{-i\left(\mu- \gamma\right) t} \int_{-\infty}^{0}ds\,\overline{\Psi}(s) e^{i s r}\,, \nonumber \\
	\Phi_{\rm out}&\equiv e^{-i\left(\mu- \gamma\right) t} \int_{0}^{+\infty}ds\,\overline{\Psi}(s) e^{i s r}\,,	
\end{align}
with
\begin{equation}
	\overline{\Psi}(s)=\frac{1}{2 \pi} \int_{-\infty}^{+\infty}dr\, \Psi(r)e^{-i s r}\,,
\end{equation}
and where we are using an even extension of $\Psi$ to negative values of $r$. Note that $\overline{\Psi}$ is a real-valued function, since $\Psi$ is real-valued. 
Now, the incoming flux of energy over a tiny spherical surface of radius $r_+\ll R$ is given by
\begin{equation}\label{Ein}
	\dot{E}_{\rm in}\simeq 4 \pi r_+^2 T_{tr}^{\rm in}(r=0)\,.
\end{equation}
At leading order, one has
\beq
T_{tr}^{\rm in}(r=0) &\simeq& \mu\, {\rm Im}\left(\Phi_{\rm in}\partial_r \Phi_{\rm in}^*\right) \nonumber \\
&=&-\frac{\mu}{2} \int_{-\infty}^{0} ds' \int_{-\infty}^{0} ds \left(s'+s\right)\overline{\Psi}(s') \overline{\Psi}(s)\,.\nonumber
\eeq
Numerical evaluation of the last expression for a fundamental NBS gives
\begin{equation}
	T_{tr}^{\rm in}(r=0) \sim 2.69\times 10^{-4}\,\mu^7 M_{\rm NBS}^5\,.
\end{equation}
Finally, the incoming flux of energy is
\begin{equation}
	\dot{E}_{\rm in}\sim 3.38\times 10^{-3}\,r_+^2 \mu^7 M_{\rm NBS}^5\,.
\end{equation}

\section{Introducing a dissipative boundary} 
This section looks at two toy models, aimed at understanding the evolution
of an NBS with a small BH at its center. The main effect that the BH produces is, naturally,
dissipation at the horizon. This dissipative boundary condition can also be mimicked with some toy models.
\subsection{A string absorptive at one end} \label{app:string_toy}
Here, we wish to study a one-dimensional model of absorption of a scalar structure
when the boundary conditions suddenly change. Consider then a string, initially fixed at $x=0,\,L$, described by the wave equation
\be
\partial^2_x\Phi-\partial^2_t\Phi=0\,.
\ee
A normal mode satisfying $\Phi(x=0)=\Phi(x=L)=0$ is
\beq
\Phi&=&e^{-i\omega_n t}\sin\omega_n x\,,\\
\omega_n&=&\frac{(n+1)\pi}{L}\,,n=0,1,2...\,.
\eeq
We take a configuration with $\omega_n=\omega_0$ and use this as initial data for a problem where the boundary condition
at the origin becomes absorptive. In particular, Laplace-transform the wave equation to find,
\beq
\frac{d^2\Psi}{dx^2}+\omega^2\Psi&=&-\dot{\Phi}(0,x)+i\omega\Phi(0,x)\,,\label{eq_inh}\\
\Psi(\omega,x)&=&\int dt e^{i\omega t}\Phi(t,x)\,.
\eeq
As boundary conditions, require that 
\be
\Psi(\omega,L)=0\,,\quad \Psi(x\sim 0)=\sin\omega x-\epsilon e^{-i\omega x}\,. 
\ee
These conditions maintain the mirror-like boundary at one extreme $x=L$, while providing an absorption
of energy at $x=0$. The flux of absorbed energy scales like $\epsilon^2\ll1$.
The solution of Eq.~\eqref{eq_inh} subjected to the above boundary conditions is
\beq
\Psi&=&i\frac{\cos^2\omega x\sin\pi x/L+\sin^2\omega x\sin \pi x/L}{\omega-\pi/L}\nonumber\\
&+&\epsilon\frac{\pi\sin\omega(L-x)}{\omega(\pi-L\omega)(i\epsilon\cos\omega L+(\epsilon-i)\sin\omega L)}\,.\label{sol_nonh_string}
\eeq

The original time-domain field is given by the inverse
\be
\Phi(t,x)=\frac{1}{2\pi}\int d\omega e^{-i\omega t}\Psi(\omega, x)\,.
\ee
The integral can be done with the help of the residue theorem. We separate the response $\Phi=\Phi_1+\Phi_2$. The first term in Eq.~\eqref{sol_nonh_string} has a simple, {\it real} pole
at $\omega=\omega_0=\pi/L$, and it evaluates to
\be
\Phi_1(t,x)=\sin(\pi x/L)e^{-i\pi t/L}\,,
\ee
i.e., it corresponds to the initial data.

The second term has poles at complex values of the frequency, which are also the QNMs of the dissipative system,
\be
\omega\approx \frac{n\pi+\epsilon -i\epsilon^2}{L}\,,
\ee
These poles lie close to the normal modes of the system, including those not present in the initial data.
They dictate an exponential decay $\sim e^{-\epsilon^2 t}$, and a consequent lifetime $\tau \sim \epsilon^{-2}$.
Note that this simple exercise shows that all modes are excited when new boundary conditions are turned on.
For NBSs, all the modes cluster around $\omega \sim \mu$, thus we expect to always be in the low-frequency regime
used to estimate the lifetime.

\subsection{A black hole in a scalar-filled sphere} \label{app:bh_bomb}
%
\begin{figure*}[ht]
\begin{tabular}{ccc}
\includegraphics[width=5cm,keepaspectratio]{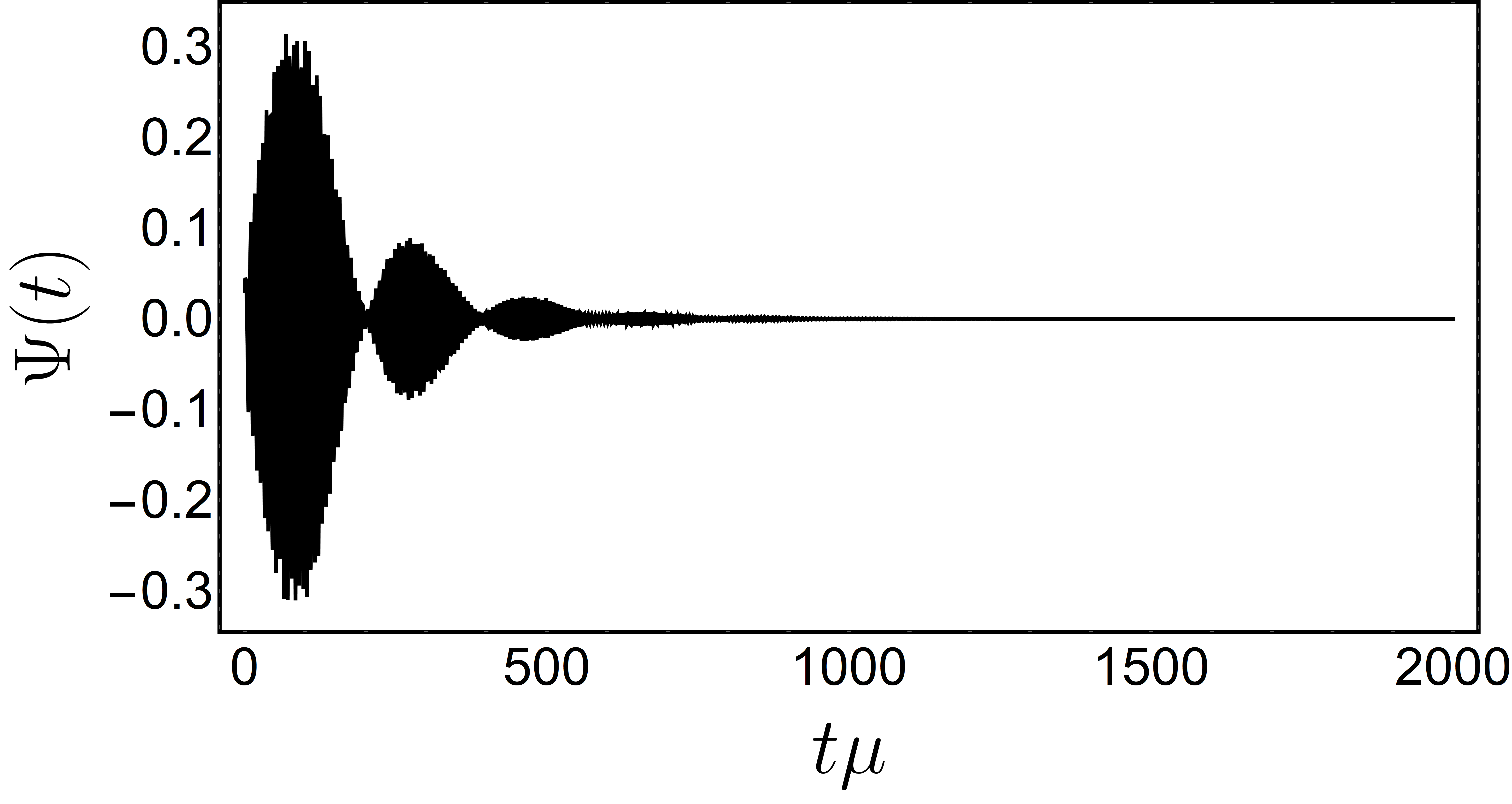} &
\includegraphics[width=5cm,keepaspectratio]{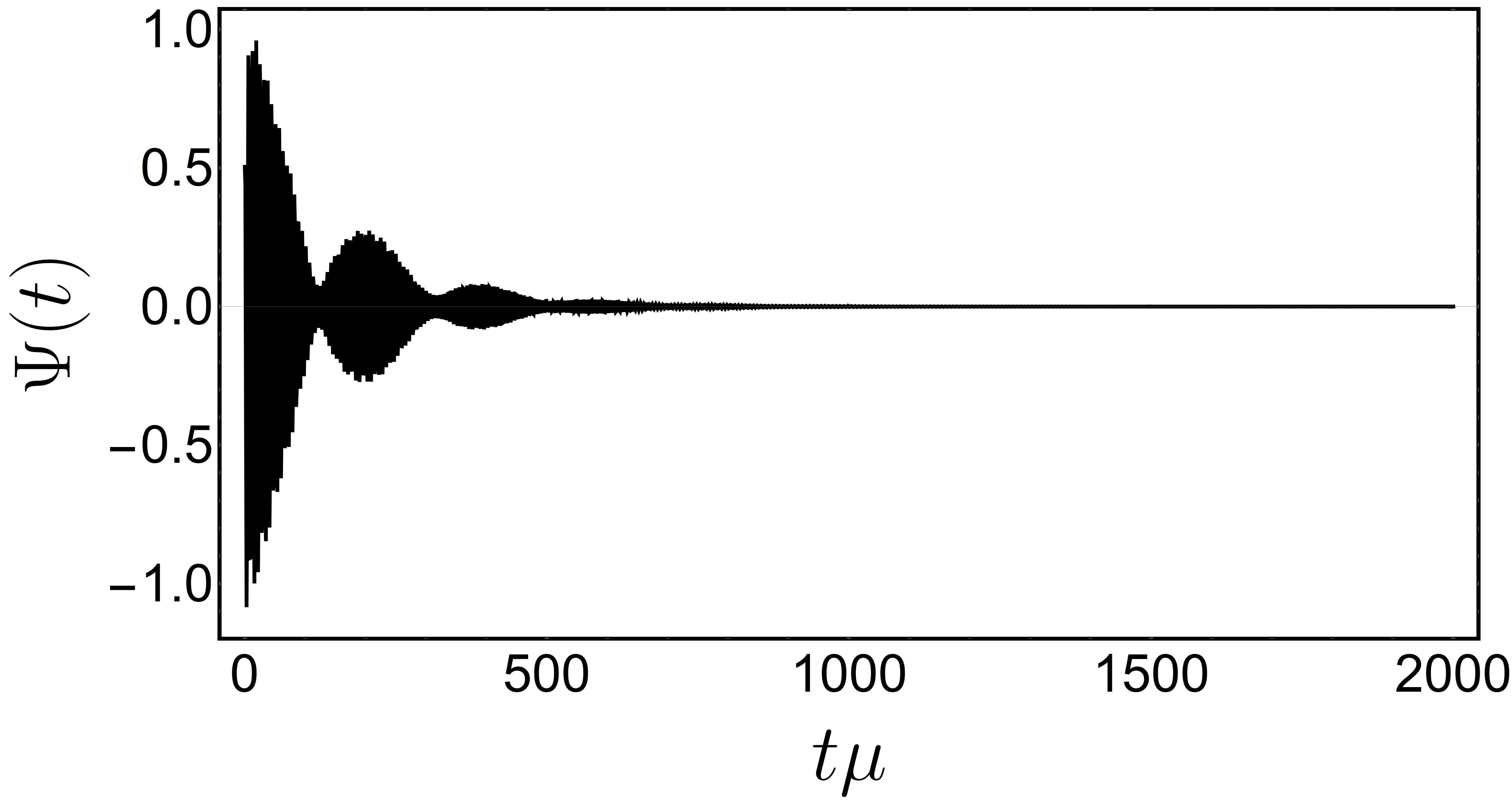}&
\includegraphics[width=5cm,keepaspectratio]{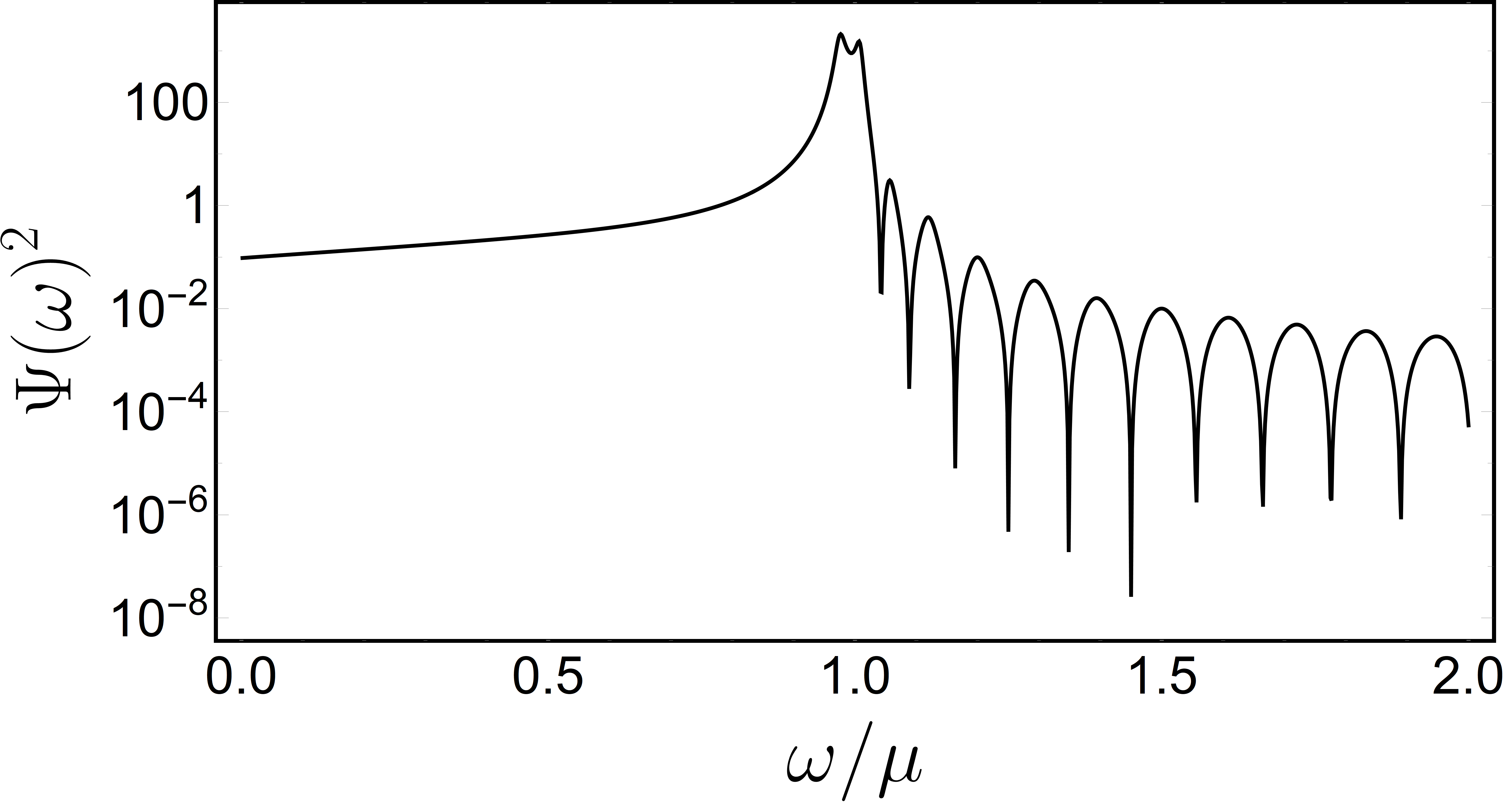}\\
\includegraphics[width=5cm,keepaspectratio]{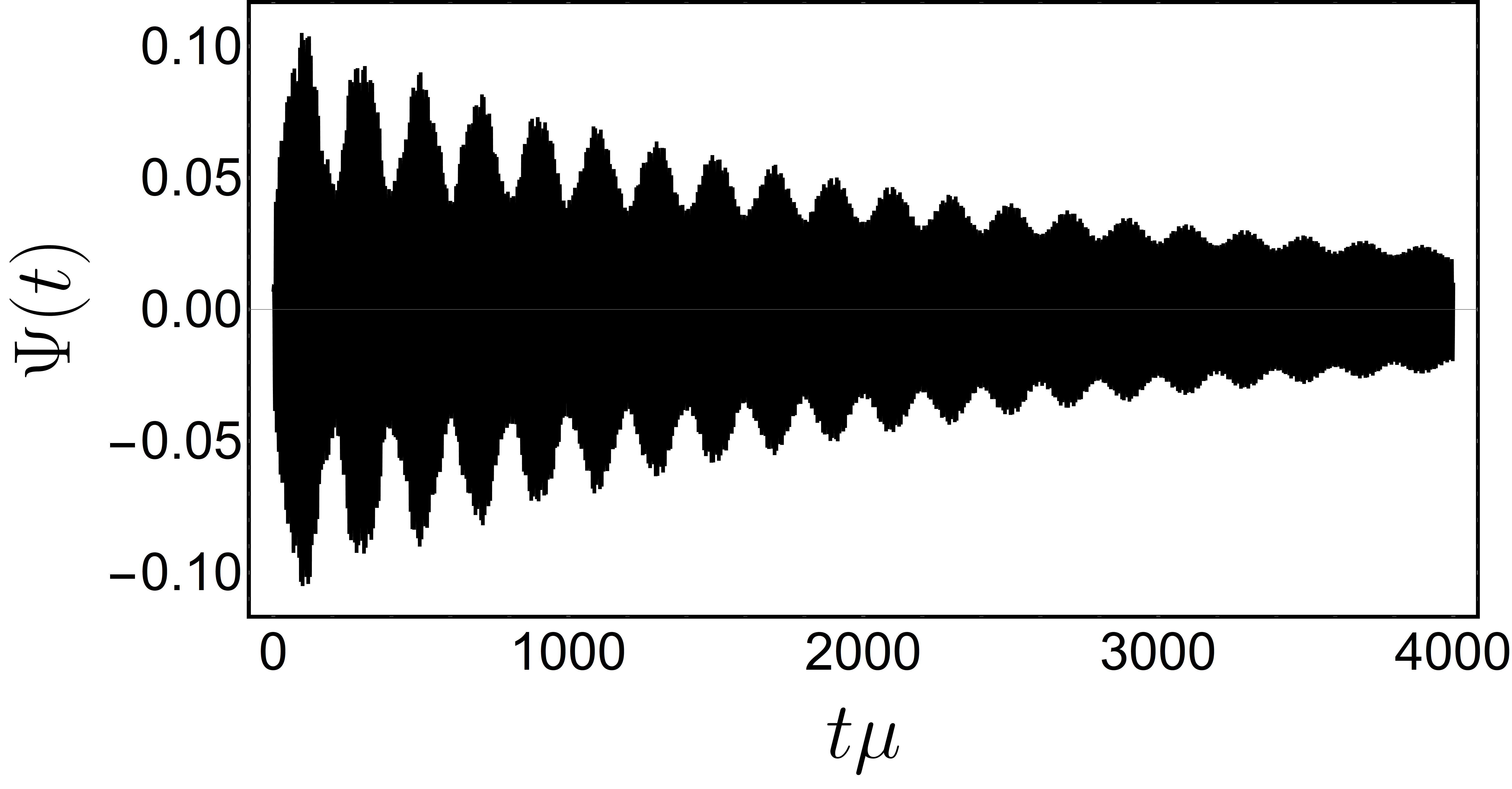} &
\includegraphics[width=5cm,keepaspectratio]{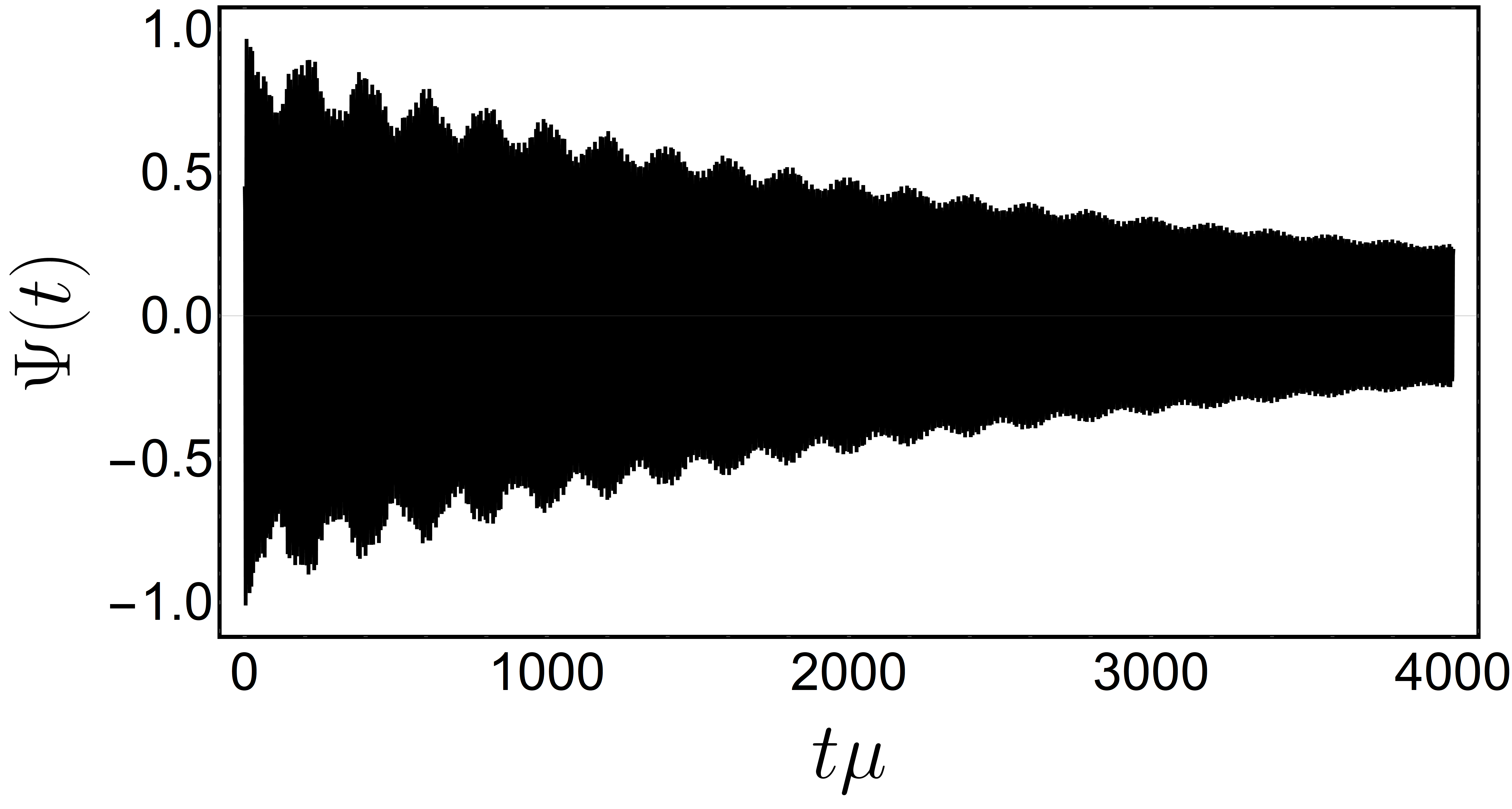}&
\includegraphics[width=5cm,keepaspectratio]{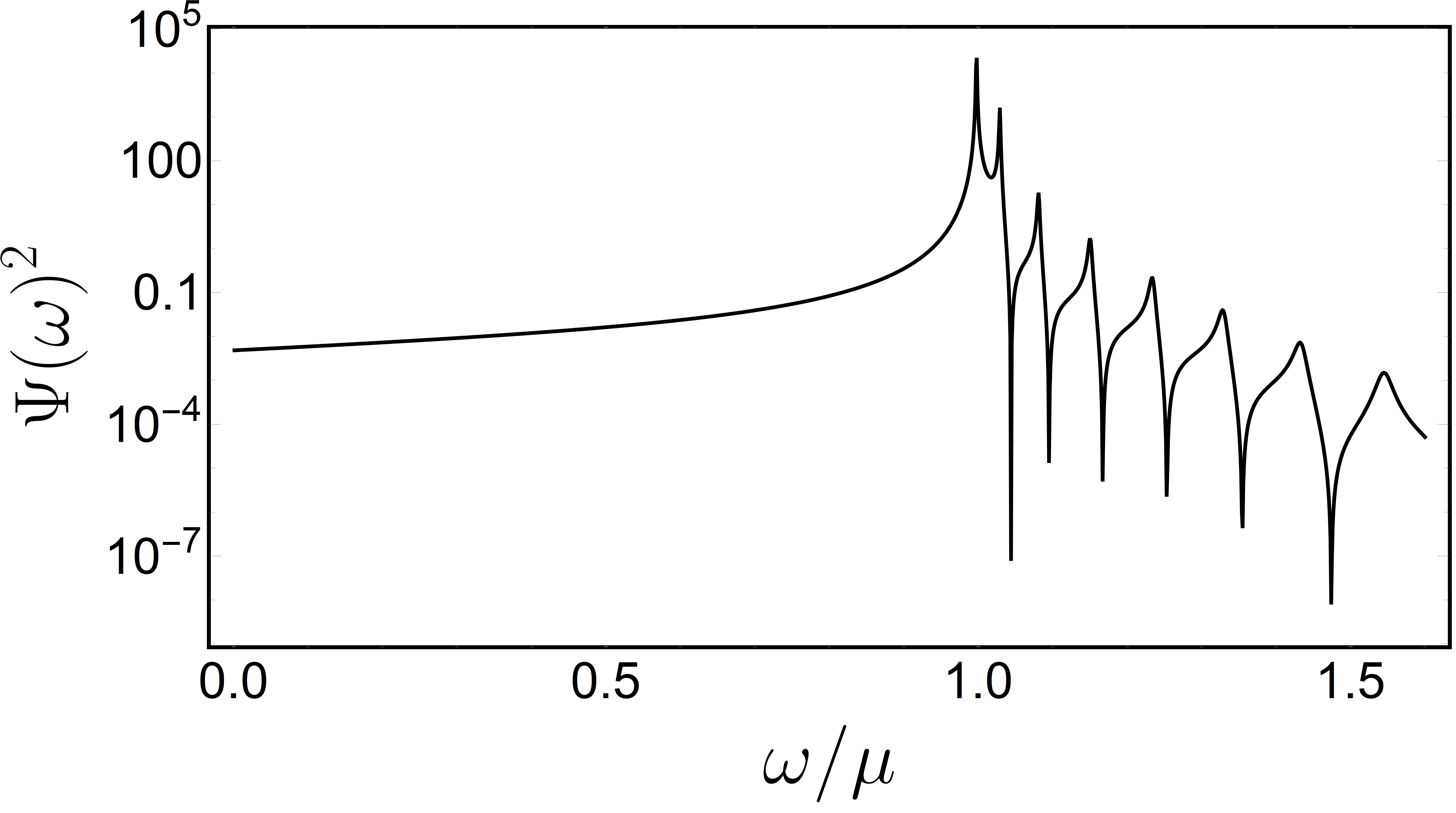}
\end{tabular}
\caption{
The evolution of a massive scalar field inside a perfectly reflecting spherical surface of radius $R\mu=20$.
In the center of such a sphere, there sits a BH of mass $M_{\rm BH}\mu=0.2$ (upper panels) and $M_{\rm BH}\mu=0.1$ (lower panels). 
{\bf Left:} Scalar field measured on the horizon. 
{\bf Center:} Scalar field measured at $r\mu=10$.
{\bf Right:} Flux measured at the horizon.
\label{fig:BHBomb_evolution}
}
\end{figure*}
A toy model more similar to the problem we wish to study is that of a BH, of mass $M_{\rm BH}$, at the center of a sphere of radius $R$ which was filled with a massive scalar field.
The profile for the scalar is, initially, that of a normal mode (the Klein-Gordon field $\Phi=\Psi/r$),
\be
\Psi=\sin \omega_0r\,,
\ee
with $\omega_0=\sqrt{\mu^2+\pi^2/R^2}$. The problem simplifies enormously when the scalar is non self-gravitating
and is a small disturbance in the background of the BH spacetime. This we assume from now onwards. In such a case all one has to do is evolve the Klein-Gordon equation
in a Schwarzschild geometry, subjected to Dirichlet conditions at the surface of the sphere. The results are summarized in Fig.~\ref{fig:BHBomb_evolution}. While they do not mimic entirely the process of accretion of a self-gravitating NBS by a central BH, these results illustrate some of the possible physics in the more realistic setup.

The figures show the scalar extracted at the horizon (left panel), at a midpoint inside the sphere (middle panel) and the flux per frequency bin (right panel).
The scalar, measured either at the horizon or somewhere within the sphere, decays exponentially. 
The first noteworthy aspect is the sensitive dependence of the decay rate on the size of the BH. Our results are consistent with a decay timescale $\tau \sim (M_{\rm BH}\mu)^{-\beta}$,
with $\beta \sim 4-5$, in agreement with our analysis in Section~\ref{sec_sitting_bh} and also with a quasinormal mode ringdown of such fields~\cite{Brito:2015oca}.
Note that such suppressed decay for small $M\mu$ couplings happens due to the filtering properties of small BHs, keeping out most of the low-frequency field.
This also explains why the ratio between the field measured at $r=10$ and at the horizon increases when the BH size decreases.
Note also that, in accordance with the simple toy model above, overtones are also excited. This is clearly seen in the Fourier analysis
(rightmost panels in Figure~\ref{fig:BHBomb_evolution}), showing local peaks at all the subsequent overtones, which were absent in the initial data. These correspond to frequencies
$\omega=\sqrt{\mu^2+\pi^2n^2/R^2}\,,n=0,\,1,...$. This is one important difference between this system and NBSs, for which overtones are all bounded in frequency.

\section{Gravitational drag by a uniform scalar field} \label{app:drag}

In this section we present the toy model considered in Ref.~\cite{Hui:2016ltb} to compute the gravitational drag acting on a point-like particle travelling through an infinite homogeneous scalar field with a constant non-relativistic velocity~$v\ll 1$. Then, we use (what we believe to be) a more realistic toy model to compute the energy and momentum lost by the infalling body through the plunge into a uniform sphere of scalar field.

Neglecting the self-gravity of the scalar field, the equations describing this process are
\begin{align} \label{SchrodingerAppend}
	i \partial_t \widetilde{\Phi}&=-\frac{1}{2 \mu} \nabla^2\widetilde{\Phi}+\mu U \widetilde{\Phi}\,, \nonumber  \\
	\nabla^2 U&=4 \pi m_p\, \delta(x)\delta(y) \delta(z+v t)\,,
\end{align}
where $m_p$ is the particle's mass.
Now, we change to the frame where the particle is stationary at the origin and the scalar field propagates with momentum $\boldsymbol{k}=\mu v \boldsymbol{e}_z$; so, the gravitational potential is simply $U=-\frac{m_p}{r}$. We consider that the scalar has a
uniform particle density $\rho_0$ in the far past -- before the interaction.  
This is the classical Coulomb scattering problem, and it is known to have the analytic solution
\begin{align} \label{CoulombScat}
	\widetilde{\Phi}&= \sqrt{\rho_0}\, e^{\frac{\pi}{2} \beta} \left|\Gamma(1-i \beta)\right|  e^{-i \left(\frac{k^2}{2 \mu}t-k z\right)} M\left[i \beta, 1, ik(r-z)\right]\,,
\end{align}
where $M$ is the confluent hypergeometric function of the first kind, $r$ is the radial distance from the particle, and the parameter~$\beta$ is
\begin{equation}
	\beta \equiv \frac{m_p \mu}{v}  \,.
\end{equation}
The Klein-Gordon scalar field~$\Phi$ can be obtained from the Schrödinger field~$\widetilde{\Phi}$ through
\begin{equation} \label{KG_Schr}
	\Phi= \frac{1}{\sqrt{\mu}}e^{-i\mu t} \widetilde{\Phi} \,, \qquad \mu \gg |\partial_t \widetilde{\Phi}|\,.
\end{equation}
As expected this solution gives~\footnote{We are using the non-relativistic limit~$k^2\ll \mu^2$.}
\begin{equation}
	T_{tt}^S(r\to \infty) = \mu \rho_0\,.
\end{equation}

A scalar field in a sphere of radius~$R$ centred at the particle exerts a gravitational drag $F_z$ on it,
\begin{equation}
	F_z=- \dot{P}_z^S-\dot{P}_z^{\rm rad}\,,
\end{equation}
using a similar reasoning to the one behind Eq.~\eqref{LossRad}. Here,~$\dot{P}_z^S$ and~$\dot{P}_z^{\rm rad}$ are, respectively, the rate of change of the momentum in the scalar inside the sphere, and the (outgoing) flux of momentum through a surface of radius~$R$; these are calculated through
\begin{align}
	&\dot{P}_z^S=-\int_{r<R} d^3\boldsymbol{r}\,  \partial_t T_{t z}^S\,, \\
	&\dot{P}_z^{\rm rad}=R^2\int_{r=R} d\theta d\varphi \sin \theta\, T_{r z}^S\,.
\end{align}
We take the radius $R$ to be the maximum characteristic length of the problem; to compare with the treatment in Section~\ref{Plunging_particle_BS} we take it to be the boson star radius. The introduction of this maximum length is necessary and serves as cutoff to the integration, since the Coulomb scattering is known to have an infrared divergence caused by the $1/r$ nature of the gravitational potential.~\footnote{In other words, the gravitational drag is known to diverge in the limit~$R \to \infty$.}
Using the divergence theorem, we can rewrite the drag force as
\begin{align}
	F_z&=- \int_{r<R} d^3\boldsymbol{r} \, \left(\partial^t T_ {t z}^S+ \partial^i T^S_ {i z}\right) \,.
\end{align}
Since we are considering a stationary regime in~\eqref{CoulombScat}, it is easy to check that~$\partial^t T_ {t z}^S$ vanishes. 
Now, using~\eqref{StressEnergy} while keeping only the leading order (Newtonian and non-relativistic) terms,
\begin{align}
	T_ {i z}^S&=\frac{1}{\mu}\,{\rm Re}\left(\partial_i \widetilde{\Phi}^*\partial_z \widetilde{\Phi}\right)\nn \\&- \frac{1}{2}g_ {i z} \left(\frac{1}{\mu}\partial^j\widetilde{\Phi}^* \partial_j \widetilde{\Phi}+i  \widetilde{\Phi}^* \partial_t \widetilde{\Phi}+2 \mu U |\widetilde{\Phi}|^2\right)\,.
\end{align}
Using Eq~\eqref{SchrodingerAppend} it is straightforward to show
\begin{align}
	\partial^i T_{i z}^S=-\mu \left(\partial_z U\right)|\widetilde{\Phi}|^2\,,
\end{align}
which implies that
\begin{equation}
	F_z=\mu\int_{r<R} d^3 \boldsymbol{r}  \left(\partial_z U\right)|\widetilde{\Phi}|^2\,.
\end{equation}
This result is symmetrical to the gravitational force that the particle exerts on the scalar and coincides with Ref.~\cite{Hui:2016ltb}. In the same reference, Hui \textit{et. al.} found that in the limit~$\beta \ll1$ the last integral can be put in the form~\footnote{In Ref.~\cite{Hui:2016ltb} the authors also obtained expressions out of the regime~$\beta\ll1$.}
\begin{align}
	&F_z= \frac{4\pi m_p^2 \,\rho_0 \mu}{v^2} C(v,R\mu)\,, \\
	&C\equiv{\rm Cin}(2 v R \mu)+\frac{\sin (2 v R \mu)}{2v R \mu}-1 \nn \,,
\end{align}
where ${\rm Cin}(x)=\int_0^x(1-\cos x') dx'/x'$ is the cosine integral. For small velocities~$v\ll1/(R\mu)$, the gravitational drag is
\begin{align}
	F_z\simeq\frac{4 \pi }{3}m_p^2 \, \rho_0 \mu^3 R^2\,.
\end{align}
This amounts to a loss of momentum by the particle~$P^{\rm lost}$ of the order 
\begin{equation}
	P^{\rm lost}\simeq F_z \frac{2 R}{v} \simeq  \frac{2}{v} m_p^2\mu^2 M \,,
\end{equation}
where~$2R/v$ is the crossing time and~$M$ is the mass of the scalar contained in the sphere of radius~$R$. Surprisingly, this expression has the same dependence on the physical quantities than the obtained for a more realistic scenario in Section~\ref{Plunging_particle_BS}; however, this result is a factor of ten larger than the one of that section.

Alternatively, one can consider a toy model closer to the treatment done in Section~\ref{Plunging_particle_BS}; this consists in linearizing the SP system with respect to an homogeneous sphere of radius $R$ made of scalar field with (particle) density $\rho_0$ and constant gravitational potential $\overline{U}_0<0$.\footnote{In fact, the assumption of a non-trivial uniform density sphere of scalar field is inconsistent with an homogeneous gravitational potential, due to the Poisson equation. Here, we assume that the Poisson equation only applies to the fluctuations of this medium.} We consider that there is no scalar field and the gravitational potential vanishes outside the sphere.
The scalar particles in this medium have energy $\Omega= \mu+\mu \overline{U}_0$, with $- \mu\ll \mu \overline{U}_0<0$. This can be readily verified by plugging the ansatz $\widetilde{\Phi}_0= e^{i\gamma t} \sqrt{\rho_0}/\mu$ in the Schrödinger equation, which inside the sphere reads
\begin{equation}
i \partial_t \widetilde{\Phi}_0=-\frac{1}{2 \mu} \nabla^2\widetilde{\Phi}_0+\mu \overline{U}_0 \widetilde{\Phi}_0\,.	
\end{equation}
That gives $\gamma=-\mu \overline{U}_0$. Then, since the KG scalar field is obtained from the Schrödinger one through~\eqref{KG_Schr}, one gets that, inside the sphere of radius~$R$, the background scalar field is $\Phi_0 = e^{-i \Omega t} \sqrt{\rho_0/\mu}$, with the energy $\Omega= \mu+\mu \overline{U}_0$ satisfying $- \mu\ll \mu \overline{U}_0<0$.

Now, we want to obtain the fluctuations caused by a point-like perturber travelling through the medium at constant (non-relativistic) velocity $v\ll 1$ along the $-\boldsymbol{e}_z$ direction. 
%
These fluctuations are described by the linearized SP system
\begin{align}
&i \partial_t \delta \widetilde{\Phi} =- \frac{1}{2 \mu} \nabla^2\delta \widetilde{\Phi} +\mu U_0 \delta \widetilde{\Phi}+ \mu \widetilde{\Phi}_ 0 \delta U\,, \label{Spert}\\
&\nabla^2\delta U= 4 \pi P \label{Ppert}\,,
\end{align}
with
\begin{equation}
	U_0=\overline{U}_0 \Theta(R-r)\,, \quad \widetilde{\Phi}_0=\frac{\sqrt{\rho_0}}{\mu} \Theta(R-r) e^{-i\mu \overline{U}_0 t}\,,
\end{equation} 
and where the source is given by
\begin{align}
	P&= m_p \frac{\delta(\varphi)}{r^2 \sin \theta} \nonumber \\
	&\times \left[\delta(r+v t) \delta(\theta) \Theta(-t)+ \delta(r-vt) \delta(\theta- \pi)\Theta(t)\right]\,.\nonumber
\end{align}
Note that the fluctuations $\delta \widetilde{\Phi}$ in the Schrödinger field are related with the fluctuation in the KG field through~$\delta \Phi= e^{-i\mu t}\delta \widetilde{\Phi}$. For simplicity, we are neglecting the self-gravity of the scalar field in the right-hand side of~\eqref{Ppert}. This is a good approximation in the region close to the particle, where the Coulombian potential is dominant.
Using the axially symmetric decompositions
\begin{align}
	P&=\sum_{l=0}^{\infty} \int \frac{d\omega}{\sqrt{2 \pi} r}e^{-i \omega t} Y_l^0(\theta) p(r)\,, \\
	\delta U&=\sum_{l=0}^{\infty} \int \frac{d\omega}{\sqrt{2 \pi} r}e^{-i \omega t} Y_l^0(\theta) u(r)\,, 
\end{align}
where
\begin{align}\label{p_source}
	p&=\sqrt{\frac{2}{\pi}} m_p \frac{Y_l^0(0)}{r v} \delta_m^0 \nonumber \\
	&\times \left[\cos\left(\frac{\omega}{v} r\right)\delta_l^{\rm even}-i \sin \left(\frac{\omega}{v}r\right)\delta_l^{\rm odd}\right]\,,
\end{align}
the Poisson equation becomes
\begin{align}
	\partial_r^2u-\frac{l(l+1)}{r^2}u=4 \pi p\,.
\end{align}
This admits the homogeneous solutions
\begin{align*}
	u^I=r^{-l}\,, \qquad u^{II}=r^{l+1}\,, 
\end{align*}
which are regular, respectively, at infinity and at the origin. Using the method of variation of parameters, one gets the inhomogeneous solution
\begin{align}
	u=-\frac{4\pi}{2l+1} \left(r^{-l} \int_{0}^{r}dr' r'^{l+1}p+r^{l+1}\int_r^{\infty}dr' \frac{p}{r'^l} \right)\,,\nonumber
\end{align}
This results in the analytical expression
\begin{align} \label{u_final}
	u&=-i\,\frac{2\sqrt{2\pi}}{2l+1}\frac{m_p}{\omega} Y_l^0(0)\delta_m^0\left(i \frac{r}{v} \omega\right)^{l+1} \nonumber \\
	&\times\Bigg\{\left(i \frac{\omega}{v}r\right)^{-2l-1} \left[\Gamma\left(l+1,i \frac{\omega}{v} r\right)-\Gamma\left(l+1,-i \frac{\omega}{v} r\right)\right]\nonumber\\
	&-\Gamma\left(-l,i \frac{\omega}{v} r\right)-\Gamma\left(-l,-i \frac{\omega}{v} r\right)\Bigg\}\,,
\end{align}
where $\Gamma(a,x)$ is the incomplete gamma function. 
Now, decomposing the scalar fluctuation as
\begin{align}
	\delta \widetilde{\Phi}=\sum_{l=0}^{\infty} \int \frac{d\omega}{\sqrt{2 \pi} r}e^{-i \left(\omega+\mu \overline{U}_0\right) t}\, Y_l^0(\theta) Z(r)\,,
\end{align}
equation~\eqref{Spert} becomes
\begin{align} \label{Zeq}
\partial_{r}^2 Z+\left[2\mu \left(\omega+\mu \overline{U}_0 \Theta(r-R) \right) - \frac{l(l+1)}{r^2}\right]Z=\nonumber \\
=2\mu \sqrt{\rho_0}\, \Theta(R-r) u\,.	
\end{align}
Outside the sphere o radius~$R$, the solution satisfying the Sommerfeld radiation condition at infinity is simply given by
\begin{equation}
	Z(r)=A \sqrt{r}\, H_{l+\frac{1}{2}}^{(1)}\left(\sqrt{2 \mu\left(\omega+ \mu \overline{U}_ 0\right)}\,r\right)\,,
\end{equation} 
where~$A$ is a complex-constant to be determined through the matching with the interior solution.
Using equation~\eqref{u_final} it is possible to see that the highest frequencies that the perturber excites (efficiently) are~$\omega \sim v/(2 R)$. This has the important consequence that for velocities $v \ll 2 R\mu |\overline{U}_0|$ the emission is strongly suppressed, because the perturber cannot excite (efficiently) waves that travel to infinity. Additionally, in the limit of small velocities~$v\ll 1/(R \mu)$, we have
\begin{align}
	Z(r \sim R)\simeq -\frac{i A}{\pi} \frac{2^{l+\frac{1}{2}}\,\Gamma\left(l+\frac{1}{2}\right)}{\left[2 \mu \left(\omega+\mu \overline{U}_0\right)\right]^{\frac{l}{2}+\frac{1}{4}}}  r^{-l}\,,
\end{align}
where we used the small argument expansion of $H^{(1)}_{l+\frac{1}{2}}$.
Inside the sphere of radius $R$, equation~\eqref{Zeq} has the independent homogeneous solutions
\begin{align}
	Z^I&=\sqrt{r}\,H^{(1)}_{l+\frac{1}{2}}(\sqrt{2 \mu \omega} \,r) \simeq -\frac{i}{\pi} \frac{2^{l+\frac{1}{2}}\,\Gamma\left(l+\frac{1}{2}\right)}{\left(2 \mu \omega\right)^{\frac{l}{2}+\frac{1}{4}}}  r^{-l}\,, \nonumber  \\
	Z^{II}&=\sqrt{r}\,J_{l+\frac{1}{2}}(\sqrt{2 \mu \omega} \,r) \simeq \frac{\left(2 \mu \omega\right)^{\frac{l}{2}+\frac{1}{4}}}{2^{l+\frac{1}{2}}\, \Gamma\left(l+\frac{3}{2}\right)} r^{l+1}\,.
\end{align}
The solution $Z^{II}$ is regular at the origin, and the solution $Z^I$ is (approximately) proportional to $r^{-l}$ everywhere inside the sphere, making it appropriate to match with the exterior solution at $r=R$.  
Using the method of variation of parameters, one obtains that the radial function $Z$ at $r=R$ is
\begin{align}
	Z(R)=-i \pi \mu \sqrt{\rho_ 0}\,  Z^I(R)\int_0^R dr' Z^{II} \,u(r')\,.
\end{align}
Then, the constant $A$ can be determined through matching between the interior and exterior solutions,
\begin{align} \label{integr}
	A=-i \pi \mu \sqrt{\rho_ 0}\left(1+\frac{\mu \overline{U}_ 0}{\omega}\right)^{\frac{l}{2}+\frac{1}{4}}\int_0^R dr' Z^{II} \,u(r')\,.
\end{align}
%
Using the large argument expansion of $H^{(1)}_{l+\frac{1}{2}}$ one gets the radial function $Z$ at infinity,
\begin{align}
	Z_\infty &\equiv Z(r\to \infty)= \nonumber \\
	&=-\frac{2^\frac{1}{4}(-i)^{l-1}}{\sqrt{\pi}} \frac{A\,\sqrt{R}\,e^{i\sqrt{2\mu \left(\omega+\mu \overline{U}_0\right)}\, r}}{(v R \mu)^\frac{1}{4}\alpha^\frac{1}{4}\left(1+\frac{\overline{U}_0 R \mu}{v \alpha}\right)^\frac{1}{4}}\,,
\end{align}
with the dimensionless parameter
\begin{equation}
\alpha \equiv \frac{\omega R}{v}\,.
\end{equation}
Evaluating the integral in~\eqref{integr} we obtain the analytical expression~\footnote{We have also solved this problem numerically (without any approximation). This analytical expression describes perfectly the exact results for the first multipoles (essentially~$l\leq3$); these account for most of the radiation.}
\begin{align}
	Z_\infty&=i \pi \delta_m^0 (R \mu)^4 \frac{m_p \sqrt{\rho_0}}{\mu^2} e^{i\sqrt{2\mu \left(\omega+\mu \overline{U}_0\right)}\, r}   \nonumber \\
	&\times\frac{(-1)^l Y_l^0(0)(v R \mu)^{\frac{l}{2}-1}}{2^{\frac{l}{2}-\frac{1}{2}}(2l+1) \Gamma\left(l+\frac{3}{2}\right)} \frac{\left(1+\frac{\overline{U}_ 0R \mu}{v \alpha}\right)^{\frac{l}{2}}}{\alpha^{\frac{l}{2}+3}} \nonumber \\
	&\times \Bigg\{\frac{2l+1}{2l+3} \left[\Gamma\left(l+3,i \alpha\right)-\Gamma\left(l+3,-i \alpha\right)\right] \nonumber \\
	&+2\frac{(i \alpha)^{2l+3}}{2l+3}\left[\Gamma\left(-l,i \alpha\right)+\Gamma\left(-l,-i \alpha\right)\right] \nonumber \\
	&+\alpha^2\left[\Gamma\left(l+1, i\alpha\right)-\Gamma\left(l+1, -i\alpha\right)\right]\Bigg\}\,.
\end{align}
The energy radiated with frequency between~$\omega$ and $\omega+d\omega$ is 
\begin{align}
	&\frac{d E^{\rm rad}}{d \omega}=\frac{\sqrt{2} }{R}\left(v R \mu \right)^{\frac{1}{2}}\left[\mu+\frac{\alpha v}{R }\left(1+\frac{\overline{U}_0R \mu}{v \alpha}\right)\right]  \nonumber \\
	&\times\alpha^\frac{1}{2}\,{\rm Re}\left[ \left(1+\frac{\overline{U}_0R \mu}{v \alpha}\right)^\frac{1}{2}\right]\sum_{l=0}^{\infty} \left|Z_ \infty\right|^2 \nn \\
	&\simeq \frac{\sqrt{2} \mu}{R}\left(v R \mu \right)^{\frac{1}{2}}  \nonumber \\
	&\times\alpha^\frac{1}{2}\,{\rm Re}\left[ \left(1+\frac{\overline{U}_0R \mu}{v \alpha}\right)^\frac{1}{2}\right]\sum_{l=0}^{\infty} \left|Z_ \infty\right|^2\,,
\end{align}  
where in the last equality we used that the scalar fluctuations are non-relativistic.
This results in the total radiated energy
\begin{align}
	&E^{\rm rad}=\frac{\sqrt{2}}{R^3}\,  \left(v R \mu\right)^{\frac{3}{2}} \nonumber \\
	&\times\sum_{l=0}^{\infty} \int_{\frac{\left|\overline{U}_0\right| R \mu}{v}}^\infty d \alpha \alpha^{\frac{1}{2}} {\rm Re}\left[ \left(1+\frac{\overline{U}_0 R \mu}{v \alpha}\right)^\frac{1}{2}\right] \left|Z_ \infty\right|^2\,.
\end{align}
The energy lost by the perturber in this process is
\begin{align}
&E^{\rm lost}=\frac{\sqrt{2}}{R^5 \mu^2}\,  \left(v R \mu\right)^{\frac{5}{2}} \nonumber \\
&\times\sum_{l=0}^{\infty} \int_{\frac{\left|\overline{U}_0\right| R \mu}{v}}^\infty d \alpha \alpha^{\frac{3}{2}} {\rm Re}\left[ \left(1+\frac{\overline{U}_0 R \mu}{v \alpha}\right)^\frac{3}{2}\right] \left|Z_ \infty\right|^2\,.
\end{align}
In the case of a vanishing gravitational potential~$\overline{U}_0=0$, we see that for small velocities the radiated energy goes with $\sim v^{-\frac{1}{2}}$ and the energy lost by the perturber with~$v^{\frac{1}{2}}$; note that~$Z_\infty \sim v^{\frac{l}{2}-1}$. In the case of a non-trivial gravitational potential, the radiated energy is highly suppressed for small velocities; this is because smaller velocities excite lower frequencies -- these may not be capable of escaping the gravitational influence of the scalar configuration.

The spectral flux of linear momentum radiated along~$z$ is given by
\begin{align}
	\frac{d P_z^{\rm rad}}{d \omega}&=\frac{4}{R^2} \left(v R \mu\right)\alpha\, \Theta\left(1+\frac{\overline{U}_0 R \mu}{v\alpha}\right)\left(1+\frac{ \overline{U}_0 R \mu}{v \alpha}\right) \nonumber \\
	&\times\sum_{l=0}^{\infty} \frac{ (l+1)\,{\rm Re}\left(Z_\infty^l\left(Z_\infty^{l+1}\right)^*\right)}{\sqrt{\left(2l+1\right)\left(2l+3\right)}} \,.
\end{align}
So, the total linear momentum radiated during this process is
\begin{align}
	&P_z^{\rm rad}=\frac{4}{R^4 \mu} \left(v R \mu\right)^2 \sum_{l=0}^{\infty} \frac{l+1}{\sqrt{(2l+1)(2l+3)}} \nonumber \\
	&\times \int_{\frac{\left|\overline{U}_0\right| R \mu}{v}}^\infty d \alpha\, \alpha \left(1+\frac{ \overline{U}_0 R \mu}{v \alpha}\right) \,{\rm Re}\left(Z_\infty^l\left(Z_\infty^{l+1}\right)^*\right)\,.
\end{align}
The loss in momentum for a small perturber~$m_p \mu \ll v$ is simply
\begin{align}
	P_z^{\rm lost}=\frac{E^{\rm lost}}{v}\,.
\end{align}
In the case of a vanishing gravitational potential, for small velocities the radiated momentum goes with~$v^\frac{1}{2}$ and perturber's loss in momentum with~$\sim v^{-\frac{1}{2}}$. Again, with a non-trivial gravitational potential both quantities are suppressed in the limit of small velocities.

Our toy model shows that: (i) the gravitational potential of a scalar configuration tends to suppress both the radiation and the loss in momentum for plunging perturbers, specially in the small velocity limit;~\footnote{Actually, although we do not present it in this work, we solved the full problem -- including the self-gravity of the scalar -- in a way similar to Section~\ref{Plunging_particle_BS} but with constant velocity. We found qualitative agreement with the toy model considered here; however, including the self-gravity of the scalar y to a larger suppression of radiation and loss of momentum.}
(ii) when neglecting the gravity of the scalar, the loss in momentum for a perturber plunging in a uniform sphere of scalar field at a constant small velocity follows~$P^{\rm lost}\sim v^{-\frac{1}{2}}$. This behavior is different than the one found in Ref.~\cite{Hui:2016ltb}; in that reference besides neglecting the gravity of the scalar, the authors study a stationary regime in an infinite scalar field medium (introducing a cut-off length~$R$ \textit{a posteriori}).

For a full realistic plunge into an NBS -- including the self-gravity of the scalar and the accelerated free fall of the perturber  -- see Section~\ref{Plunging_particle_BS}.
\section{Exciting a spherical box} \label{app:internal_modes}
Here we study the scalar field inside a spherical box of radius $R$, which is sourced by a particle in circular orbital motion.
Our approach to study the resonances follows the treatment of forced oscillations in Ref.~\cite{landau1982mechanics}.

Consider a $U(1)$-invariant scalar field theory described by the action
\begin{equation} \label{theory}
	S\equiv \frac{1}{2}\int d^4 x \sqrt{-\eta} \,\partial^\mu \Phi \partial_\mu \Phi^*\,,
\end{equation}
on a Minkowski background. Moreover, let us consider that this scalar field is sourced by a point particle through the action~\eqref{coupling_Qball}. 
Let us first start with the sourced equation of motion
\begin{equation}\label{eq:box}
	\nabla_\mu\partial^\mu  \Phi= T_p\,,
\end{equation}
which is obtained through the variation of the total action, and where $T_p$ is the trace of the particle's stress-energy tensor. Assuming the particle motion to be in the equatorial plane, with $r_{\rm orb}$ and $\omega_{\rm orb}$, the orbital radius and angular frequency, respectively, the trace of the particle's stress-energy tensor is given by Eq.~\eqref{T_p_orbiting}. Consider then the decompositions
\begin{align*}
	\Phi&=\sum_{l,m}\int \frac{d\omega}{\sqrt{2\pi}}e^{-i\omega t}\frac{\phi(r;\omega,l,m)}{r}Y_{l}^m(\theta,\varphi)\,, \\
	T_p&=\sum_{l,m}\int \frac{d\omega}{\sqrt{2\pi}}e^{-i\omega t}\frac{T_{l}^m(r;\omega)}{r^2}Y_{l}^m(\theta,\varphi) \,.
\end{align*}
Thus, one can use the method of variation of parameters to obtain the general inhomogeneous solution of Eq.~\eqref{eq:box}
\begin{equation}
\phi=\phi^h-\phi^I \int_0^r dr'\left(\frac{\phi^{II} T_l^m}{r' W}\right)-\phi^{II} \int_r^R dr'\left(\frac{\phi^{I} T_l^m}{r' W}\right)\,,\nonumber
\end{equation}
with $0 \leq r\leq R$, where $\phi^h$ is the general homogeneous solution
\begin{equation*}
	\phi^h=A_I \phi^I+A_{II} \phi^{II}\,,
\end{equation*}
with
\begin{align*}
\phi^I&=\sqrt{r} \left[J_{l+\frac{1}{2}}(\omega r)-\frac{J_{l+\frac{1}{2}}(\omega R)}{Y_{l+\frac{1}{2}}(\omega R)}Y_{l+\frac{1}{2}}(\omega r)\right]\,,\\
\phi^{II}&=\sqrt{r} J_{l+\frac{1}{2}}(\omega r)\,.
\end{align*}
The Wronskian of $\phi^I$ and $\phi^{II}$ is $W=2 J_{l+\frac{1}{2}}(\omega R)/\left(\pi Y_{l+\frac{1}{2}}(\omega R)\right)$. 

Now, we want to impose regularity at the origin $\phi(0)=0$, and Dirichlet conditions at the surface of the box $\phi(R)=0$. Notice that the homogeneous solutions $\phi^{II}$ satisfy these boundary conditions for discrete values of $\omega=\omega_n$ (with $n \in \mathbb{Z} \setminus \{0\}$), such that $J_{l+\frac{1}{2}}(\omega_n R)=0$; the frequencies $\omega_n$ are the normal modes of a scalar field in a spherical box. The wronskian above vanishes for these modes. We use the notation $\omega_{-n}=-\omega_n$.

For $\omega \neq \omega_n$, the boundary conditions imply that $\phi^h=0$. Thus,
\beq
\phi&=&A_l^m(\omega) \phi^{II} \delta(\omega-\omega_n)\nonumber\\
&-&\phi^I \int_0^r dr'\left(\frac{\phi^{II} T_l^m}{r' W}\right)-\phi^{II} \int_r^R dr'\left(\frac{\phi^{I} T_l^m}{r' W}\right)\,,\nonumber
\eeq
with arbitrary complex coefficients $A_l^m(\omega_n)=A_{l,n}^m$.
In particular, at the origin,
\beq 
\phi(r\to 0)&\sim&\left(\frac{\omega}{2}\right)^{l+\frac{1}{2}} \frac{r^{l+1}}{\Gamma\left(l+\frac{3}{2}\right)} \nonumber\\
&\times&\left[A_l^m(\omega) \delta(\omega-\omega_n)-\int_0^R dr'\left(\frac{\phi^{I} T_l^m}{r' W}\right)\right]\,.\nonumber
\eeq
The total field at the origin is then
\begin{align}
	&\Phi(r\to 0)\sim \sum_{l,m}\frac{ Y_l^m(\theta,\varphi)r^{l}}{\Gamma\left(l+\frac{3}{2}\right)2^{l+\frac{1}{2}}} \left[\sum_{n \neq 0}\frac{(\omega_n)^{l+\frac{1}{2}}}{\sqrt{2\pi}} A_{l,n}^me^{-i\omega_n t}+ \right. \nonumber \\
	&\left. m_p(m \omega_{\rm orb})^{l+\frac{1}{2}} \sqrt{1-(r_{\rm orb} \omega_{\rm orb})^2} Y_l^m\left(\frac{\pi}{2},0\right) \frac{\phi^I e^{-i m \omega_{\rm orb} t}}{r_0 W(m\omega_{\rm orb})} \right]\,,\nonumber
\end{align}
with $\phi^I=\phi^I(r_{\rm orb}; m \omega_{\rm orb})$.
For simplicity, take for now $A_{l,n}^m=0$ (\textit{i.e.} we neglect the free normal mode part of the solution).
Moreover, notice that in this theory the energy density of the scalar field is
\begin{equation}
	T_{t t}=\frac{1}{2}\left(|\partial_t \Phi|^2+|\partial_r \Phi|^2+\frac{1}{r^2} |\partial_\theta \Phi|^2 +\frac{1}{r^2 \sin^2 \theta} |\partial_\varphi \Phi|^2\right)\,.\nonumber
\end{equation}
Thus, at the origin, the energy density is
\be
T_{t t}(0)\sim	\frac{\omega_{\rm orb}^3}{16 \pi^3}\left(\frac{m_p \phi^I(r_{\rm orb};\omega_{\rm orb})}{r_{\rm orb}W(\omega_{\rm orb})}\right)^2\left[1-\left(r_{\rm orb} \omega_{\rm orb}\right)^2\right]\,.\nonumber
\ee
Since the Wronskian $W$ vanishes for $\omega_{\rm orb}=\omega_{n'}$, the energy density $T_{t t}(r\to 0)$ diverges for these frequencies. Thus, resonances are produced for particles with orbital frequencies $\omega_{\rm orb}=\omega_{n'}$. 

It is interesting to pose the question of how $T_{t t}(r\to 0)$ increases with time when the system is in resonance with $\omega_{\rm orb}=\omega_{n'}$. To answer this, let us choose
\beq
A_{l,n}^m&=&-\sqrt{2 \pi} m_p \sqrt{1-(r_{\rm orb} \omega_{\rm orb})^2}(1-\delta_0^m)\delta_n^{\text{sign}(m)\,n'}\nonumber\\
&\times& Y_l^m\left(\frac{\pi}{2},0\right) \frac{\phi^I(r_{\rm orb}; m \omega_{\rm orb})}{r_{\rm orb} W(m\omega_{\rm orb})}\left(\frac{m \omega_{\rm orb}}{\omega_n}\right)^{l+\frac{1}{2}}.\nonumber
\eeq
With this choice, the total field at the origin is
\beq
\Phi&\sim& \sum_{l,m\neq 0}\frac{ Y_l^m(\theta,\varphi)r^{l}}{\Gamma\left(l+\frac{3}{2}\right)2^{l+\frac{1}{2}}}m_p(m \omega_{\rm orb})^{l+\frac{1}{2}}  Y_l^m\left(\frac{\pi}{2},0\right) \nonumber \\
&\times&\sqrt{1-(r_{\rm orb} \omega_{\rm orb})^2} \frac{\phi^I}{r_{\rm orb}} \left[ \frac{e^{-i m \omega_{\rm orb} t}-e^{-i\text{sign}(m) \omega_{n'} t}}{W(m\omega_{\rm orb})}\right] \,,\nonumber
\eeq
again with $\phi^I=\phi^I(r_{\rm orb}; m \omega_{\rm orb})$. 
So, in this case, the energy density at the origin is
\beq
T_{t t}(r\to 0)&\sim& \left|\frac{e^{-i \omega_{\rm orb} t}-e^{-i\omega_{n'} t}}{W(\omega_{\rm orb})}\right|^2\nonumber \\
&\times&\frac{\omega_{\rm orb}^3}{16 \pi^3}\left(\frac{m_p \phi^I}{r_{\rm orb}}\right)^2\left[1-\left(r_{\rm orb} \omega_{\rm orb}\right)^2\right]\,.\nonumber
\eeq
Notice that, with this choice of coefficients $A_{l,n}^m$, the energy density $T_{t t}(r\to 0)$ is well-defined in the limit $\omega_{\rm orb}\to \omega_{n'}$, and is equal to
\beq
\lim_{\omega_{\rm orb}\to \omega_{n'}}&& T_{t t}(r\to 0)\sim \frac{t^2}{\left[W'(\omega_{n'})\right]^2}\times \nonumber \\
&&\frac{\omega_{n'}^3}{16 \pi^3}\left(\frac{m_p \phi^I(r_{\rm orb};\omega_{n'})}{r_{\rm orb}}\right)^2\left[1-\left(r_{\rm orb} \omega_{n'}\right)^2\right]\,,\nonumber
\eeq
where $W'(\omega)$ is the Wronskian derivative with respect to the frequency. Thus, one concludes that in a resonance the energy density at the origin increases quadratically with time.

\newpage

\bibliographystyle{apsrev4}

\bibliography{References}

\end{document}